\documentclass[useAMS,usenatbib]{mn2e}
\usepackage{graphicx}
\usepackage{psfig,epsfig}
\usepackage{deluxetable}

\usepackage[letterpaper,total={17.8cm,24.0cm},centering]{geometry}

\newcommand{\metal}{$[Z/H]$}
\newcommand{\thalf}{$t_{50}$}
\newcommand{\atlas}{{{ATLAS$^{\mathrm{3D}}$}}}
\newcommand{\sauron}{{\tt {SAURON}}}

\newcommand{\hb}{H$\beta$}
\newcommand{\mgb}{Mg$\,b$}
\newcommand{\mgfe}{[MgFe50]$^\prime$}
\newcommand{\afe}{[$\alpha$/Fe]}
\newcommand{\lambdar}{$\lambda_{\rm R}$}
\newcommand{\Nid}{[N{\small I}]$\lambda\lambda 5198,5200$}
\newcommand{\Ni}{[N{\small I}]}
\newcommand{\oiii}{[O{\small III}]}
\newcommand{\oiiid}{[O{\small III}]$\lambda\lambda 4959,5007$}
\newcommand{\sigmae}{$\sigma_{e}$}

\newcommand{\mjam}{$M_{\mathrm{JAM}}$}

\newcommand{\kms}{~\textrm{km}\,\textrm{s}$^{-1}$}
\newcommand{\kmsm}{~\textrm{km}\,\textrm{s}^{-1}}

\newcommand{\Rmaj}{$R^{\rm maj}_{e}$}
\newcommand{\Reff}{$R_{e}$}
\newcommand{\Reffe}{$R_{e}/8$}
\newcommand{\Refft}{$R_{e}/2$}

\newcommand{\lsim}{\mathrel{\hbox{\rlap{\hbox{\lower4pt\hbox{$\sim$}}}\hbox{$<$}}}}
\newcommand{\gsim}{\mathrel{\hbox{\rlap{\hbox{\lower4pt\hbox{$\sim$}}}\hbox{$>$}}}}

\newcommand{\msun}{$M_{\odot}$}

\title[]
  {The \atlas\/ Project - XXX. Star formation histories and stellar population scaling relations of early-type galaxies}

\author[R.\ M.\ McDermid et al.] {Richard M. McDermid$^{1,2}$\footnote{richard.mcdermid@mq.edu.au}, Katherine Alatalo$^{3}$, Leo Blitz$^{4}$, Fr\'ed\'eric Bournaud$^{5}$, \newauthor
Martin Bureau$^{6}$, Michele Cappellari$^{6}$, Alison F. Crocker$^{7}$, Roger L. Davies$^{6}$,  \newauthor
Timothy A. Davis$^{8}$, P. T. de Zeeuw$^{9,10}$, Pierre-Alain Duc$^{5}$, Eric Emsellem$^{9,11}$, \newauthor
Sadegh Khochfar$^{12}$, Davor Krajnovi\'c$^{13}$, Harald Kuntschner$^{9}$, Raffaella Morganti$^{14,15}$, \newauthor
Thorsten Naab$^{16}$, Tom Oosterloo$^{14,15}$, Marc Sarzi$^{16}$, Nicholas Scott$^{17}$, \newauthor
Paolo Serra$^{14,18}$, Anne-Marie Weijmans$^{19}$ and Lisa M. Young$^{20,21}$\\
$^{1}$Department of Physics and Astronomy, Macquarie University, Sydney NSW 2109, Australia\\
$^{2}$Australian Gemini Office, Australian Astronomical Observatory, PO Box 915, Sydney NSW 1670, Australia\\
$^{3}$Infrared Processing and Analysis Center, California Institute of Technology, Pasadena, California 91125, USA \\
$^{4}$Department of Astronomy, Campbell Hall, University of California, Berkeley, CA 94720, USA\\
$^{5}$Laboratoire AIM Paris-Saclay, CEA/IRFU/SAp -- CNRS -- Universit\'e Paris Diderot, 91191 Gif-sur-Yvette Cedex, France\\
$^{6}$Sub-Department of Astrophysics, Department of Physics, University of Oxford, Denys Wilkinson Building, Keble Road, Oxford, OX1 3RH, UK\\
$^{7}$Ritter Astrophysical Observatory, University of Toledo, Toledo, OH 43606, USA\\
$^{8}$Centre for Astrophysics Research, University of Hertfordshire, Hatfield, Herts AL1 9AB, UK\\
$^{9}$European Southern Observatory, Karl-Schwarzschild-Str. 2, 85748 Garching, Germany\\
$^{10}$Sterrewacht Leiden, Leiden University, Postbus 9513, 2300 RA Leiden, the Netherlands\\
$^{11}$Universit\'e Lyon 1, Observatoire de Lyon, Centre de Recherche Astrophysique de Lyon and Ecole Normale Sup\'erieure de Lyon, \\
\hspace{0.5cm} 9 avenue Charles Andr\'e, F-69230 Saint-Genis Laval, France\\
$^{12}$Institute for Astronomy, University of Edinburgh, Royal Observatory, Edinburgh, EH9 3HJ, UK\\
$^{13}$Leibniz-Institut f\"ur Astrophysik Potsdam (AIP), An der Sternwarte 16, D-14482 Potsdam, Germany\\
$^{14}$Netherlands Institute for Radio Astronomy (ASTRON), Postbus 2, 7990 AA Dwingeloo, The Netherlands\\
$^{15}$Kapteyn Astronomical Institute, University of Groningen, Postbus 800, 9700 AV Groningen, The Netherlands\\
$^{16}$Max-Planck-Institut f\"ur Astrophysik, Karl-Schwarzschild-Str. 1, 85741 Garching, Germany\\
$^{17}$Sydney Institute for Astronomy (SIfA), School of Physics, The University of Sydney, NSW 2006, Australia\\
$^{18}$CSIRO Astronomy \& Space Science, PO Box 76, Epping, NSW 1710, Australia\\
$^{19}$School of Physics and Astronomy, University of St Andrews, North Haugh, St Andrews KY16 9SS, UK\\
$^{20}$Physics Department, New Mexico Institute of Mining and Technology, Socorro, NM 87801, USA\\
$^{21}$Academia Sinica Institute of Astronomy \& Astrophysics, PO Box 23-141, Taipei 10617, Taiwan, R.O.C.\\
$^*$richard.mcdermid@mq.edu.au}

\date{Accepted 2015 January 15. Received 2015 January 15; in original form 2014 October 14\hfill}
 
\pagerange{\pageref{firstpage}--\pageref{lastpage}} \pubyear{0000}

\begin{document}

\label{firstpage}

\maketitle

\clearpage

\begin{abstract}

We present the stellar population content of early-type galaxies from the \atlas\/ survey. Using spectra integrated within apertures covering up to one effective radius, we apply two methods: one based on measuring line-strength indices and applying single stellar population (SSP) models to derive SSP-equivalent values of stellar age, metallicity, and alpha enhancement; and one based on spectral fitting to derive non-parametric star-formation histories, mass-weighted average values of age, metallicity, and half-mass formation timescales. Using homogeneously derived effective radii and dynamically-determined galaxy masses, we present the distribution of stellar population parameters on the Mass Plane (\mjam, \sigmae, \Rmaj), showing that at fixed mass, compact early-type galaxies are on average older, more metal-rich, and more alpha-enhanced than their larger counterparts.

From non-parametric star-formation histories, we find that the duration of star formation is systematically more extended in lower mass objects. Assuming that our sample represents most of the stellar content of today's local Universe, approximately 50\% of all stars formed within the first 2\,Gyr following the big bang. Most of these stars reside today in the most massive galaxies ($>10^{10.5}$\msun), which themselves formed 90\% of their stars by $z \sim$2. The lower-mass objects, in contrast, have formed barely half their stars in this time interval. Stellar population properties are independent of environment over two orders of magnitude in local density, varying only with galaxy mass. In the highest-density regions of our volume (dominated by the Virgo cluster), galaxies are older, alpha-enhanced and have shorter star-formation histories with respect to lower density regions.

\end{abstract}
%
%
\begin{keywords}
galaxies: abundances
galaxies: elliptical and lenticular, cD
galaxies: evolution
galaxies: stellar content
\end{keywords}
%
%
%
%
\section{Introduction}
\label{sec:intro}

Our understanding of the stellar populations in early-type galaxies (hereafter ETGs) has made significant progress in recent years, through a combination of advancements in both observations and stellar population models. These systems are often cited as ideal laboratories for uncovering the fossil record of galaxy formation processes thanks to the apparent lack of dust or star formation, which obfuscates the stellar continuum in objects of later type. However, the modern observational picture is not quite so straightforward. ETGs are now known to host some degree of young stellar populations \cite[e.g.][]{trager2000,kaviraj07,kuntschner10}. More surprising is the recent finding that sizable reservoirs of molecular gas can be found residing inside ETGs \cite[]{young2008}, coincident with evidence of PAH features in the infrared domain \cite[]{shapiro2010}, both of which strongly indicate ongoing star-formation within these galaxies, albeit at a rather low level \citep{davis14}.

In addition to a large range in ages, ETGs are also found to show strong correlations between their population parameters and mass-sensitive quantities such as luminosity and velocity dispersion \cite[e.g.][]{jorgensen99,trager2000,thomas05}. In general, these studies find that high-mass galaxies appear older, metal rich and with enhanced abundance ratios with respect to lower mass objects - a picture emphatically confirmed with very large samples of galaxies from modern surveys \cite[][]{bernardi06,gallazzi06,graves09a,thomas10}. In addition to this general behaviour, specific co-dependencies between population parameters after controlling for the general mass-dependency have also been proposed \cite[]{trager2000,smith2008,graves10}. Understanding what drives the large intrinsic scatter in the stellar population scaling relations is key to understanding the evolutionary processes involved.

Despite a deepening clarity in our observational view of ETGs, many open questions remain. For example, the role of young stars in these galaxies is still not clear. Although often referred to as `rejuvenated' galaxies, it is currently unclear if these objects have actually rekindled their star formation in recent times, or contrarily, are in the process of recently shutting down star formation for the first time. The role of environment is also somewhat unclear. Previously considered as important for ETGs in general \cite[][]{thomas05,schawinski07}, the interplay of mass and environment may also play a role \citep{thomas10}.

Here we present the stellar population properties of the \atlas\/ sample \citep{cappellari11a} using two approaches: one which uses line strengths to derive single stellar population- (SSP-) equivalent parameters to describe the age, metallicity and alpha-enhancement; and one which uses spectral fitting to obtain a non-parametric star formation history that can be used to derive mass-weighted average age and metallicity, and estimate the duration of star formation. We compare both approaches, which yield complementary information, and check for self-consistency between them. We then examine the trends of stellar populations with various other properties. The \atlas\/ project comprises a well-characterized sample with various multi-wavelength data in addition to the \sauron\/ integral field spectroscopy, making it a powerful data set to explore correlations between multiple properties.

This paper proceeds as follows: Section 2 describes the observations and analysis methods; Sections 3 and 4 present our population parameters on the mass-size plane, and against velocity dispersion and mass, respectively; Section 5 presents trends from our empirical star formation histories, and Section 6 concludes.

%
%
\section{Observations, line-strengths and analysis}
\label{sec:method}

Here we describe the optical spectroscopy used in our analysis. The same spectral data cubes were used for deriving the line strengths and for the spectral fitting, so calibration and removal of emission applies to both techniques.

%
%
\subsection{\sauron\/ observations and basic calibration}
\label{sec:observations}

The \sauron\/ spectrograph and basic data reduction is described in detail in \citet{bacon01}. Further details of the \atlas\/ \sauron\/ observations are given in \cite{cappellari11a}.  The details of our calibrations and line-strength measurements follow closely the techniques developed and described by \citet{kuntschner06}. Most of the 260 \atlas\/ \sauron\/ spectra were obtained after the spectrograph was upgraded with a volume-phase holographic (VPH) grating in 2004, except for 64 objects observed as part of the \sauron\/ Survey \citep{dezeeuw02} and other related projects. The pre- and post-VPH data are reduced and analysed self-consistently to respect the slightly different dispersion characteristics. The wavelength coverage of the pre- and post-VPH data are very similar, with only a slight difference in resolution. The adopted spectral resolution for pre-VPH data is 110\kms\/ compared to 99\kms\/ for VPH data. Observations of the almost featureless white dwarf EG131 were used to verify that the detailed transmission characteristics, including the medium-scale variations described in \citet{kuntschner06}, were present in the pre- and post-VPH data, agreeing to better than 1\%. Corrections for airmass were applied using the average La Palma extinction curve \citep{king85}.

A total of 67 different standard stars were observed during both the \sauron\/ and \atlas\/ campaigns in order to calibrate our line-strengths onto the standard Lick system \citep{worthey94}, as well as to monitor stability and reproducibility in our measurements (160 separate observations were made). Offsets to the Lick system are derived in Appendix \ref{app:stars}, along with a comparison of pre- and post-VPH values, a comparison with the MILES library \citep{blazquez06}, and discussion of the intrinsic uncertainties in our measurements. Appendix \ref{app:literature} presents a comparison of our line-strength measurements on the Lick system with \citet{gonzalez93} and \citet{trager98} for overlapping galaxies, showing good agreement given the uncertainties of matching apertures and differing treatment of emission.

%
%

\subsection{Emission correction}

The \sauron\/ spectral range includes several nebular emission lines, all of which lie close to or within the bandpasses of our line-strength indices. Namely: \hb, present in both emission and absorption; \oiiid, which strongly affects the Fe5015 index; and \Nid, which falls within the \mgb\/ index. As in \citet{kuntschner06}, we employ the technique developed by \cite{sarzi06}, which involves simultaneously finding the best linear combination of absorption- and emission-line templates whilst fitting non-linearly for the emission line kinematics. The kinematics of the various lines are tied to that of \oiii, being the more robustly measured line. The resulting emission-line fits are subtracted from the galaxy spectrum if the AoN (ratio of the amplitude of the fitted line to the local noise level in the spectrum, which includes fit residuals) is significant. Following \cite{sarzi06}, the AoN thresholds used are AoN$>4$ for \oiii, and then if this is detected, AoN$>3$ for \hb\/ and AoN$>4$ for \Ni.

The nebular emission lines are fitted and subtracted on the spatially-resolved data-cubes \citep[after spatially binning to a threshold signal-to-noise ratio of 40 per pixel with the Voronoi binning software of][]{cappellari03} before generating the aperture spectra. This largely avoids strongly non-Gaussian emission-line profiles (due to the rotation and/or distribution of the gas), which can leave significant residuals.

To estimate how strongly the various indices are affected by residual under- or over-subtracted emission, we use the spectral template fits to provide a good representation of the spectral lines in the absence of any emission. We then compare the observed indices with those measured on the template fits. The differences between the observed and template indices are typically similar to or smaller than the statistical errors on the line measurements. However, a few galaxies show significantly larger differences, due to imperfect emission correction in the presence of very strong emission lines with complex, non-Gaussian line profiles. Rather than use the index measured on the template, which is known to be biased toward solar abundance ratios, we adopt the {\em difference} between the observed and best-fit template indices as an estimate of the systematic errors. We furthermore take the statistical uncertainties from repeat observations of standard stars as a lower limit on the errors (see Table \ref{tab:errors} given in Appendix \ref{app:stars}).

%
%
\subsection{Aperture line strength measurements}
\label{sec:indices}

Figure 1 of \citet{emsellem11} shows the distribution of radii covered by our integral-field spectroscopic data, defined as the maximum circular radius, $R_{\rm max}$, that is at least 85\%\/ filled with spectra. Note that this fill factor includes lost spaxels removed due to e.g. foreground stars. The median maximal radius is $\sim$1.2 \Reff, with only 10 galaxies having coverage of less than \Reff/2. In terms of aperture coverage, more than half the sample has a filling factor of 85\% or more within a circle of \Reff, and all but 11 galaxies have this coverage at \Reff/2.

\begin{figure*}
 \begin{center}
  \epsfig{file=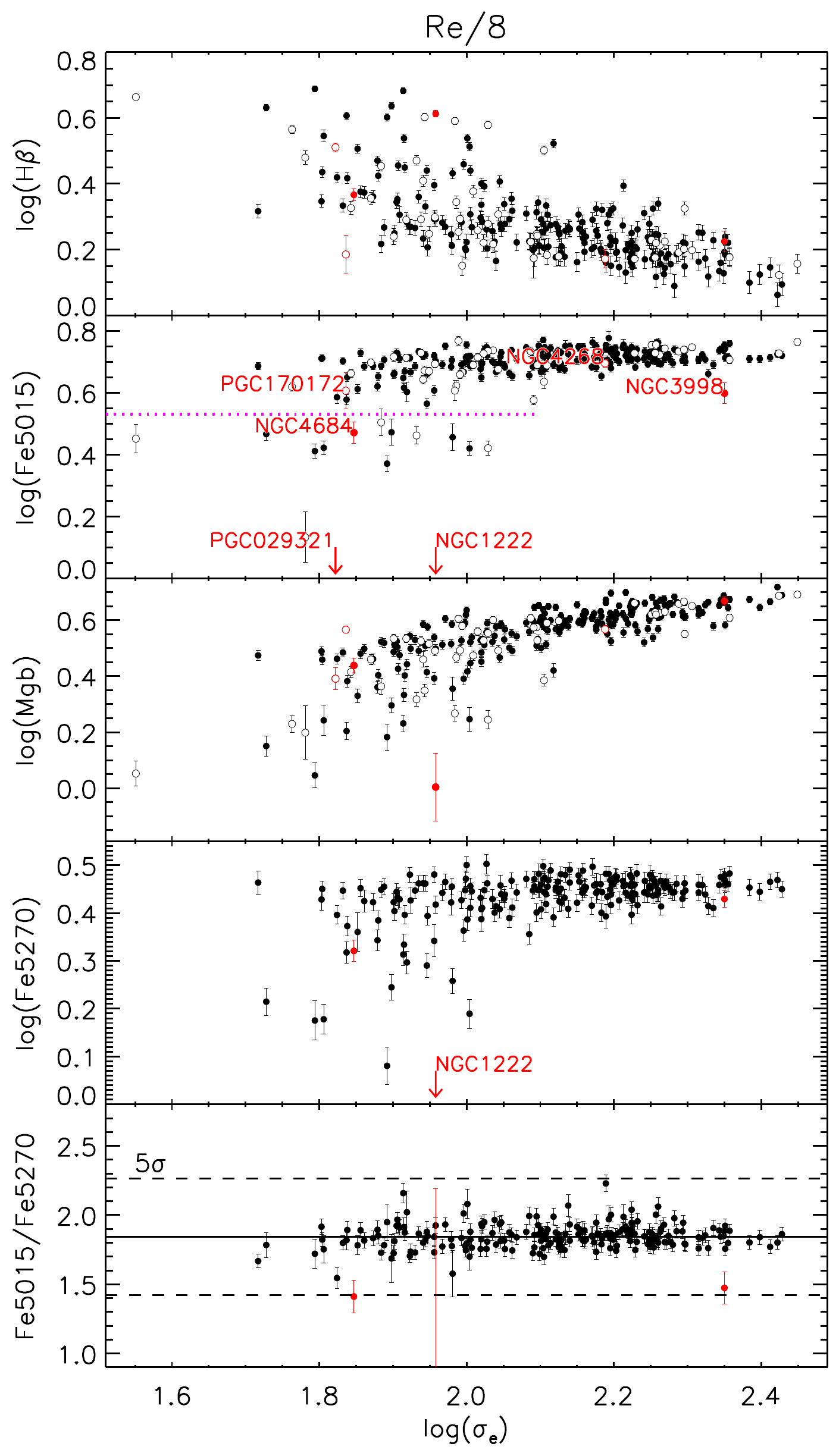, angle=0, width=12cm}
  \caption[]{Absorption line indices measured within \Reff/8 against the effective velocity dispersion, \sigmae\/ \cite[measured within one \Reff, taken from][]{cappellari13a}. Red symbols and text indicate problematic galaxies as described in the main text. The dotted magenta line in the Fe5015 plot indicates the separation of `low-Fe' objects discussed in the text, given here as $\log($Fe5015$)=0.53$. The solid line in the bottom panel indicates the biweight mean value, 1.84, with dashed lines showing the $\pm 5$ times the standard deviation around this value. Closed and open symbols in the upper panels correspond to galaxies with and without Fe5270 measurements respectively.}
  \label{fig:ls_sigma}
  \end{center}
\end{figure*}

\begin{figure}
 \begin{center}
  \epsfig{file=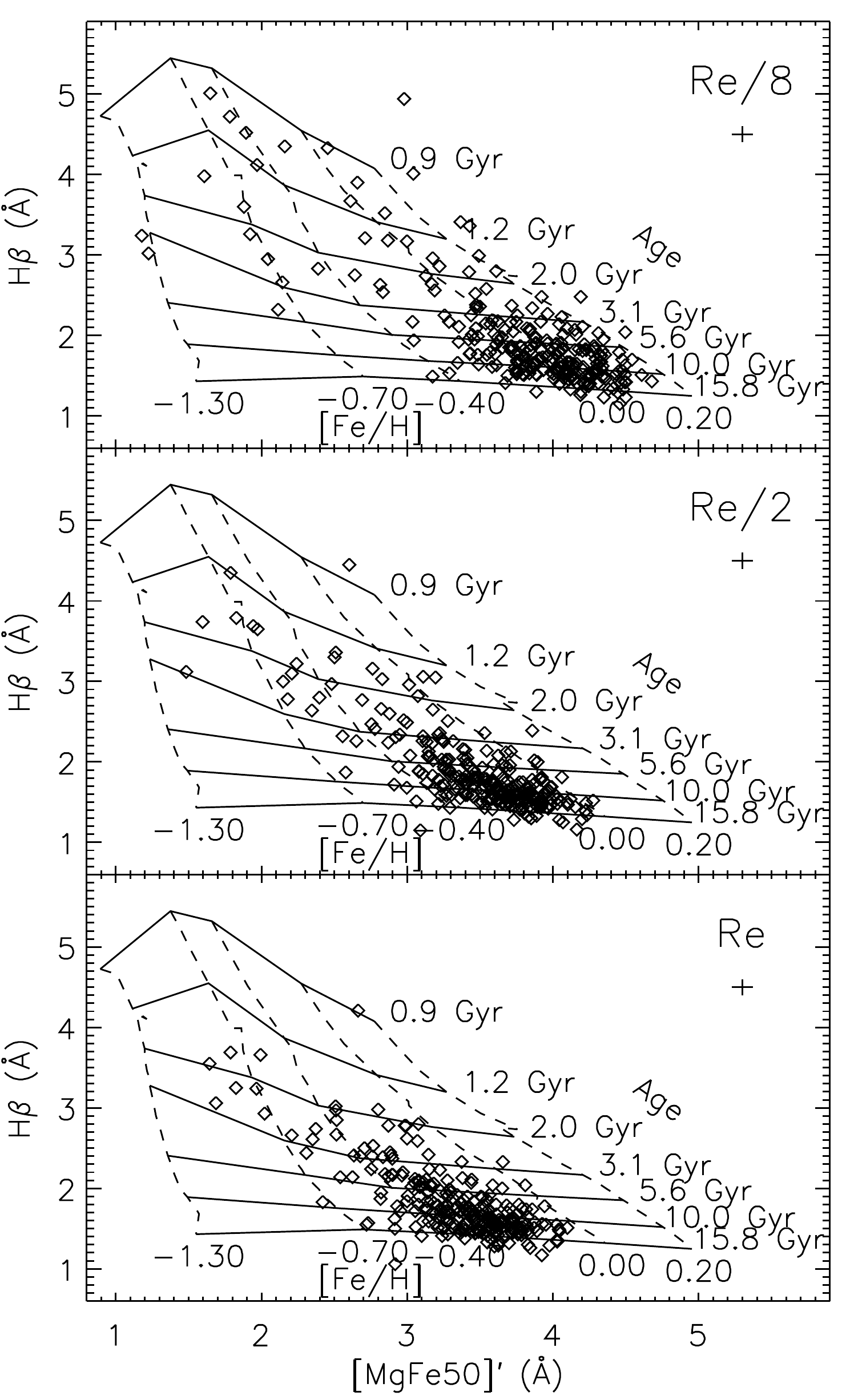, angle=0, width=8cm}
  \caption[]{Index-index diagram for our sample using the indices \hb\/ versus \mgfe, overplotted with model predictions from \citet{schiavon07} for single stellar populations with constant ages, metallicities and solar abundance ratios. From top to bottom, each panel uses indices measured within different circular apertures, with radii of \Reff/8, \Reff/2 and \Reff\/ as indicated. An error bar in each panel indicates the statistical errors.}
  \label{fig:index_plane}
  \end{center}
\end{figure}

We measure line strengths in three different circular aperture sizes: \Reff/8, \Reff/2 and \Reff. The smallest of these is comparable to aperture sizes commonly used in conventional long-slit investigations \citep[e.g.][]{trager2000,thomas05}. The \Reff/2 aperture is our largest common aperture that requires minimal aperture corrections in a small number of objects. The \Reff\/ aperture is the most representative `global' value that we can derive based on our data.

To generate the aperture measurements from the data-cubes, we use an analogy of `growth curves', co-adding spectra within increasing circular aperture sizes and measuring the indices at each step. We assign a characteristic radius to each aperture spectrum that represents the radius of a circle having the same area as that of the included \sauron\/ data, computed from the actual bins within the aperture. Thus, where the aperture is well-filled by the \sauron\/ data, this radius is similar to the radius of the desired circular aperture. But as the aperture becomes more poorly filled near the edge of the field, the characteristic radius will tend to a fixed value (the area of the full observed field) regardless of the target aperture size. By using a threshold filling factor of 85\%, the difference between the target aperture and delivered area is small.

For the galaxies where the maximum characteristic radius is smaller than the target aperture radius, an aperture correction is applied to the measured indices. Due to the presence of age gradients that systematically vary in strength with galaxy age, we derive age-dependent aperture corrections. Appendix \ref{app:appcor} gives details of this correction and of the coefficients to apply such corrections. The aperture corrections for our sample are typically 2\% or less, with a maximum of 8\% for our largest correction, and our overall results do not depend on these corrections.

The line indices themselves are measured on the Lick/IDS system \citep{worthey94} following the standard treatment of broadening the spectra as per the Lick broadening function. We account for the dilution of the line strength due to velocity broadening by measuring the equivalent indices on the broadened and unbroadened optimal template fit, thus accounting for line-strength and LOSVD dependencies of the velocity dispersion corrections \citep{kuntschner06,mcdermid06}. We measure the following indices in all galaxies and aperture sizes: \hb, Fe5015 and \mgb. The \sauron\/ wavelength range has a red cut-off wavelength that changes as a function of position in the field, causing \cite{kuntschner06} to define a new index, Fe5270$_{\rm S}$, which linearly traces the Fe5270 Lick index. We use this modified index to estimate the standard Lick Fe5270 index for a subset of objects and apertures where the \sauron\/ wavelength coverage allows.\footnote{We derive Fe5270 using the empirical relation  ${\rm Fe5270} = 1.28 \times {\rm Fe}5270_{\rm S} + 0.03$ from \cite{kuntschner06}} The line strengths measured within the three aperture sizes considered are given in Tables \ref{tab:re8}--\ref{tab:re}, which are available in full via the online journal, and also via our project webite http://purl.com/atlas3d.

Figure \ref{fig:ls_sigma} presents our line indices measured within one of the three apertures, \Reff/8, plotted as a function of the velocity dispersion measured within one \Reff, $\log($\sigmae) taken from \cite{cappellari13a}. Filled symbols denote the galaxies where Fe5270 could be measured within the given aperture size, with the remainder plotted as open symbols.

As found by numerous other authors \cite[e.g.][]{trager2000, thomas05}, there are rather tight, positive correlations between the metal-sensitive lines and velocity dispersion, and the age-sensitive \hb\/ index shows a negative trend with \sigmae. All indices show an increased scatter at lower \sigmae\/ values, implying a greater variety of apparent stellar population properties at lower masses. 

The Fe5015 index lies on top of the strong \oiii\/ emission line at 5007\AA. For this index, the apparent scatter is driven by a number of galaxies sitting below the main relation. To test whether this is the result of infill from poorly-subtracted \oiii\/ emission or genuinely weaker iron, we use the fact that some galaxies have a measurable Fe5270 index, which is free from nebular emission contamination. Figure \ref{fig:ls_sigma} shows several galaxies where Fe5270 also lies below the general relation, but the ratio of the two iron indices (bottom panel of Figure \ref{fig:ls_sigma}) is consistent with the sample average, demonstrating that the weak Fe5015 measurements in these cases are not due to poor emission-line correction. We discuss these `low-Fe' objects further in Section \ref{sec:sfhcomp}.

We highlight a few objects in Figure \ref{fig:ls_sigma} that need special treatment. Firstly, two objects had significantly lower signal-to-noise ratio than the rest of the sample, namely NGC\,4268 and PGC\,170172. These two objects are not considered further in our analysis. In addition, the following objects show large errors in their Fe5015 values due to imperfect subtraction of the very strong and non-Gaussian \oiii\/ emission line, as indicated by their spuriously low Fe5015/Fe5270 ratio in Figure \ref{fig:ls_sigma}: NGC\,1222 \citep[starburst galaxy,][]{balzano83}, NGC\,4684, NGC\,5273, and NGC\,3998. These latter two objects are known LINERs with a broad nuclear emission line component \citep{veroncetty10} that only affects the small \Reffe\/ aperture size, and so we include them when considering the larger aperture measurements. NGC\,1222 is excluded for all aperture sizes. PGC\,029321 has no Fe5270 measurement, but shows large discrepancies between the Fe5015 index measured on the observed and fitted spectrum due to emission residuals, and so is also excluded in our analysis. For all the problematic cases mentioned here, we still provide measured values in Tables \ref{tab:re8}--\ref{tab:re} where possible, but flag them in the `quality' column.

\begin{table*}
\caption{Measured Lick index measurements and SSP parameters measured within \Reffe.}
\begin{tabular}{lcccccccc}
\hline \hline
Name & \hb\/ [\AA] & Fe5015\/ [\AA]  & \mgb\/ [\AA]  & Fe5270\/ [\AA] & Age$_{\rm SSP}$\/ [Gyr] & [Z/H]$_{\rm SSP}$  & \afe$_{\rm SSP}$ &  Quality \\
 &  \Reffe & \Reffe &  \Reffe &  \Reffe &  \Reffe &  \Reffe &  \Reffe &   \\
(1) & (2) & (3) & (4) & (5) & (6) & (7) & (8) & (9) \\
\hline
IC0560 &  2.95 $\pm$ 0.10 &  4.17 $\pm$ 0.14 &  2.29 $\pm$ 0.12 &  2.20 $\pm$ 0.11 &  1.85 $\pm$ 0.21 & -0.11$ \pm$ 0.05 &  0.09 $\pm$ 0.05 &  1 \\ 
IC0598 &  2.88 $\pm$ 0.10 &  4.64 $\pm$ 0.14 &  2.46 $\pm$ 0.12 &  2.31 $\pm$ 0.12 &  1.70 $\pm$ 0.22 &  0.04 $\pm$ 0.05 &  0.05 $\pm$ 0.04 &  1 \\ 
IC0676 &  3.51 $\pm$ 0.15 &  2.64 $\pm$ 0.14 &  1.75 $\pm$ 0.23 &  1.51 $\pm$ 0.11 &  2.02 $\pm$ 0.23 & -0.64 $\pm$ 0.08 &  0.37 $\pm$ 0.11 &  2 \\ 
IC0719 &  2.16 $\pm$ 0.10 &  5.14 $\pm$ 0.14 &  3.26 $\pm$ 0.12 &  2.75 $\pm$ 0.11 &  3.53 $\pm$ 0.68 &  0.03 $\pm$ 0.05 &  0.11 $\pm$ 0.04 &  1 \\ 
IC0782 &  2.27 $\pm$ 0.10 &  4.98 $\pm$ 0.14 &  2.88 $\pm$ 0.12 &    -- $\pm$   -- &  3.10 $\pm$ 0.36 & -0.03 $\pm$ 0.06 &  0.06 $\pm$ 0.05 &  1 \\ 
IC1024 &  4.00 $\pm$ 0.10 &  2.35 $\pm$ 0.14 &  1.52 $\pm$ 0.17 &  1.20 $\pm$ 0.11 &  1.59 $\pm$ 0.18 & -0.75 $\pm$ 0.06 &  0.41 $\pm$ 0.09 &  2 \\ 
\hline
\vspace{-0.2cm}
\label{tab:re8}
\end{tabular}

Columns: (1) Galaxy name. (2)-(5) Lick index measurements within \Reffe, measured on the Lick/IDS system. Fe5270 is measured indirectly via the Fe5270$_{\rm S}$ by applying the relation ${\rm Fe5270} = 1.28 \times {\rm Fe}5270_{\rm S} + 0.03$ \citep{kuntschner06}. Due to field coverage issues, this index cannot be measured for all objects. (6)-(8) Estimates of SSP-equivalent age, metallicity and abundance ratio within a circular aperture of radius \Reffe\/ using the SSP models of \cite{schiavon07} and using the three indices \hb, Fe5015, and \mgb. (9) Quality flag: 1 = Data are of good quality; 2 = Object with weak Fe5015 index in the \Reffe\/ aperture; 3 = Strong emission line residuals; 4 = low signal-to-noise ratio. The full version of this table with all 260 galaxies is made available via the online journal, and via our project website http://purl.com/atlas3d.\newline

\caption{Measured Lick index measurements and SSP parameters measured within \Refft.}
\begin{tabular}{lcccccccc}
\hline \hline
Name & \hb\/ [\AA] & Fe5015\/ [\AA]  & \mgb\/ [\AA]  & Fe5270\/ [\AA] & Age$_{\rm SSP}$\/ [Gyr] & [Z/H]$_{\rm SSP}$  & \afe$_{\rm SSP}$ &  Quality \\
 &  \Refft & \Refft &  \Refft &  \Refft &  \Refft &  \Refft &  \Refft &   \\
(1) & (2) & (3) & (4) & (5) & (6) & (7) & (8) & (9) \\
\hline
IC0560 &  2.53 $\pm$ 0.10 &  3.86 $\pm$ 0.14 &  2.32 $\pm$ 0.12 &  2.15 $\pm$ 0.11 &  2.67 $\pm$ 0.31 & -0.30 $\pm$ 0.05 &  0.14 $\pm$ 0.06 &  1 \\ 
IC0598 &  2.43 $\pm$ 0.10 &  4.20 $\pm$ 0.14 &  2.47 $\pm$ 0.12 &  2.22 $\pm$ 0.11 &  2.84 $\pm$ 0.33 & -0.22 $\pm$ 0.05 &  0.11 $\pm$ 0.05 &  1 \\ 
IC0676 &  3.02 $\pm$ 0.10 &  3.25 $\pm$ 0.14 &  1.92 $\pm$ 0.12 &  1.75 $\pm$ 0.11 &  2.20 $\pm$ 0.25 & -0.48 $\pm$ 0.06 &  0.22 $\pm$ 0.07 &  2 \\ 
IC0719 &  2.43 $\pm$ 0.10 &  4.16 $\pm$ 0.14 &  2.69 $\pm$ 0.12 &  2.46 $\pm$ 0.11 &  2.84 $\pm$ 0.33 & -0.18 $\pm$ 0.04 &  0.18 $\pm$ 0.05 &  1 \\ 
IC0782 &  2.08 $\pm$ 0.10 &  4.51 $\pm$ 0.20 &  2.58 $\pm$ 0.12 &    -- $\pm$   -- &  4.77 $\pm$ 0.85 & -0.25 $\pm$ 0.05 &  0.03 $\pm$ 0.06 &  1 \\ 
IC1024 &  3.59 $\pm$ 0.15 &  2.25 $\pm$ 0.14 &  1.66 $\pm$ 0.20 &  1.30 $\pm$ 0.11 &  2.06 $\pm$ 0.24 & -0.78 $\pm$ 0.07 &  0.50 $\pm$ 0.05 &  2 \\ 
\hline
\vspace{-0.2cm}
\label{tab:re2}
\end{tabular}

Columns are as per Table \ref{tab:re8} but for an \Refft\/ aperture size. The full version of this table with all 260 galaxies is made available via the online journal, and via our project website http://purl.com/atlas3d.\newline

\caption{Measured Lick index measurements and SSP parameters measured within \Reff.}
\begin{tabular}{lcccccccc}
\hline \hline
Name & \hb\/ [\AA] & Fe5015\/ [\AA]  & \mgb\/ [\AA]  & Fe5270\/ [\AA] & Age$_{\rm SSP}$\/ [Gyr] & [Z/H]$_{\rm SSP}$  & \afe$_{\rm SSP}$ &  Quality \\
 &  \Reff & \Reff &  \Reff &  \Reff &  \Reff &  \Reff &  \Reff &   \\
(1) & (2) & (3) & (4) & (5) & (6) & (7) & (8) & (9) \\
\hline
IC0560 &  2.37 $\pm$ 0.10 &  3.75 $\pm$ 0.16 &  2.31 $\pm$ 0.12 &  2.10 $\pm$ 0.10 &  3.53 $\pm$ 0.58 & -0.38 $\pm$ 0.06 &  0.19 $\pm$ 0.07 &  1 \\ 
IC0598 &  2.30 $\pm$ 0.10 &  4.03 $\pm$ 0.14 &  2.41 $\pm$ 0.12 &  2.16 $\pm$ 0.11 &  3.38 $\pm$ 0.57 & -0.30 $\pm$ 0.05 &  0.12 $\pm$ 0.05 &  1 \\ 
IC0676 &  2.66 $\pm$ 0.10 &  3.37 $\pm$ 0.14 &  1.96 $\pm$ 0.13 &  1.79 $\pm$ 0.11 &  2.84 $\pm$ 0.38 & -0.51 $\pm$ 0.06 &  0.21 $\pm$ 0.07 &  2 \\ 
IC0719 &  2.35 $\pm$ 0.10 &  3.84 $\pm$ 0.14 &  2.56 $\pm$ 0.12 &  2.33 $\pm$ 0.10 &  3.38 $\pm$ 0.60 & -0.29 $\pm$ 0.05 &  0.22 $\pm$ 0.06 &  1 \\ 
IC0782 &  2.03 $\pm$ 0.10 &  4.47 $\pm$ 0.27 &  2.50 $\pm$ 0.12 &    -- $\pm$   -- &  5.31 $\pm$ 0.85 & -0.30 $\pm$ 0.06 &  0.01 $\pm$ 0.07 &  1 \\ 
IC1024 &  3.33 $\pm$ 0.18 &  2.47 $\pm$ 0.14 &  1.71 $\pm$ 0.16 &  1.39 $\pm$ 0.12 &  2.29 $\pm$ 0.42 & -0.73 $\pm$ 0.06 &  0.43 $\pm$ 0.07 &  2 \\ 
\hline
\vspace{-0.2cm}
\label{tab:re}
\end{tabular}

Columns are as per Table \ref{tab:re8} but for an \Reff\/ aperture size. The full version of this table with all 260 galaxies is made available via the online journal, and via our project website http://purl.com/atlas3d.

\end{table*}

\subsection{Deriving SSP parameters}

As in \cite{mcdermid06}, we make use of a customized version of the \citet{schiavon07} models where [Ti/Fe] $= 0.5 \times$ [Mg/Fe]. These models were found to give better agreement between abundance ratios computed with Fe5015 and Fe5270 \cite[see][for details]{kuntschner10}, and this relative scaling of [Ti/Fe] and [Mg/Fe] is also approximately consistent with the measured abundances for early-type galaxies from \citet{johansson12}. The models make predictions of the Lick indices for a grid of various ages, metallicities and alpha-element enhancements. Each position in this grid represents the prediction for a single population of stars sharing these parameters.

We show in Figure \ref{fig:index_plane} the distribution of our full sample in the indices \hb\/ versus \mgfe\footnote{$[\mathrm{MgFe50}]^\prime = \frac{0.69 \times \mathrm{Mg}b + \mathrm{Fe5015}}{2}$ \citep{kuntschner10}} for the three aperture sizes, overplotted with these model predictions using \afe$=0$. The effects of abundance ratio are largely mitigated with this choice of indices \citep{thomas03}. The models have been adapted for zero-point differences between the Lick system and the Schiavon models using Table 1 of \citet{schiavon07}.

From this figure, it can be seen how the changing distribution of \hb\/ for the different apertures implies that the young stars  in general are centrally concentrated. It can also be seen that the inferred metallicity becomes lower for larger apertures, due to the inclusion of the metal-poor outer regions. We also note that several objects have implied ages that are older than the canonical age of the Universe from 
Planck \cite[$13.798\pm0.037$\,Gyr;][]{planckXVI}. In Appendix \ref{app:old_ages} we show that these points are consistent with objects close to this upper age limit considering our observational uncertainties.

To derive the best-fitting SSP parameters for our data, we use the three indices that are measured across the full field for all galaxies, namely \hb, Fe5015 and \mgb. We employ the straightforward yet robust technique of finding the SSP model that best predicts our observed line indices simultaneously. As per \citet{mcdermid06} and \citet{kuntschner10}, we do this by means of chi-squared fitting: effectively computing the `distance' from our measured indices to all the predicted values of those indices, and finding the model with the minimum total distance. To reduce the effects of grid discretization, we oversample the original models using linear interpolation. Measurement errors on the indices are included as weighting factors in the sum. This particular technique has been used by several authors \citep{proctor04,thomas10}, although there are other methods for inverting the model grids to derive parameter values \citep[e.g.][]{graves08}.

We estimate the parameter errors by taking the range of points at which the difference in chi-square from the minimum is equal to unity, giving an estimate of the projected dispersion on each parameter. This method is computationally efficient, and we verified that it agrees well with a Monte-Carlo approach. All SSP values are given in Tables \ref{tab:re8}--\ref{tab:re}, which are available in full via the online journal, and via our project website: http://purl.com/atlas3d..

\subsection{Mass-weighted parameters from spectral fitting}
\label{sec:specfit}

We use the publicly-available\footnote{http://purl.org/cappellari/idl} penalized pixel fitting code \cite[pPXF, ][]{cappellari04} to fit a linear combination of SSP model spectra from the MIUSCAT model library \citep{vazdekis12}, spanning a regular grid of $\log$(age) (0.1 - 14\,Gyr) and metallicity ($[Z/H]=-1.71 - 0.22$, equivalent to $Z=0.0004 - 0.03$), giving 264 templates in total. Due to the degeneracy of inferring the star formation history from the integrated galaxy light, we make use of linear regularization \citep{press92} to impose a smoothness constraint on the solution, using the `REGUL' option of pPXF. The solution is defined by the weights applied to each model in a regularly-sampled grid of age and metallicity, and regularisation imposes a constraint in this parameter-space such that models with neighbouring age and metallicity must have weights that vary smoothly. The relative importance of this smoothness constraint compared to the actual fit quality is controlled via a simple `regularisation parameter', which essentially controls how smooth the solution is. The SSP model spectra are provided to an initial birth cloud mass of 1 solar mass, and so the relative `zero-age' mass-to-light ratios are included in the fitted weights. The distribution of weights from pPXF therefore defines the relative contributions in (zero-age) mass of the different populations: effectively the star formation history (SFH) for that galaxy.

For each galaxy, we fit the spectrum integrated within one effective radius as per the largest aperture considered for the SSP analysis. We derive solutions for a range of regularization parameter values, and select the value that gives a resulting best-fit solution with $\Delta \chi^{2} \sim \sqrt{2N}$, where $N$ is the number of pixels in the spectrum being fitted, and $\Delta\chi^2$ is computed as the difference between the $\chi^2$ of the current solution with that of the non-regularized case. This criterion is suggested by \citet{press92} as a robust indication of when the fit has become unacceptable. In this way, we are deriving the {\it smoothest} star formation history that still reproduces the \sauron\/ spectrum. The actual star formation histories of galaxies are likely not smooth. Star formation is inherently stochastic when considered with sufficient temporal and spatial resolution, and specific events such as mergers and interactions may trigger or truncate the star formation process on short timescales.  In fact, we have also shown that the colors and \hb\/ indices of the bluest and most H2-rich galaxies can be reproduced with a small amount of current or recent star formation activity superposed on an old stellar population \citep{young14}. However, inferring such events from the integrated light is highly degenerate for typical observational data, and the smoothness constraint offers a robust and objective result that can be used to study systematic trends in the entire sample. We also note that non-smooth solutions are in no way excluded by our approach, but if a smooth solution can give a fit of similar quality (as determined via $\Delta\chi^2$), then the smooth solution will be adopted.

The mass-weighted age and metallicity are given as:

\begin{equation}
\label{eq:age_sfh}
\log(\mathrm{Age}_{\rm SFH}) = \frac{\sum w_i \log(t_{{\rm SSP},i})}{\sum w_i} \hspace{0.3cm}
\end{equation}

and

\begin{equation}
 \mathrm{[Z/H]}_{\rm SFH} = \frac{\sum w_i \mathrm{[Z/H]}_{{\rm SSP},i}} {\sum w_i} \mathrm{,}
\end{equation}

\noindent respectively, where $w_i$ denotes the weight given by pPXF to the $i$'th template, which has age $t_{{\rm SSP},i}$ and metallicity $[Z/H]_{{\rm SSP},i}$. The grid of model ages is sampled logarithmically, hence the logarithmic form of equation \ref{eq:age_sfh}. Errors on these quantities were computed via Monte-Carlo simulations {\it without} regularisation.

The SSP spectral templates we use have fixed (solar) abundance ratios, and so we cannot derive a mass-weighted \afe. We can, however, derive a measure of the star formation timescale directly from the SFH itself. The measure we use is the time, \thalf, taken to form 50\% of the current day stellar mass within one \Reff, and is derived from the cumulative function of the SFH (see section \ref{sec:sfh}). All mass-weighted quantities are given in Table \ref{tab:mwvalues}.

\begin{table}
\caption{Mass-weighted stellar population properties measured within \Reff. The full version of this table with all 260 galaxies is made available via the online journal, and via our project website http://purl.com/atlas3d.}
\begin{tabular}{lccc}
\hline \hline
Name & Age$_{\rm SFH}$\/ [Gyr] & [Z/H]$_{\rm SFH}$  & \thalf\/ [Gyr]  \\
 &  \Reff & \Reff &  \Reff \\
(1) & (2) & (3) & (4) \\
\hline
IC0560 &  6.73 $\pm$ 1.07 & -0.46 $\pm$ 0.06 &  6.55 $\pm$ 1.07 \\ 
IC0598 &  5.37 $\pm$ 0.99 & -0.40 $\pm$ 0.05 &  8.60 $\pm$ 0.99 \\ 
IC0676 &  6.00 $\pm$ 1.92 & -0.55 $\pm$ 0.18 &  7.22 $\pm$ 1.92 \\ 
IC0719 &  9.21 $\pm$ 0.86 & -0.25 $\pm$ 0.08 &  4.23 $\pm$ 0.86 \\ 
IC0782 &  6.16 $\pm$ 0.83 & -0.19 $\pm$ 0.11 &  7.09 $\pm$ 0.83 \\ 
IC1024 &  4.39 $\pm$ 0.63 & -0.77 $\pm$ 0.15 &  8.65 $\pm$ 0.63 \\ 
\hline
\label{tab:mwvalues}
\end{tabular}

Columns: (1) Galaxy name. (2) Mass-weighted age. (3) Mass-weighted metallicity. (4) Half-mass formation time. These properties were derived from a regularised spectral fit to a spectrum integrated within \Reff. Errors were computed via Monte-Carlo realisations {\em without} regularisation constraints. See text for details. 

\end{table}

\subsection{Comparison of SFHs with SSPs and SDSS}
\label{sec:sfhcomp}

\begin{figure}
\begin{center}
  \epsfig{file=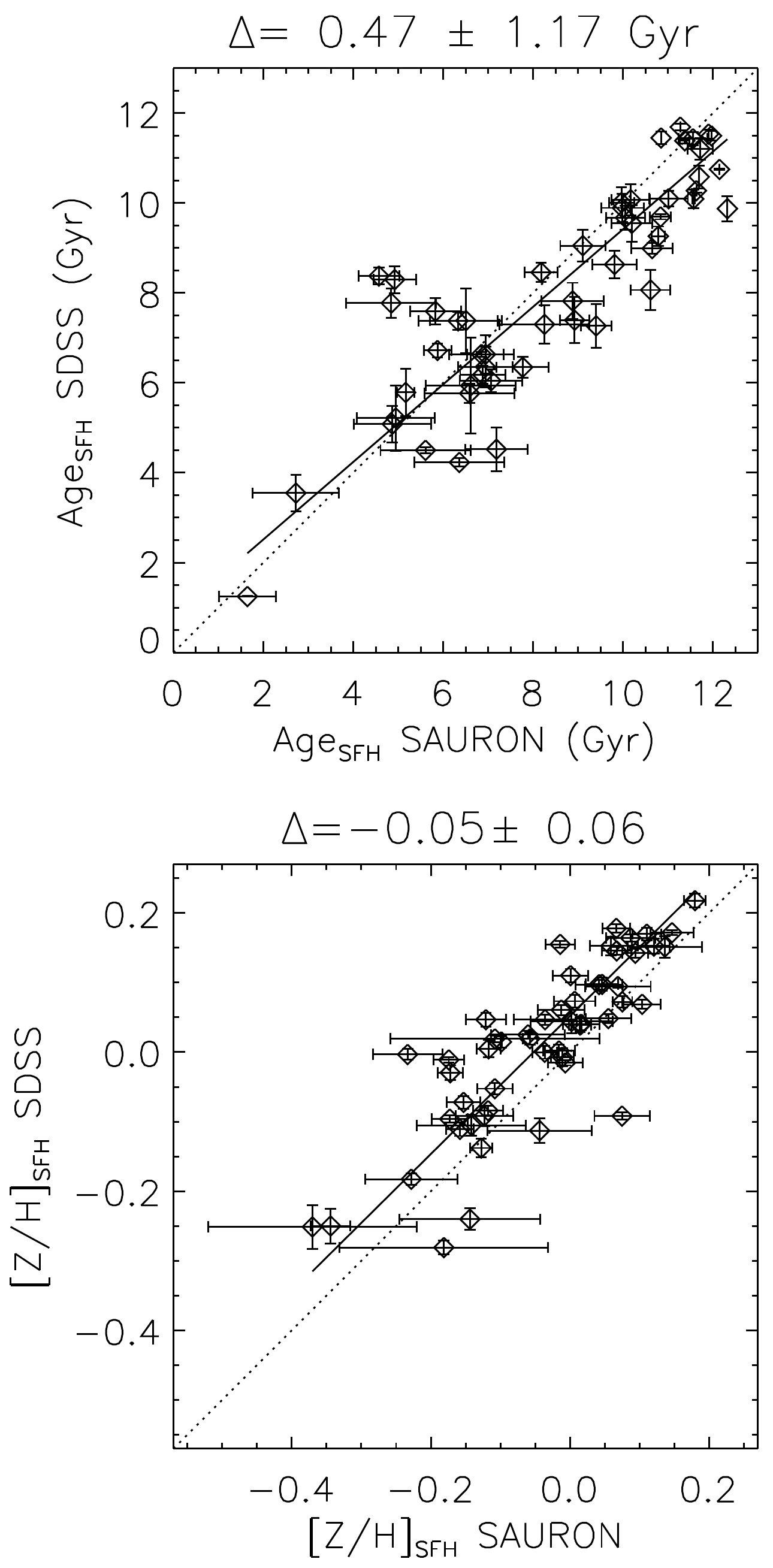, angle=0, width=6cm}
  \caption[]{Comparison of mass-weighted age (top) and metallicity (bottom) derived from fitting \sauron\/ and SDSS spectra within the same spatial aperture. The mean difference (\sauron\/-SDSS) and the standard deviation are given in the plot titles. The dotted lines indicate the 1:1 identity line, with a thick line showing the best linear fit. Errors were computed via Monte-Carlo simulations {\it without} regularisation.}
  \label{fig:comp_sfh_sdss}
  \end{center}
\end{figure}

\begin{figure*}
 \begin{center}
  \epsfig{file=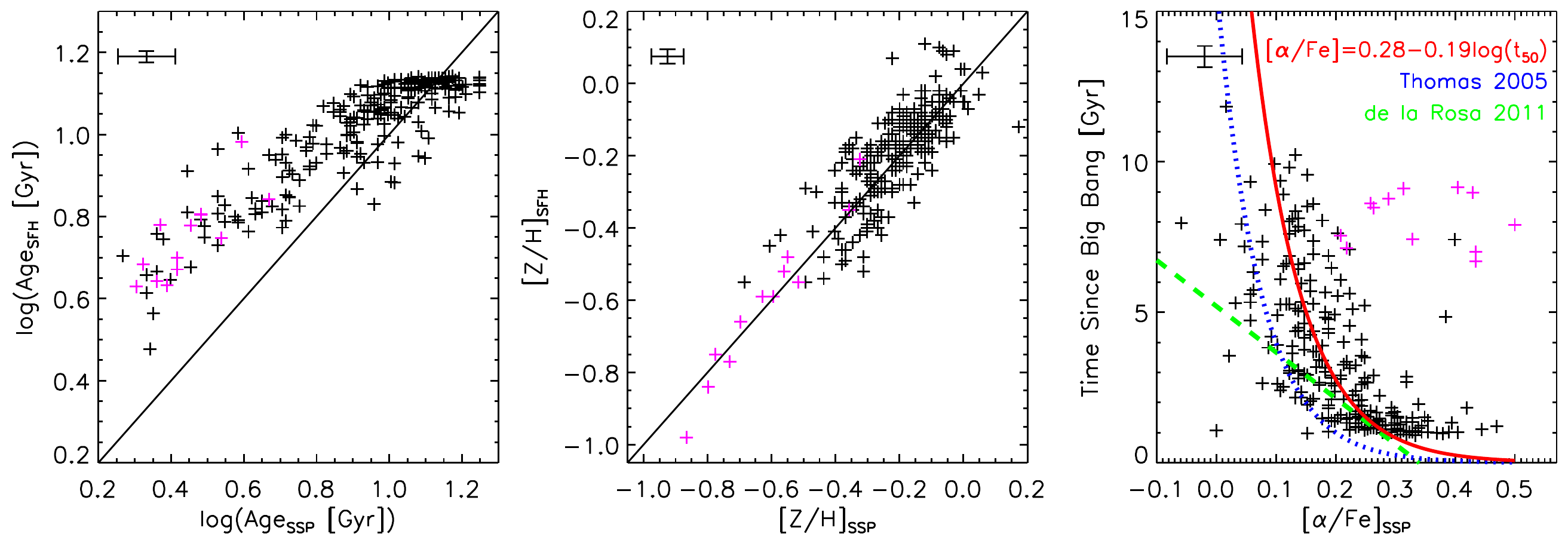, angle=0, width=17.5cm}
  \caption[]{Comparison of SSP-equivalent parameters (x-axis) and mass-weighted parameters derived using spectral fitting (y-axis) within the same aperture of one effective radius \Reff. Magenta points highlight the objects with low Fe as discussed in Section \ref{sec:indices}. Identity lines are given for age and metallicity. The right panel shows \afe\/ versus \thalf, the time taken to form half the current-day mass. The relation $t_{50}=-15.3[\alpha/Fe]+5.2$ given by eqn. 2 in \cite{delarosa11} is over plotted on our data with a green dashed line. The blue dotted line shows the canonical conversion of abundance ratio to star formation timescale from equation (4) of \cite{thomas05} based on chemical evolution models: $[\alpha/Fe]\sim\frac{1}{5} - \frac{1}{6} \log\Delta t$. The red solid line and text indicate our best-fitting function of this same form, giving an empirical relation between \afe\/ and \thalf\/ for the \atlas\/ sample.}
  \label{fig:comp_ssp_sfh}
  \end{center}
\end{figure*}

\begin{figure*}
 \begin{center}
  \epsfig{file=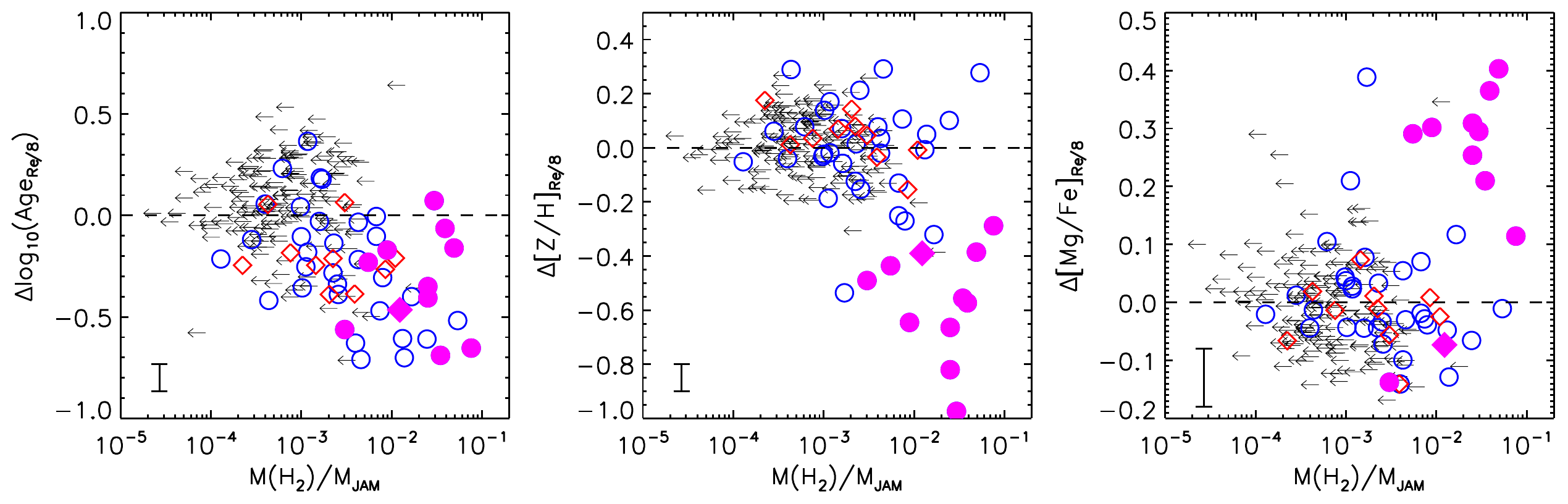, angle=0, width=17.5cm}
  \caption[]{Residuals from the SSP-velocity dispersion trends plotted as a function of molecular gas fraction. Median errors are indicated in the bottom left corner of each panel, with H$_2$ mass fraction errors within the plotting symbol. Arrows indicate 3$\sigma$ upper-limits for the H$_2$ mass. Diamonds denote CO detections in the Virgo cluster. Open circles show detections of non-Virgo objects. Filled magenta symbols correspond to the `low-Fe' objects discussed in Section \ref{sec:indices}. Their weak Fe5015 indices result in low metallicity and high \afe. We speculate that these objects have had recent gas accretions, giving rise to young, metal poor stars that have not yet been enriched with ejecta from Type 1a supernovae. Only one of these low-Fe objects is found in the Virgo cluster (filled magenta diamond).}
  \label{fig:co_pop}
  \end{center}
\end{figure*}

An obvious concern in deriving the SFH from \sauron\/ spectra is the short wavelength range available. To test this, we compare the SFHs derived from \sauron\/ with those derived from fitting SDSS spectroscopy \citep[DR8,][]{york2000} of the galaxies in common, using the entire SDSS wavelength range to constrain the SFH - a more than ten-fold increase in the wavelength coverage of the spectrum. For this test, we adapt the \sauron\/ aperture to match the 3\arcsec\/ diameter SDSS fiber, thus minimizing the effects of population gradients. Figure \ref{fig:comp_sfh_sdss} shows a comparison of the \sauron\/ and SDSS mass-weighted average ages and metallicities derived by applying the above procedure. The mass-weighted parameters compare very well, showing no significant systematic biases across the range of age and metallicity considered despite the dramatically different spectral ranges used. Age shows a slight trend to older ages derived by \sauron, but the oldest case is still consistent with SDSS within the 1$\sigma$ scatter. Metallicity is also very consistent, albeit with a small average offset, but again within the 1$\sigma$ scatter. Overall, the agreement is remarkably good, and suggests that the short wavelength range of \sauron\/ does not significantly bias our derived SFHs.

We compare our mass-weighted and SSP-equivalent parameters in Figure \ref{fig:comp_ssp_sfh}, this time using a 1\Reff\/ aperture. The mass-weighted ages are systematically older for all but the oldest ages, where we are biased by the lower maximum age of our spectral template library. Including templates up to 17\,Gyr reduces the `bunching' of points at oldest ages, but for clarity of interpreting our SFHs, we use a maximum age of 14\,Gyr in our template library, consistent with current cosmological models \citep{planckXVI}. The older mass-weighted ages derived from spectral fitting are a direct result of the well-known bias of SSP-equivalent ages by small mass fractions of young populations, and our findings agree well with the exploration of this issue by \cite{serra07}. Metallicities show good agreement, also as expected following the findings of \citet{serra07}, who showed that SSP-equivalent metallicities are close to the mass-weighted values, being less biased by the presence of young stars. We also note that there is no obvious bias caused by using models limited to solar abundance ratios.

Finally, we find a clear trend between the SSP-equivalent alpha enhancement and \thalf. We over-plot the linear relation given by \cite{delarosa11}, who conducted a similar analysis on SDSS spectra using the {\tt STARLIGHT} code of \cite{cidfernandes05}. We also show the canonical relation between abundance ratio and star formation timescale from \cite{thomas05}, equation (4): $[\alpha/Fe]\sim\frac{1}{5} - \frac{1}{6} \log\Delta t$, where $\Delta t$ is the FWHM of the assumed Gaussian SFH. This relation reproduces the non-linearity of our measured timescale against \afe. We also note that the limited age resolution of the SSP models at the oldest ages creates a `floor' in the \thalf\/ values at the shortest timescales. This important region, corresponding to the earliest epochs of massive galaxy formation, is where abundance ratios can yield higher fidelity than our SFH. We therefore use our empirical half-mass formation time to constrain a relation of the same form as that of \cite{thomas05} to relate \afe\/ to \thalf. The resulting expression is indicated by the red line in Fig. \ref{fig:comp_ssp_sfh}, and is given by:

\begin{equation}
\label{eq:afe_thalf}
[\alpha/Fe] = 0.28 - 0.19 \log(t_{50}) \qquad .
\end{equation}

The objects plotted with magenta symbols in Figure \ref{fig:comp_ssp_sfh} (the low-Fe objects) are mostly outliers to the \afe-\thalf\/ trend, having  {\it large} values of \thalf\/ - i.e. long formation timescales, contrary to what is expected from their high \afe\/ values. The relatively weak iron indices result in low metallicities and subsequently {\it high} \afe, but since they also have low velocity dispersions, they are strong outliers in the scaling relations shown in Figure \ref{fig:ssp_sigma}. Figure \ref{fig:co_pop} shows the residuals of the SSP parameter - velocity dispersion relations for an \Reffe\/ aperture, plotted against molecular gas mass fraction, with the low-Fe objects marked with filled symbols. This shows that, as well as having low metallicity, the low-Fe objects also have high molecular gas fractions and are among the youngest in the sample. This suggests their apparently high SSP-equivalent \afe\/ may indicate the presence of stars that have not yet been appreciably enriched by Type 1a supernovae - the main contributor of iron to the gas involved in star-formation. Their young age causes them to outshine the older, less $\alpha$-enhanced stars, and dominate the SSP abundance measurement. They do not show any preferred kinematic misalignment \citep{davis13}, and are not Virgo cluster members except for one, NGC\,4694, which is a low-mass object in the outskirts of Virgo. We speculate that these low-Fe galaxies have recently accreted a new supply of cold, low-metallicity gas, which is currently in the process of forming stars. These twelve low-Fe objects are shown as magenta symbols in subsequent plots unless stated otherwise, and are indicated in Tables \ref{tab:re8}-\ref{tab:re} via the `quality' flag.

%
%

%
%

\section{Population parameters on the Mass Plane}
\label{sec:mass_size}

Galaxy stellar populations have been found to correlate better with stellar surface density \citep{kauffmann03,franx08} or with velocity dispersion \citep{graves10} than with mass or galaxy size. The same was found to hold for the dynamical M/L, which mainly traces age variations \citep{cappellari06}. In \cite{cappellari13b} we used accurate masses and velocity dispersions to show that indeed the velocity dispersion is the best simple tracer of multiple galaxy properties. We showed that velocity dispersion is such a good predictor of galaxy properties because it traces the bulge fraction, which appears to be the main driver of the M/L, \hb\/ and colour, as well as the molecular gas fraction in ETGs. Here we focus on trends in the stellar population parameters.

\begin{figure*}
 \begin{center}
  \epsfig{file=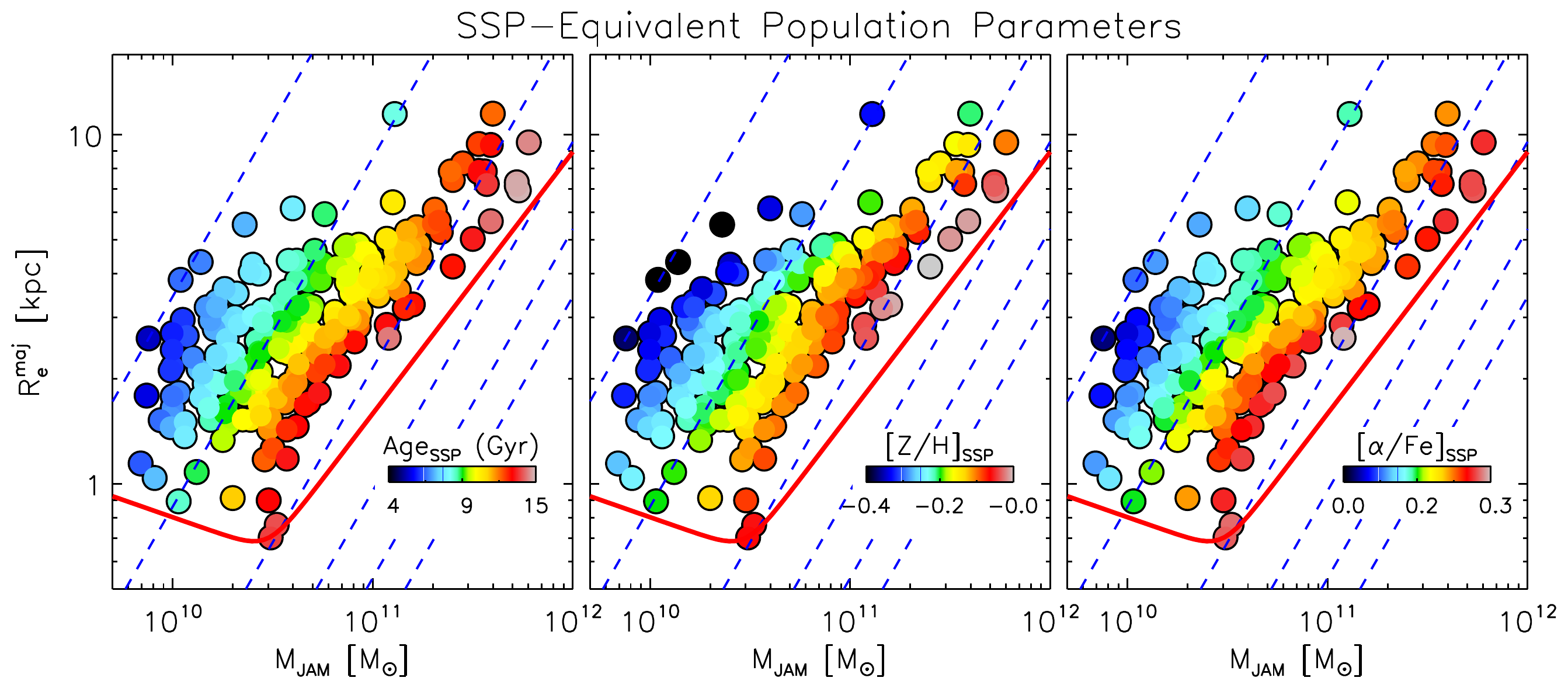, angle=0, width=17cm}
  \caption[]{SSP-equivalent population parameters measured within one effective radius, plotted on the plane of effective radius \Rmaj\/ versus dynamical mass, \mjam. Colours indicate the population parameters as indicated in each plot after spatially averaging with the LOESS technique described in the text. For a fixed range in \mjam, the smallest galaxies are generally the oldest, most metal rich and alpha-enhanced. Dashed lines show lines of constant velocity dispersion: 50, 100, 200, 300, 400 and 500\kms\/ from left to right, as implied by the virial mass estimator \mjam$=5R^{\rm maj}_\mathrm{e}\sigma^2/\mathrm{G}$. The red curve shows the `zone of exclusion' (ZOE) defined in \cite{cappellari13b}.}
  \label{fig:pop_size}
  \end{center}
\end{figure*}

\begin{figure*}
 \begin{center}
  \epsfig{file=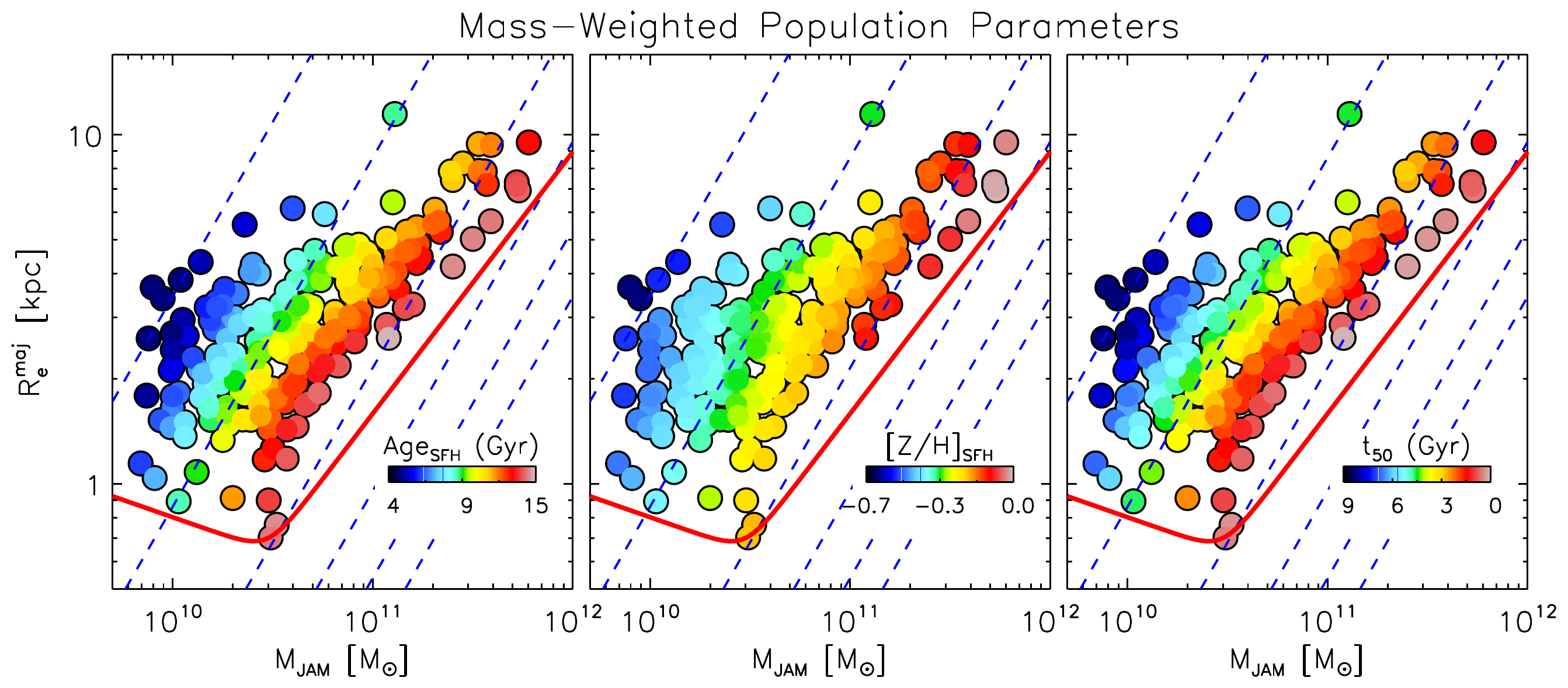, angle=0, width=17cm}
  \caption[]{Same as Figure \ref{fig:pop_size}, but using the mass-weighted age and metallicity, and \thalf\/ - the time taken to form half the stars inside 1\Reff.}
  \label{fig:mwpop_size}
  \end{center}
\end{figure*}

Figure \ref{fig:pop_size} and \ref{fig:mwpop_size} show the distribution of population parameters as a function of mass and size (which we hereafter refer to as the Mass Plane) for our sample, using the SSP-equivalent and mass-weighted parameters respectively. The mass \mjam\/ is derived applying Equation 3 of \cite{cappellari13b}, using the mass-to-light ratio $(M/L)_{\rm JAM}$ and $r$-band luminosity given in Table 1 of \cite{cappellari13a}, columns 6 and 15 respectively. We use the robust size parameter \Rmaj, corresponding to the major axis of the half-light isophote \citep{hopkins10} as derived in \cite{cappellari13a}, and taken from Table 1 there.

The points in Figs. \ref{fig:pop_size} and \ref{fig:mwpop_size} are coloured according to the population parameters as indicated. The distributions have been adaptively averaged by applying an implementation\footnote{available from http://purl.org/cappellari/idl} of the LOESS algorithm \citep{cleveland88}, using a local linear approximation, and a regularization factor $f=0.6$. The same approach is taken in \citet{cappellari13b} and is found to give a good approximation to the underlying intrinsic distribution, mimicking the average behaviour that would be derived from much larger samples \citep[e.g.][]{gallazzi06}. We note that, since we are adaptively averaging the data points, the range of values is reduced in accordance with the reduced effects of errors and intrinsic scatter.

As in \citet{cappellari13b}, we over-plot lines of constant velocity dispersion, as implied by the virial mass estimator \mjam$=5R^{\rm maj}_\mathrm{e}\sigma^2/\mathrm{G}$. We find that the SSP-equivalent population parameters follow trends of nearly constant velocity dispersion on the Mass Plane, as indicated by the tilt of the iso-colour regions. The mass-weighted age and formation timescale follow even more tightly the lines of constant dispersion, which reflects the lower scatter in the mass-weighted properties as a result of lower formal uncertainties, and less susceptibility to small fractions of young stars being present. The mass-weighted metallicity is also closely linked to the velocity dispersion, though the iso-colour regions show an appreciable dependence on \mjam\/ also (i.e. steeper iso-colour regions).

These figures show that, for fixed mass, more compact galaxies are on average older, more metal rich, have higher abundance ratios and shorter formation timescales than larger galaxies. The trends with SSP age and metallicity agree with previous results from averaged properties of stacked SDSS spectra \citep{vdwel09,shankar10}. We show here that these trends hold for measurements of individual galaxies. Moreover, the trend with abundance ratio, mass-weighted properties, and formation timescale measured directly from the star-formation history can only be obtained from a spectroscopic analysis, and are reported here for the first time.

\begin{figure*}
 \begin{center}
  \epsfig{file=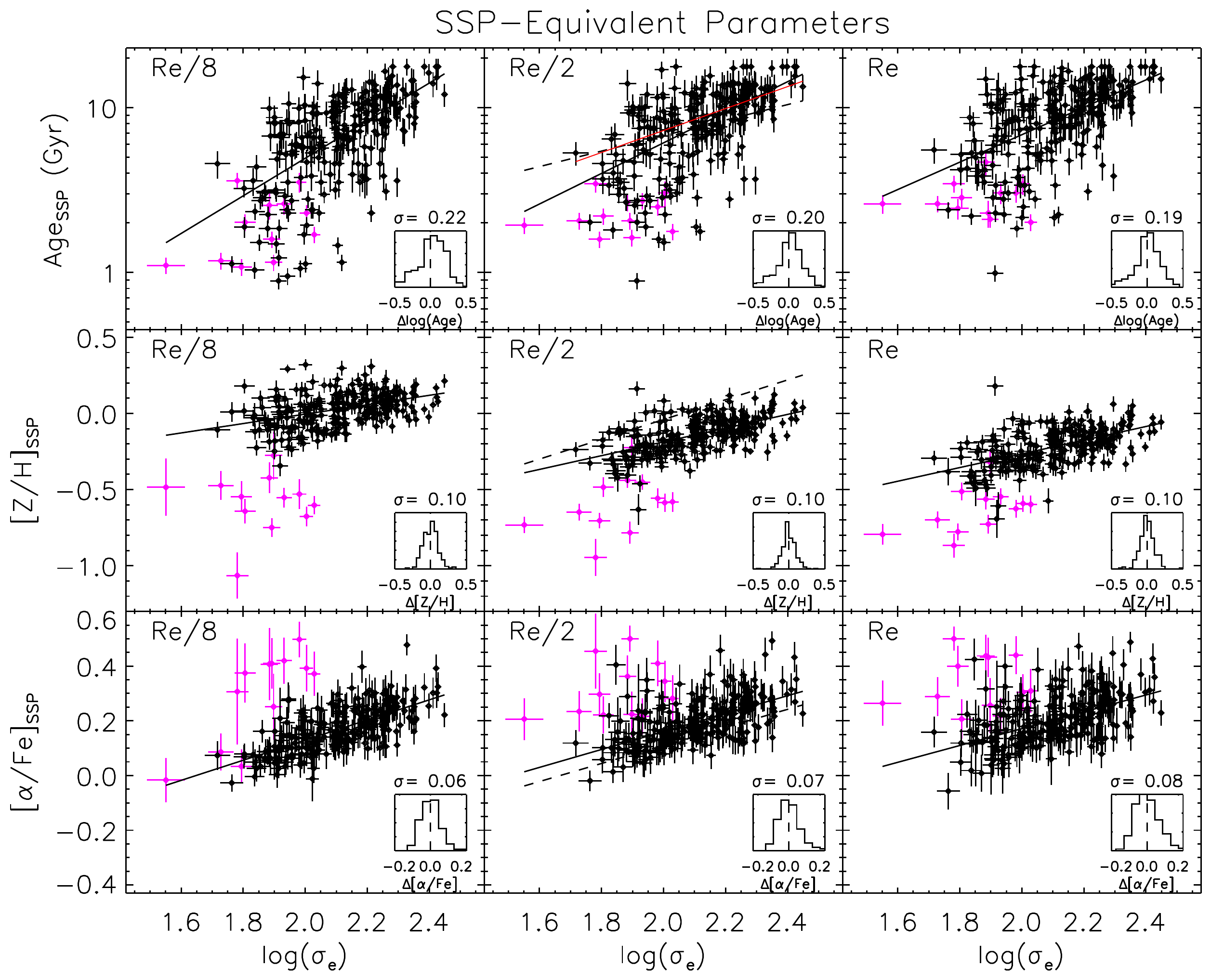, angle=0, width=17cm}
  \caption[]{SSP parameters for the three apertures as a function of $\log$(\sigmae). Magenta points are the galaxies exhibiting low metallicity as described in Section \ref{sec:indices}. Dashed lines on the \Reff/2 aperture plots show the relations from \citet{thomas10}, which have a comparable aperture size (see text for details). The red line shows the \Reff/2 age trend when fitting only galaxies older than 2.5\,Gyr, as done by \citet{thomas10}. Solid lines are robust linear fits. Magenta symbols indicate the low-Fe objects highlighted in section \ref{sec:specfit}. Solid lines are robust linear fits to the black points. Parameters of these fits are given in Table \ref{tab:params}. Insert histograms show the residuals about the global best fit.}
  \label{fig:ssp_sigma}
  \end{center}
\end{figure*}

\cite{graves09b} found that \Reff\/ does not correlate with either population parameters or their residuals from the population-$\sigma$ relations. This can be understood from the mass-size plane. At each \Rmaj, there is a similar range of population parameters, thus showing no particular correlation. Likewise, as $\sigma$\/ traces the parameters closely, the residuals of the population-$\sigma$\/ trends show no significant residual correlation with \Rmaj. Indeed, Figures \ref{fig:pop_size} and \ref{fig:mwpop_size} show that velocity dispersion is an excellent predictor of all three stellar population parameters, with the smoothed population parameters varying essentially parallel to the lines of constant (virial-predicted) velocity dispersion.

We advocate that the Mass Plane is a novel and useful tool for understanding galaxy properties. Comparing our Figures \ref{fig:pop_size} and \ref{fig:mwpop_size} to Figures 3--8 of \cite{cappellari13b} reveals the close connections between the stellar population parameters and other important quantities, in particular  colour, molecular gas mass fraction, and bulge fraction, such that, at a given mass, smaller galaxies are redder in colour, have lower fractions of molecular gas, and have a second moment $V_{\rm rms} \equiv \sqrt{V^2 + \sigma^2 }$ dominated by a dynamically hot central `bulge' component. There are no particular trends with intrinsic shape other than the fact that the population of round objects above \mjam$\gsim 2\times10^{11}$\msun\/ are typically old, metal rich and $\alpha-$enhanced. Nor is there a close connection to the specific angular momentum, \lambdar, which supports the findings of \cite{naab14} where \lambdar\/ is shown to trace a galaxy's evolution, but that there are many evolutionary paths to given current-day value. We explore the connection between the kinematic classes and stellar populations in Section \ref{sec:lambda_pop}.

%
%

\section{Population trends with velocity dispersion and mass}

In this section we present a more traditional two-dimensional analysis of the stellar population parameters, examining their scaling properties with velocity dispersion and mass, for SSP-equivalent and mass-weighted properties. This allows a more quantitative study of these relations and their scatter, as well as providing context for our findings on the Mass Plane, allowing direct comparison to previous authors. All linear fit parameters are given in Table \ref{tab:params}.

\subsection{SSP-equivalent parameters versus velocity dispersion}
\label{sec:ssp_sigma}

Figure \ref{fig:ssp_sigma} shows the resulting SSP-equivalent parameters for our three aperture radii, plotted against the effective velocity dispersion, $\log$(\sigmae), given in column 2 of Table 1 in \citet{cappellari13a}. As shown in the previous section, the population parameters correlate tightly with velocity dispersion, confirming the trends from previous authors. For comparison, in Figure \ref{fig:ssp_sigma} we overplot with dashed lines the relations from \citet{thomas10} based on analysis of 3,360 galaxies observed with a single fibre by the Sloan Digital Sky Survey, selected as part of the `MOSES' project. We make this comparison on the \Reff/2 aperture, which should be the most comparable to the effective aperture of that study.

The \citet{thomas10} lines show general agreement, but there are some important differences. In particular, our slope with age is significantly steeper, which is trivially explained by the fact that \citet{thomas10} exclude objects younger than 2.5\,Gyr (at z=0.05-0.06) from their fit. We indicate with a red line our fit applying the same selection, and the agreement is excellent. There is also a significant difference in slopes for the metallicity relation. Low velocity dispersion objects are similar, but the \citet{thomas10} best-fit linear relation over-estimates the high velocity dispersion objects by 0.3 dex -- several times larger than the total scatter.  In Appendix \ref{app:literature} we show that the difference in slope is consistent with the lack of any aperture correction applied to the MOSES data to account for the fixed 3\arcsec\/ diameter aperture used. This steepens the relation by including more of the outer regions in lower-mass (smaller) objects relative to the high-mass (larger) objects, which are dominated by the high metallicity inner regions. Even considering the offset in slope, the MOSES best-fit linear relation is still $\sim 0.2$ dex higher in $[Z/H]$. \afe\/ also shows a small, less significant, offset such that our values are $\sim 0.05$ dex higher than from MOSES. We have verified that these offsets are not caused by differences in the models of \cite{schiavon07} and \cite{thomas03,thomas04}, which, when applied consistently to our line-strengths, show average offsets smaller than the 1$\sigma$ scatter, and which would anyway exacerbate the offsets shown in Figure \ref{fig:ssp_sigma}. The differences between our findings and \cite{thomas10} may arise from the different combination of indices used by the two studies, but it is beyond the scope of this paper to explore this further.

Contrary to \citet{thomas10}, we do not find strong evidence for bimodality in our distribution of age and velocity dispersion. Our sample size may be too small to constrain this, though we note that splitting our sample using their age threshold of 2.5\,Gyr and our \Reff/2 values yields a similar fraction of `young' galaxies \citep[23 out of 260, or 8.8\%, for our sample; 10.15\% in][]{thomas10}. 

From Figure \ref{fig:ssp_sigma}, the effect of different aperture sizes can be seen. Most notably, as we `zoom out' from the \Reffe\/ aperture to \Reff, the age-$\sigma$ relation becomes systematically more shallow, showing again that young stars, when present, are preferentially found in the central regions of the low velocity dispersion galaxies. The [Z/H]-$\sigma$ relation becomes steeper and the average metallicity decreases from around solar for \Reff/8 to -0.2 within \Reff, showing that early-type galaxies are only metal rich in their central regions, and are actually rather metal-poor objects when considered on the scales of one effective radius. Enhancement of alpha elements does not show much change considering larger apertures, although we remark that the scatter of the \afe-$\sigma$ relation increases by 15\% as we go from \Reff/8 to \Reff, unlike for age and metallicity, which become $15$\% and $10$\% {\em tighter}, respectively.

\begin{figure}
 \begin{center}
  \epsfig{file=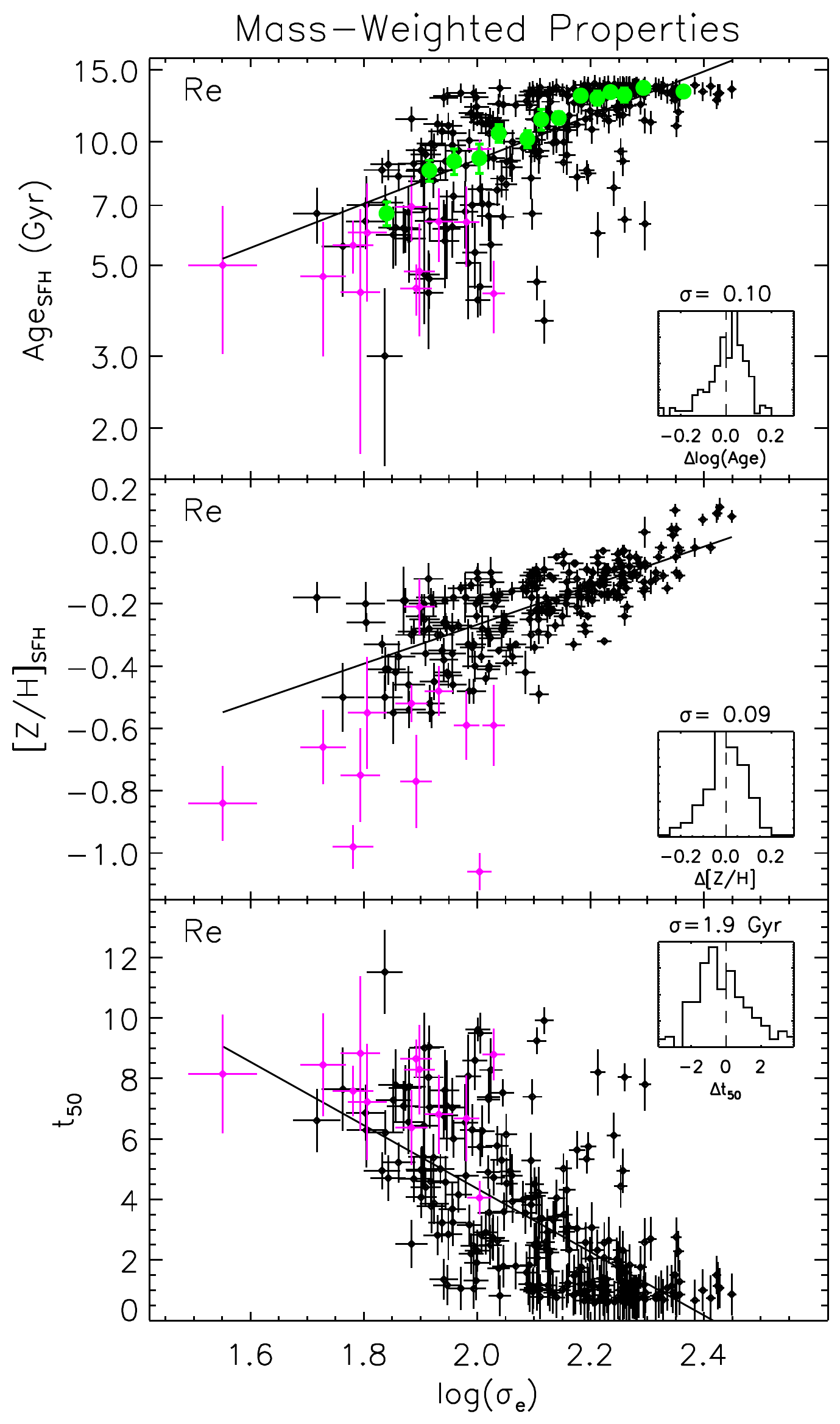, angle=0, width=8cm}
  \caption[]{Mass-weighted age, metallicity and half-mass formation time of the \atlas\/ sample plotted against velocity dispersion. Magenta points are the Fe-poor objects as per Fig. \ref{fig:ssp_sigma}. Insert histograms show the residuals about the global best fit. Solid lines are robust linear fits to the black points. Parameters of these fits are given in Table \ref{tab:params}.}
  \label{fig:mw_sigma}
  \end{center}
\end{figure}

\subsection{Mass-weighted parameters versus velocity dispersion}
\label{sec:sfh_sigma}

Figure \ref{fig:mw_sigma} presents the mass-weighted age and total metallicity derived from the spectral fitting technique as described in Section \ref{sec:specfit}. The spectrum fitted is integrated within an effective radius, making these values directly comparable to the \Reff\/ aperture of the previous sections. The general trends are similar: age and metallicity increase with velocity dispersion. However, the correlation with mass-weighted age is significantly tighter than with SSP age, showing a factor 2.6 less scatter (though with a somewhat smaller range in age). 

We also show in Figure \ref{fig:mw_sigma} the median-binned mass-weighted ages to illustrate better the general trend, which suggests a possible break in the slope of the relation at velocity dispersions higher than $\approx 160$\kms. We caution, however, that this is largely a result of limiting the maximum age of the stellar population models to 14\,Gyr. Including the maximum-aged MIUSCAT templates (17.8\,Gyr) in the star formation history removes any significant deviation from a linear relation of mass-weighted age with $\log$(\sigmae). In addition, by fitting a linear relation to points below the apparent break, the extrapolated age and scatter at the highest velocity dispersion is $18\pm4$\,Gyr - only mildly inconsistent with the fiducial age of the Universe. This suggests that a linear relation is indeed an appropriate choice. We return to this issue in Section \ref{sec:sfh_mass}.\looseness-2

The metallicity trend is remarkably similar considering the vastly different approaches and models used. These results give direct empirical confirmation of the fact that early-type galaxies are overwhelmingly composed of old stars, but in general have a small (but non-negligible) mass-fraction of younger stars that contribute disproportionately to the integrated light of the galaxy - a fact inferred from early SSP analyses \cite[]{gonzalez93,trager2000} and shown unambiguously with GALEX UV colours \citep{kaviraj07}. We consider this more quantitatively in Section \ref{sec:sfh}.

The almost negligible change in the metallicity trend also confirms the insensitivity of SSP-equivalent metallicity to mixed populations as discussed above. The main differences are a slight reduction in scatter and a steepening of the relation, likely due to reduced uncertainties by fitting the whole spectrum, and a reduced bias in the contribution of young stars responsible for boosting the metallicity in the SSP relations.
 
%
%

\begin{figure*}
 \begin{center}
  \epsfig{file=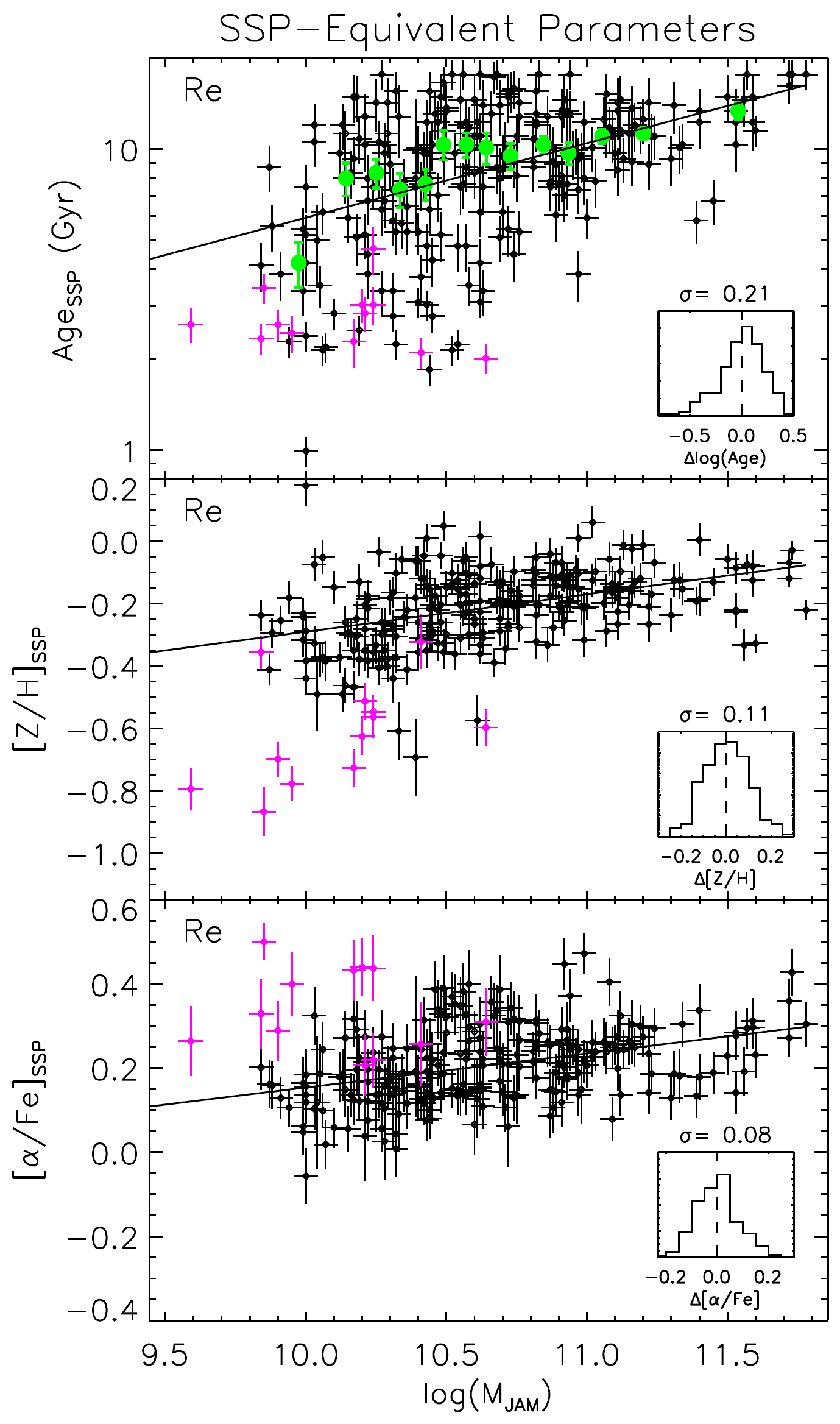, angle=0, height=14cm}
  \hspace{1cm}
  \epsfig{file=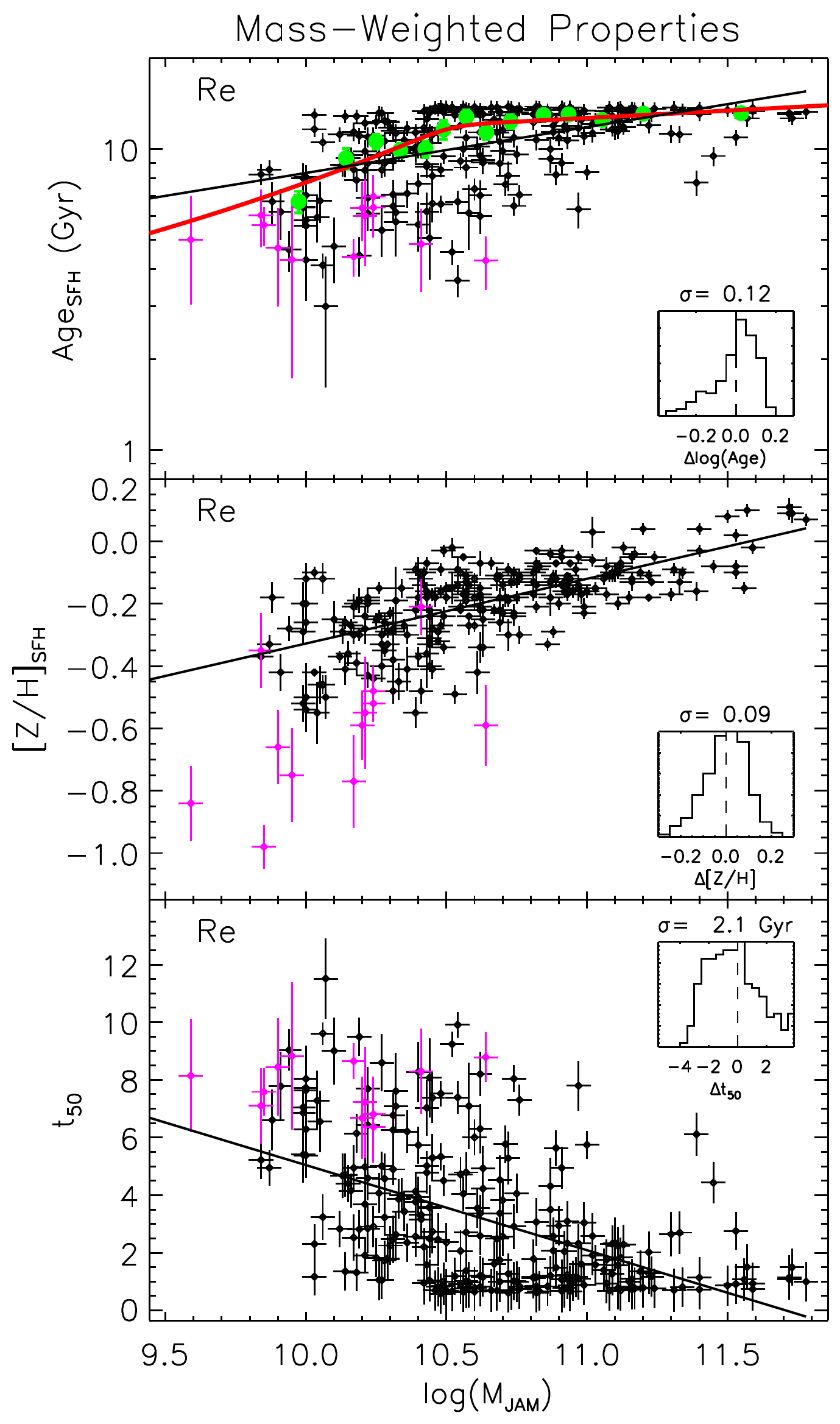, angle=0, height=14cm}
  \caption[]{{\it Left:} SSP parameters within one effective radius as a function of dynamical mass, \mjam. {\it Right:} Mass-weighted age, metallicity, and half-mass formation time, as a function of \mjam. Magenta points indicate the Fe-poor objects as in previous plots. Solid lines are robust linear fits to the black points. Parameters of these fits are given in Table \ref{tab:params}.  Insert histograms show the residuals about the global best fit. Green points indicate the median values within bins of 20 neighbouring points, with error bars indicating the `error' on the median, computed as $\sigma/\sqrt{N}$, where $\sigma$ is the dispersion within the bin, and $N=20$. The red line indicates a double power law fit to the binned values, as described in Section \ref{sec:sfh_mass}.}
  \label{fig:ssp_mass}
  \end{center}
\end{figure*}

\subsection{SSP parameters versus dynamical mass}

Figure \ref{fig:ssp_mass} ({\em left}) presents the SSP population trends as measured against \mjam\/ for the 1\Reff\/ aperture (the smaller apertures follow similar relative differences as those mentioned in the previous section). We find that the general trends of population parameters are consistent with the picture derived from velocity dispersion alone, such that lower mass galaxies tend to show younger ages, lower metallicities, and lower abundance ratios. 

The relations with mass are generally more shallow compared to the trends with \sigmae, especially in age and metallicity, owing to two key differences. Firstly, galaxies with the lowest masses span almost the full range of ages and metallicities, whereas at the lowest velocity dispersion, galaxies are generally the youngest and most metal poor of the sample. Secondly, the high mass regime where galaxies span a narrow range of age and metallicity (\mjam$> 10^{11}$\msun) accounts for about one third of the mass range in the sample. For velocity dispersion, this low-scatter region ($\log(\sigma) > 2.3$) is less than a quarter of the sample range. When we move to the mass-based relations, the velocity dispersion distributions are essentially `stretched out' among the older objects, reducing the slope of the relations. Both effects are a direct consequence of the distribution of galaxies on the Mass Plane.

Previous studies have found tighter relations of stellar population parameters with velocity dispersion than with mass \citep{graves10,wake12}, as we also qualitatively highlighted on the Mass Plane. Insert histograms in Figures \ref{fig:ssp_sigma} through \ref{fig:ssp_mass} show the dispersions around our linear fits and quantify their standard deviations (see Table \ref{tab:params} for the fit parameters and standard deviations). We likewise find that the SSP residuals are somewhat tighter using \sigmae\/ than when using mass, although we note that the formal difference is small (0.01 dex in age and 0.02 in \metal). We note that the improvement of \sigmae\/ over mass as a linear indicator of population parameters is significantly more apparent on the Mass Plane directly than by comparing the residuals of it's edge-on projections, where intrinsic scatter makes the interpretation less clear.

\subsection{Mass-weighted parameters versus dynamical mass}
\label{sec:sfh_mass}

Figure \ref{fig:ssp_mass} ({\em right}) presents the mass-weighted age, metallicity, and half-mass formation time \thalf\/, as a function of \mjam. We see similar trends as before, but there is now a rather clear change in the behaviour of age as a function of mass. Rather than being approximated as a linear increase of age with mass, albeit with a large scatter at lower masses, the trend of mass-weighted age and mass shows a distinct non-linearity. This is made more apparent by considering the median value within mass bins, as indicated by the green points in Figure \ref{fig:ssp_mass}. The binned mass-weighted ages show a distinct departure from the linear fit, whereas the binned SSP values remain reasonably approximated by the best fit line. 

To characterise this bimodal trend in more detail, we fit a double power law on the median-binned mass-weighted values of the form:

\begin{equation}
\log({\rm Age})  = 2^{(\beta-\gamma)/\alpha} \log({\rm Age}_b) \left( \frac{M}{M_\mathrm{b}} \right) ^\gamma \biggl[ 1 + \left( \frac{M_\mathrm{b}}{M} \right) ^\alpha \bigg] ^\frac{\gamma - \beta}{\alpha}
\end{equation}

\noindent based on the `Nuker' law for surface brightness profile fitting \citep{lauer95}. The best fit curve has a low-mass slope $\gamma=-0.18$, a high-mass slope of $\beta=-0.016$, and a characteristic mass of $M_\mathrm{b} = 3.1 \times 10^{10}$\,\msun\/ at a `break' age of Age$_{\rm b}=11.8$\,Gyr that marks the transition of the two slopes, with a `sharpness' parameter for the break in slope given by  $\alpha=10$. This bimodal trend is also evident in the other age relations, but to a lesser extent. Using the mass-weighted age, the presence of a narrow sequence of old galaxies can be seen above the transition mass, with a distinct change in the mass-scaling behaviour below this mass. This transition mass is the same as the inflection point of the `zone of exclusion' (ZOE) determined by \cite{cappellari13b}, which defines the low-\Rmaj\/ envelope of the Mass Plane as shown in Figs. \ref{fig:pop_size} and \ref{fig:mwpop_size}. It also corresponds to the characteristic mass marking the transition between star-forming and passive galaxies found by \cite{kauffmann03}, as well as a possible change in slope of the relation between supermassive black hole masses and their host galaxy mass \citep{scott13b}. Considering the population trends on the Mass Plane, the correspondence of this `knee' in the age-mass relation to the inflection of the ZOE on the mass-size plane becomes obvious: below this mass, the low-\Rmaj\/ envelope turns toward larger galaxies, with correspondingly younger ages and longer formation timescales.

We note again that we impose an upper age limit of 14\,Gyr on the model templates used, in accordance with the fiducial age of the Universe  \cite[$13.798\pm0.037$\,Gyr;][]{planckXVI}. This puts a `ceiling' on the mass-weighted age values, as discussed previously in Section \ref{sec:sfh_sigma} for the trend of mass-weighted age with $\log$(\sigmae) (see Figure \ref{fig:mw_sigma}). However, unlike the trend with $\log$(\sigmae), including the full available age range of the MIUSCAT spectral models (up to 17.8\,Gyr) in the spectral fit does {\it not} remove the bimodality. There is a slightly increased spread in mass-weighted age at the oldest values, but the mass at which the ages transition to a narrow range of old values is unchanged, and the clear bimodal shape remains. Moreover, extrapolating the linear trend in Figure \ref{fig:ssp_mass} from masses below the transition mass results in strongly unphysical ages ($\approx 30$\,Gyr) at our highest masses, suggesting that the departure from a simple linear trend is not artificial.

Mass-weighted metallicity shows a tight relation with \mjam, with only slightly higher scatter than with velocity dispersion. This supports the interpretation from the Mass Plane that iso-surfaces of constant metallicity run steeply with mass, especially for mass-weighted metallicity.

The half-mass formation time is distributed similarly to the mass-weighted age, being strongly related to that property. We include it here for completeness.

\subsection{Kinematic class}
\label{sec:lambda_pop}

\begin{figure}
 \begin{center}
  \epsfig{file=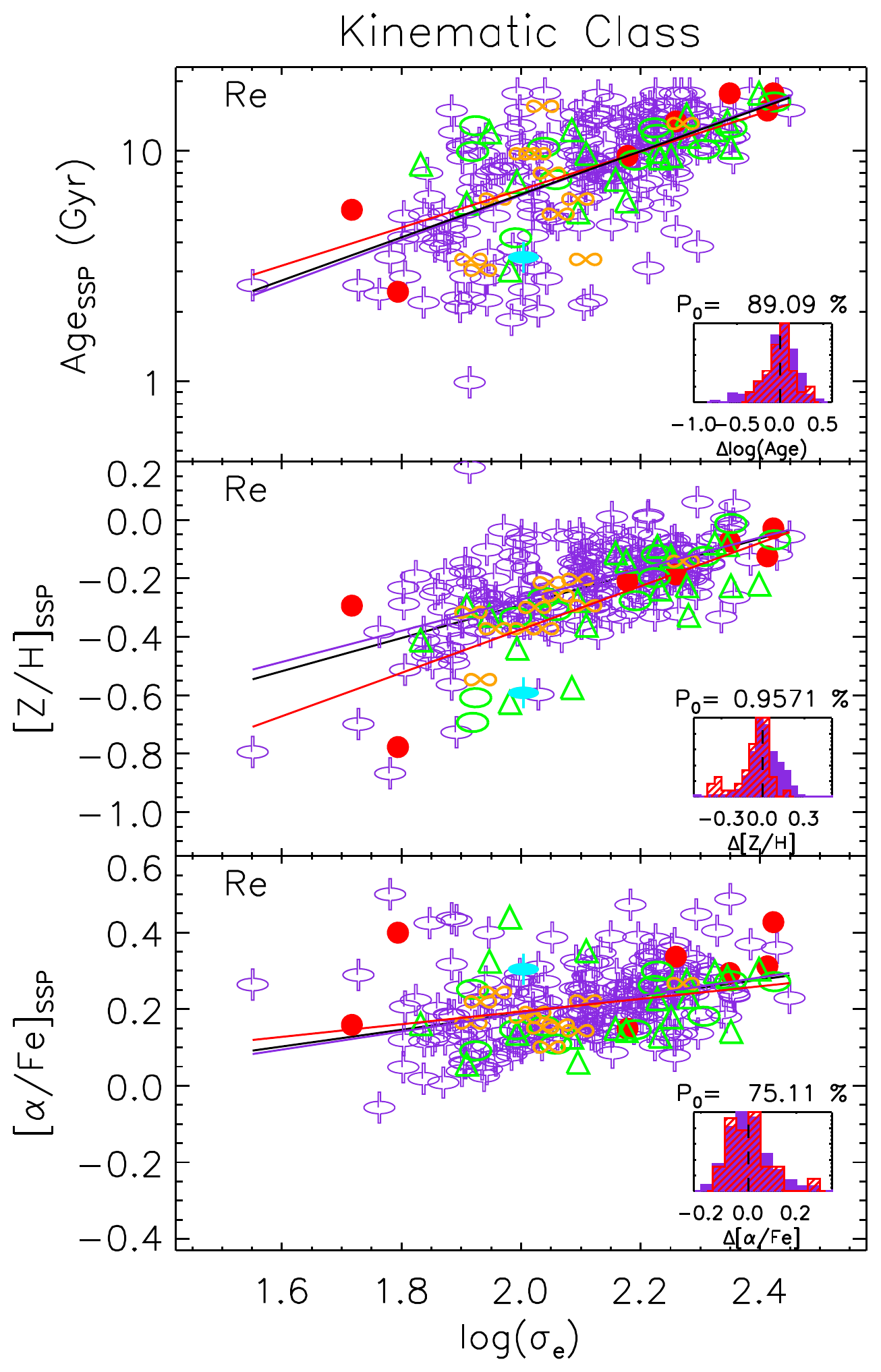, angle=0, height=14.cm}
  \caption[]{As per Fig \ref{fig:ssp_sigma}, but with symbols indicating kinematic classes as defined in \cite{krajnovic11}. Red circles show little or no rotation, green ellipses have complex velocity maps with no specific features, green triangles are KDCs, orange lemniscates have off-centre dispersion peaks indicative of superposed counter-rotating disks. These correspond to the respective classes {\it a, b, c,} and {\it d} in \cite{krajnovic11}, referred to collectively as `non-regular rotators'. Purple spindles are the class {\it e} from \cite{krajnovic11}, or `regular rotators'. Filled spindles have obvious bars. Red and purple lines show fits to only the non-regular and regular rotators respectively. The solid black lines show fits to all classes together. Parameters of these linear fits are given in Table \ref{tab:params}. Insert histograms show the residuals of the regular (purple) and non-regular (red) rotators about the global best fit to the combined population, quoting also $P_0$: the probability that the chance deviation between the two samples should be larger than that measured, as derived from a K-S test of the two histograms. Small values of $P_0$ imply the distributions are different. The only significant difference is in the metallicity relation, where the average offset of the non-regular rotators is consistent with the mass increase expected from a single major (equal mass ratio) merger.}
  \label{fig:ssp_sigma_kinem}
  \end{center}
\end{figure}

One of the key findings of the \sauron\/ survey was the discovery that ETGs have essentially two categories of global dynamics, characterised as `fast' and `slow' rotators \citep{emsellem07}. Fast rotators have extended regular rotation fields well described as inclined disks, though they are clearly not `thin-disk' systems, having significant velocity dispersion also. Slow rotators do not show regular rotation, and instead show counter-rotating structures, decoupled cores, or no rotation at all. In \cite{emsellem11}, we expand on this work with the complete \atlas\/ sample, using the quantity \lambdar\/ as a way of separating these two classes of ETG. In \cite{krajnovic11}, we have shown that, applying the Fourier analysis technique of kinemetry \citep{krajnovic06}, we can further classify the velocity fields of galaxies, giving `regular' rotators (corresponding very closely to the fast rotators, which show disk-like rotation), and `non-regular' rotators (corresponding closely to the slow rotators, which show various substructures that cannot be described by simple disk-like rotation).

Figure \ref{fig:ssp_sigma_kinem} presents the SSP-\sigmae\/ relations as before, but this time indicating the main kinematic classes described in \cite{krajnovic11}. From this figure it can be seen that the various kinematic classes span broad ranges in velocity dispersion and stellar population properties. Non-regular rotators are on average slightly older than regular rotators, although we note there are several examples of non-regular rotators among the youngest galaxies in our sample. There is little discernible difference in abundance ratios, though we note that the massive non-rotators are among the most enhanced objects in our sample. The most noticeable difference between the regular and non-regular rotators is with metallicity, where the non-regular rotators appear to fall generally on or below the global trend, giving rise to a moderately significant difference in the residual distributions around the global linear fit as quantified by the K-S test probability, $P_0$, indicated in the insert histograms. This implies that non-regular rotators have experienced less metallicity enrichment, on average, than their regular-rotating counterparts of the same mass. In \cite{young11}, we also found that non-regular rotators typically have lower mass fractions and detection rates of molecular gas (though there are notable exceptions at low velocity dispersion values, e.g. NGC1222, NGC3073). Less molecular gas implies less star formation, which is the mechanism for enriching the stellar population with metals.

The offset in metallicity at fixed velocity dispersion can be interpreted conversely as an offset in velocity dispersion for a fixed metallicity, which could arise as a result of a major merger. The metallicity offset is also seen in the relation with \mjam. Interpreting the offset as one of increased mass and velocity dispersion at fixed metallicity, we find that the linear-fit mass- ($\sigma$-) metallicity relation is offset by 0.3\,dex in $\log$(\mjam) corresponding to a doubling of the mass, with a corresponding shift of 0.1\,dex in $\log$($\sigma$). These shifts are consistent with the predictions of \cite{hilz12} for a single 1:1 dry merger, where the velocity dispersion is found to increase by a factor 1.2-1.5 following a 1:1 merger. This is different to the behaviour expected from the idealised virial predictions of \cite{naab09}, where velocity dispersion remains unchanged. The idealised case, however, ignores the effects of violent relaxation, which acts to unbind a non-negligible fraction of the mass and increase the central density and velocity dispersion. This interpretation implies that, on average, non-regular rotators are consistent with having experienced a single dissipationless major merger that could have shaped their current kinematic characteristics. This picture is consistent with \cite{khochfar11}, where semi-analytic models show that slow/non-regular rotators have on average experienced more major mergers than fast/regular rotators. It is also consistent with \cite{scott13a}, where we showed that the radial relation of \mgb\/ line strength and escape velocity for our sample limits the number of possible major mergers to $\lsim 1.5$, beyond which objects would become inconsistent with the observed trends.

We add a note of caution to this simple interpretation. Firstly, if the offset is in the velocity dispersion, one may reasonably expect to see some sign of a corresponding offset in the other population parameters, which is not observed. This is complicated, however, by the differing response of the parameters to mass- and luminosity-weighting. Metallicity is essentially a mass-weighted property, suggesting that the offset is driven by a process that impacts most of the stellar mass in these galaxies. Age, however, is sensitive to small amounts of recent star formation, making trends difficult to interpret. Figure \ref{fig:comp_ssp_sfh} showed that ages $\gsim 8$\,Gyr are close to the mass-weighted values. Considering only galaxies with ages older than this, the offset in metallicity remains apparent in the K-S test, but the relations with age and abundance become sufficiently shallow that any offset in age or velocity dispersion is poorly constrained, so we cannot firmly rule out a shift in the velocity dispersion axis on this basis.

Secondly, as we have shown in \cite{naab14}, the net angular momentum of a galaxy is a sensitive tracer of its evolution, but with a broad range of possible paths. Some of the non-regular rotators are among the very metal-poor (Fe-poor) objects, which have high mass fractions of cold gas and young ages, consistent with a gas-rich accretion or merger event. On an individual galaxy basis, the specific assembly history can be varied. However, here we highlight the offset in the mass- ($\sigma$-) metallicity relation as the only significant {\it average} difference between the stellar population properties of these broad kinematic classes.

\subsection{The role of environment}
\label{sec:environment}

The \atlas\/ sample spans a range of environment, from sparse fields to the central regions of the Virgo cluster. Differences in the galaxy populations of high and low density environments are well documented \cite[e.g.][]{thomas05}. However, reported trends and results are surprisingly weak considering the predictions of galaxy formation simulations, where dense regions differ in the star-formation properties and assembly histories \citep{dekel06,khochfar08,cattaneo08}.

\begin{figure}
 \begin{center}
  \epsfig{file=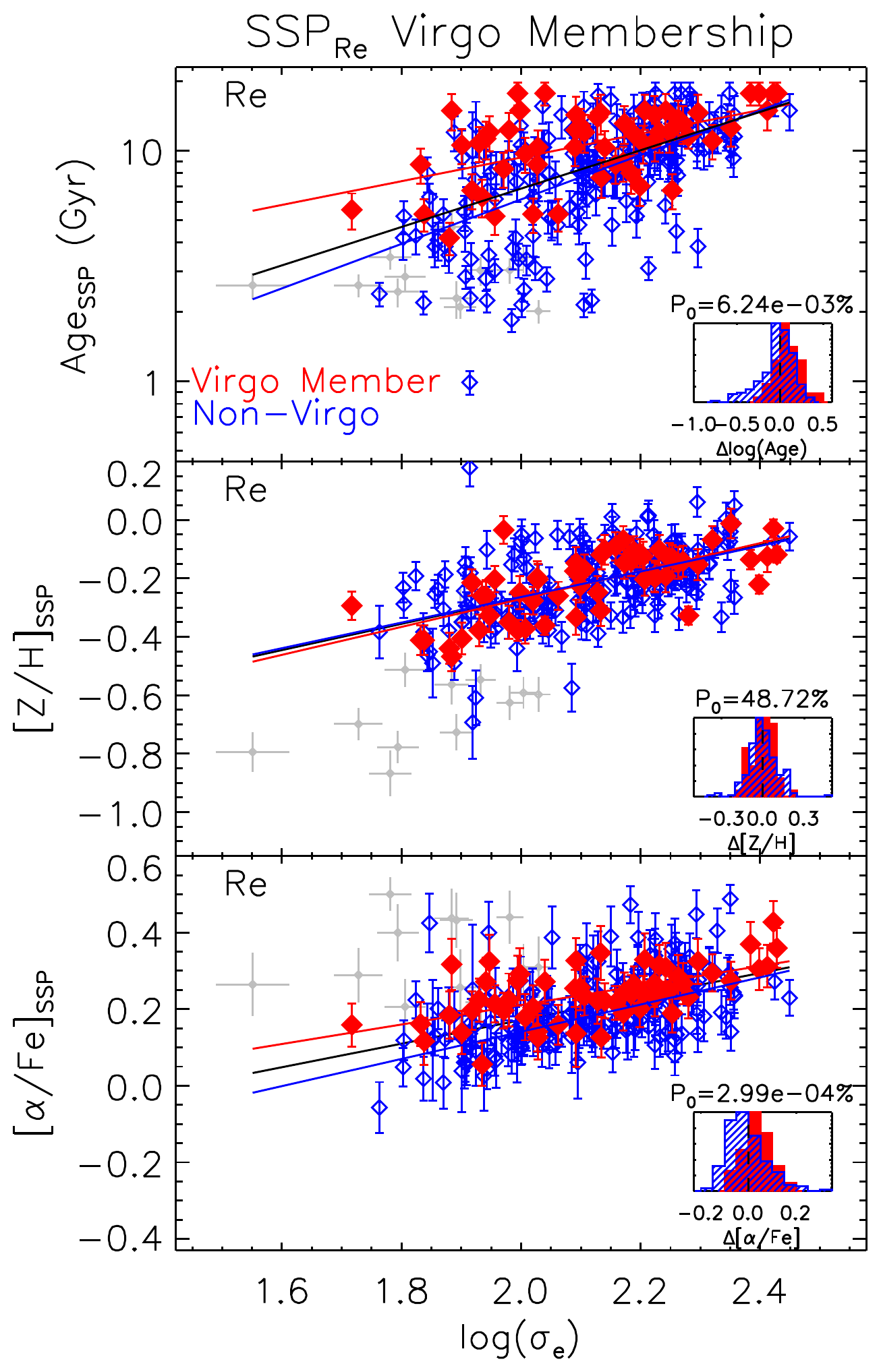, angle=0, height=14.cm}
  \caption[]{Same as Fig. \ref{fig:ssp_sigma}, but separated into Virgo member galaxies (red symbols) and non-Virgo (blue). Magenta cross-bars show the outlying Fe-poor galaxies. Black, red and blue lines show the error-weighted best fit straight line to all galaxies (excluding Fe-poor objects), Virgo, and non-Virgo galaxies respectively. Insert histograms show the residuals of the best fit line to all galaxies, separated into Virgo (red) and non-Virgo (blue) members. $P_0$ gives the probability that the two histograms are drawn from the same populations, as derived from a K-S test. Metallicities are very similar, but age and \afe\/ show a high probability of sampling two different distributions.}
  \label{fig:ssp_sigma_Virgo}
  \end{center}
\end{figure}

Figure \ref{fig:ssp_sigma_Virgo} shows the SSP-equivalent population parameters and velocity dispersion for our \Reff\/ aperture, but this time we colour Virgo members red, and non-members blue. Insert histograms show the residuals of these two sub-samples about the global best fit to the combined sample, quoting also $P_0$: the probability that the chance deviation between the two samples should be larger than that measured, as derived from a Kolmogorov-Smirnov (K-S) test of the two histograms. Small values of $P_0$ imply the distributions are different.

Both Virgo and non-Virgo galaxies show similar trends of population properties with velocity dispersion. However Virgo objects are older by approximately 2.5\,Gyr on average. The two samples deviate in age most significantly when comparing the largest aperture size. This reflects the fact that when Virgo galaxies have young stars, they tend to be centrally concentrated, whereas outside the cluster environment, recent star formation can be widespread across the galaxy on \Reff\/ scales. Differences in abundance ratios are also statistically most significant within the \Reff\/ aperture, showing a very small probability that the measured deviation of the two distributions occurred by chance. Virgo galaxies have systematically higher abundance ratios on these large spatial scales than their counterparts in the field at the {\it same} velocity dispersion. At face value, this indicates a truncated star formation history on large scales compared to outside the cluster, indicative of a process acting to suppress star formation preferentially on larger scales, such as ram-pressure stripping \citep{gunn72}. The suppression of spatially extended star formation in Virgo cluster early-type galaxies is also supported by the study of current-day star formation in early-type galaxies by \cite{shapiro2010}, who found that all early-type galaxies with `widespread' star formation are found outside of dense environments. Likewise, in previous papers of the \atlas\/ series we have reported on truncated distributions of molecular gas \citep{davis13} and a lack of extended HI disks \citep{serra12} within Virgo compared to lower-density regions, further supporting this picture.

The metallicity distributions are essentially the same between Virgo and non-Virgo objects in all aperture sizes, though we note that the `low-Fe' objects are not present in Virgo, suggesting that the cluster environment specifically prohibits the accretion of low metallicity gas, even though the incidence of molecular gas detections within Virgo is statistically indistinguishable from the field \citep{young11}. We return to this point when considering the empirical star formation histories in Section \ref{sec:sfh_env}

Our findings here are consistent with our spatial gradient analysis presented in \cite{scott13a}. Interestingly, although the global properties show differences between Virgo and non-Virgo objects, objects in both environments show the same radial gradient of \mgb\/ with local escape velocity (though Virgo is offset to higher \mgb\/ values). This implies that the mechanism acting to modify the star formation process in the cluster galaxies maintains the close relationship between the stellar population and the local gravitational potential.

Virgo is the largest bound group of galaxies in our sample. However, there are regions of our volume that have small-scale densities comparable to parts of Virgo. Likewise the outer regions of Virgo have densities comparable to that in the field. In \cite{cappellari11b}, we showed that using a local estimator of environmental density gives a well-defined relation of kinematic morphologies with environment. Here we explore if such similar trends exist for the stellar populations.

\begin{figure*}
 \begin{center}
  \epsfig{file=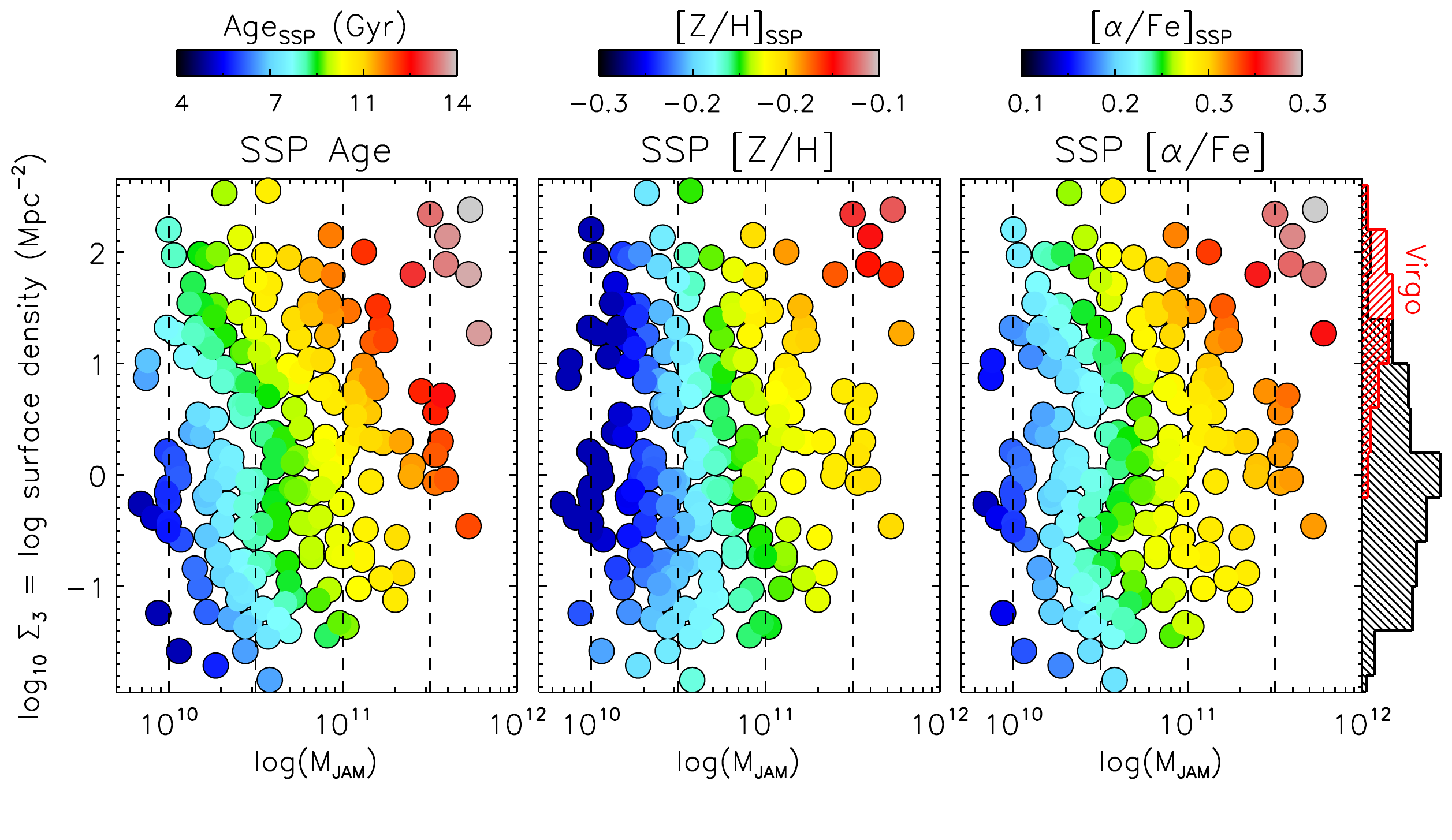, angle=0, width=17cm}
  \epsfig{file=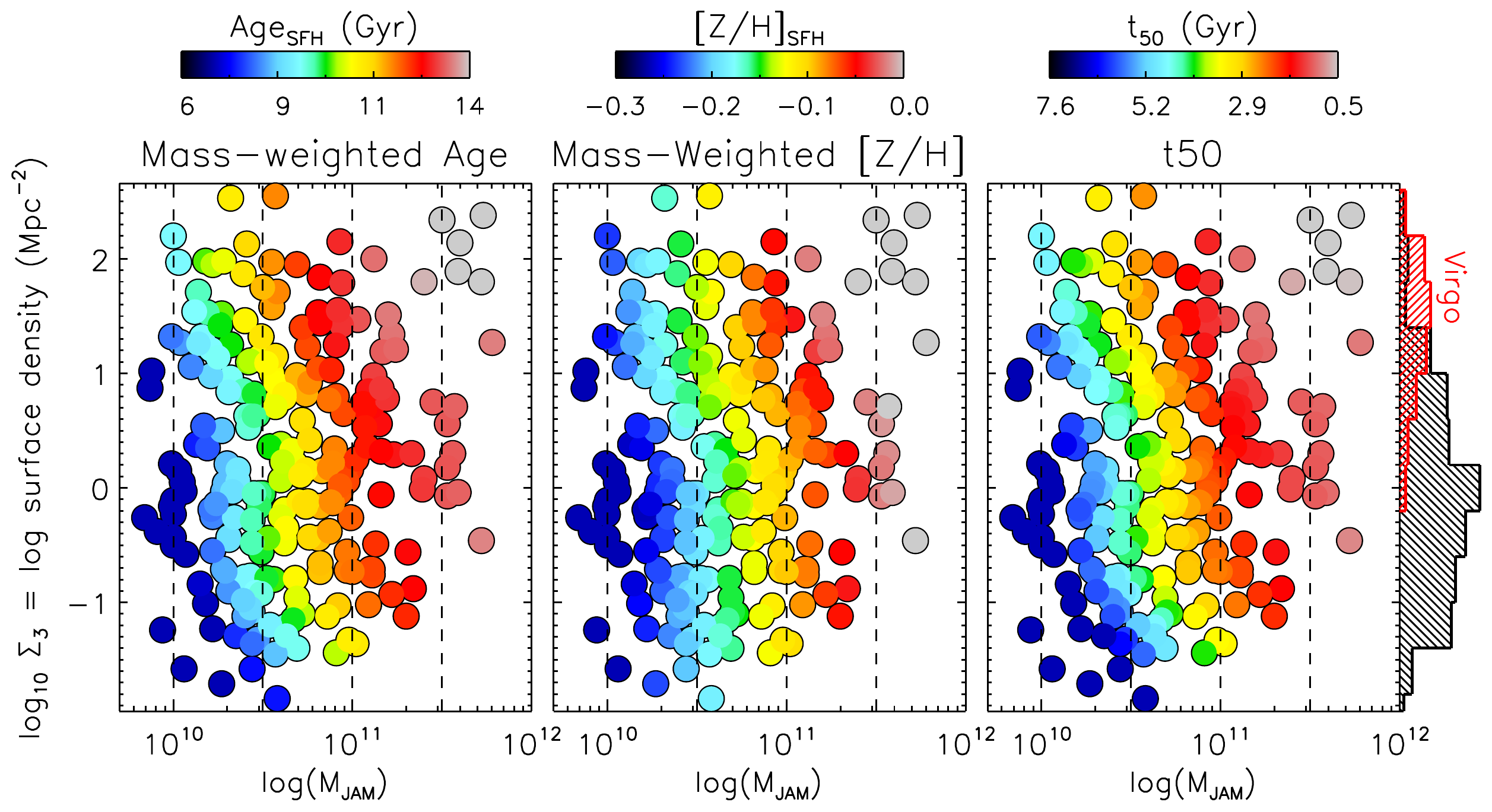, angle=0, width=17cm}
  \caption[]{Distribution of dynamical mass $M_{\rm JAM}$ versus environment density tracer $\Sigma_3$, defined in the main text. In each panel, points are coloured according to the stellar population parameters indicated in the plot title, after locally averaging the measured values using the LOESS technique. The top row provides the SSP-equivalent parameters of age, metallicity and abundance ratio; the bottom row gives the mass-weighted age, metallicity and half-mass formation time. The far-right histograms indicate how Virgo cluster members (red) and non-members (black) are distributed with $\Sigma_3$.}
  \label{fig:ssp_mass_environment}
  \end{center}
\end{figure*}

Figure \ref{fig:ssp_mass_environment} shows the LOESS-smoothed distribution of SSP-equivalent and mass-weighted population parameters (both use \Reff\ apertures) as a function of dynamical mass and environmental density. Environment is quantified by $\Sigma_3 = N_{\mathrm{gal}}/(\pi R_3^2)$, which gives the surface density in Mpc$^{-2}$ of galaxies inside a cylinder of radius $R_{3}$ and height $h = 600$\kms\/ (i.e. $\Delta V_{\mathrm{hel}} < 300\kmsm)$ centered on the galaxy, which includes the $N_{\mathrm{gal}} = 3$ nearest neighbours from the parent sample, as described in \cite{cappellari11a}.

\begin{table*}
\begin{center}
\caption{Line fit parameters.}
\begin{tabular}{ccccccccccccc}
\multicolumn{13}{c}{(i) SSP = $a\log$(\sigmae)$+b$ (Figure \ref{fig:ssp_sigma})} \\
\hline
\multicolumn{3}{c}{$\log$(Age$_{\rm SSP}$)} & & \multicolumn{3}{c}{[$Z/H$]$_{\rm SSP}$} & & \multicolumn{3}{c}{[$\alpha/Fe$]$_{\rm SSP}$ } & & \\
 a & b & $\sigma$ & & a & b & $\sigma$ & & a & b & $\sigma$ &  & Notes \\
\cline{1-3}\cline{5-7}\cline{9-11}\cline{13-13}
 1.14$\pm$ 0.09 & -1.59$\pm$ 0.13 &  0.22 & &  0.31$\pm$ 0.04 & -0.62$\pm$ 0.06 &  0.10 & &  0.37$\pm$ 0.03 & -0.61$\pm$ 0.04 &  0.06 & & \Reffe \\
 0.93$\pm$ 0.08 & -1.07$\pm$ 0.11 &  0.20 & &  0.45$\pm$ 0.04 & -1.09$\pm$ 0.06 &  0.10 & &  0.33$\pm$ 0.03 & -0.49$\pm$ 0.04 &  0.07 & & \Refft \\
 0.83$\pm$ 0.08 & -0.83$\pm$ 0.11 &  0.19 & &  0.45$\pm$ 0.04 & -1.16$\pm$ 0.06 &  0.10 & &  0.31$\pm$ 0.03 & -0.44$\pm$ 0.05 &  0.08 & & \Reff \\

\cline{1-3}\cline{5-7}\cline{9-11}\cline{13-13}
\hline
\\
\\
\\
\multicolumn{13}{c}{(ii) SFH = $a\log$(\sigmae)$+b$ (Figure \ref{fig:mw_sigma})}\\
\hline
\multicolumn{3}{c}{$\log$(Age$_{\rm SFH}$)} & & \multicolumn{3}{c}{[$Z/H$]$_{\rm SFH}$} & & \multicolumn{3}{c}{ t$_{50}$ } & &   \\
 a & b & $\sigma$ & & a & b & $\sigma$ & & a & b & $\sigma$ & &  Notes\\
\cline{1-3}\cline{5-7}\cline{9-11}\cline{13-13}
 0.54$\pm$ 0.04 & -0.12$\pm$ 0.06 &  0.10 & &  0.63$\pm$ 0.04 & -1.52$\pm$ 0.06 &  0.09 & & -10.47$\pm$ 0.76 & 25.30$\pm$ 1.11 &  1.93 & & \Reff \\
\cline{1-3}\cline{5-7}\cline{9-11}\cline{13-13}
\hline
\\
\\
\\
\multicolumn{13}{c}{(iii) SSP = $a\log$(\mjam)$+b$ (Figure \ref{fig:ssp_mass}, left)} \\
\hline
\multicolumn{3}{c}{$\log$(Age$_{\rm SSP}$)} & & \multicolumn{3}{c}{[$Z/H$]$_{\rm SSP}$} & & \multicolumn{3}{c}{[$\alpha/Fe$]$_{\rm SSP}$ } & &   \\
 a & b & $\sigma$ & & a & b & $\sigma$ & & a & b & $\sigma$ & &Notes\\
\cline{1-3}\cline{5-7}\cline{9-11}\cline{13-13}
 0.25$\pm$ 0.03 & -1.70$\pm$ 0.10 &  0.21 & &  0.12$\pm$ 0.02 & -1.49$\pm$ 0.05 &  0.11 & &   0.08$\pm$ 0.01 & -0.66$\pm$ 0.04 &  0.08 & & \Reff \\
\cline{1-3}\cline{5-7}\cline{9-11}\cline{13-13}
\hline
\\
\\
\\
\multicolumn{13}{c}{(iv) SFH = $a\log$(\mjam)$+b$ (Figure \ref{fig:ssp_mass}, right)} \\
\hline
\multicolumn{3}{c}{$\log$(Age$_{\rm SFH}$)} & & \multicolumn{3}{c}{[$Z/H$]$_{\rm SFH}$} & & \multicolumn{3}{c}{t$_{50}$ } & &   \\
 a & b & $\sigma$ & & a & b & $\sigma$ & & a & b & $\sigma$ & &Notes\\
\cline{1-3}\cline{5-7}\cline{9-11}\cline{13-13}
 0.15$\pm$ 0.02 & -0.60$\pm$ 0.05 &  0.12 & &  0.21$\pm$ 0.01 & -2.39$\pm$ 0.05 &  0.09 & &  -2.94$\pm$ 0.32 & 34.45$\pm$ 1.04 &  2.13 & & \Reff \\
\cline{1-3}\cline{5-7}\cline{9-11}\cline{13-13}
\hline
\\
\\
\\
\multicolumn{13}{c}{(v) SSP = $a\log$(\sigmae)$+b$ (Figure \ref{fig:ssp_sigma_kinem}, kinematic classes)} \\
\hline
\multicolumn{3}{c}{$\log$(Age$_{\rm SSP}$)} & & \multicolumn{3}{c}{[$Z/H$]$_{\rm SSP}$} & & \multicolumn{3}{c}{[$\alpha/Fe$]$_{\rm SSP}$ } & &   \\
 a & b & $\sigma$ & & a & b & $\sigma$ & & a & b & $\sigma$ & &Notes\\
\cline{1-3}\cline{5-7}\cline{9-11}\cline{13-13}
 0.97$\pm$ 0.09 & -1.12$\pm$ 0.12 &  0.20 & &  0.53$\pm$ 0.05 & -1.34$\pm$ 0.07 &  0.10 & &  0.24$\pm$ 0.04 & -0.28$\pm$ 0.06 &  0.09 & & Regular Rotators \\
 0.82$\pm$ 0.14 & -0.81$\pm$ 0.20 &  0.17 & &  0.74$\pm$ 0.10 & -1.86$\pm$ 0.14 &  0.12 & &  0.17$\pm$ 0.07 & -0.14$\pm$ 0.10 &  0.09 & & Non-Regular Rotators \\
 0.93$\pm$ 0.07 & -1.06$\pm$ 0.11 &  0.20 & &  0.57$\pm$ 0.04 & -1.43$\pm$ 0.06 &  0.12 & &  0.22$\pm$ 0.03 & -0.25$\pm$ 0.05 &  0.09 & & All (inc. low-Fe) \\
\cline{1-3}\cline{5-7}\cline{9-11}\cline{13-13}
\hline
\\
\\
\\
\multicolumn{13}{c}{(vi) SSP = $a\log$(\sigmae)$+b$ (Figure \ref{fig:ssp_sigma_Virgo}, Virgo members)} \\
\hline
\multicolumn{3}{c}{$\log$(Age$_{\rm SSP}$)} & & \multicolumn{3}{c}{[$Z/H$]$_{\rm SSP}$} & & \multicolumn{3}{c}{[$\alpha/Fe$]$_{\rm SSP}$ } & &   \\
 a & b & $\sigma$ & & a & b & $\sigma$ & & a & b & $\sigma$ & &Notes\\
\cline{1-3}\cline{5-7}\cline{9-11}\cline{13-13}
 0.51 $\pm$ 0.11 & -0.06 $\pm$ 0.16 &  0.14 & &  0.48 $\pm$ 0.07 & -1.23 $\pm$ 0.10 &  0.08 & &  0.25 $\pm$ 0.05 & -0.30 $\pm$ 0.07 &  0.06 & & Members \\
 0.96 $\pm$ 0.09 & -1.14 $\pm$ 0.13 &  0.18 & &  0.44 $\pm$ 0.05 & -1.14 $\pm$ 0.07 &  0.10 & &  0.36 $\pm$ 0.04 & -0.57 $\pm$ 0.06 &  0.08 & & Non-Members \\
\cline{1-3}\cline{5-7}\cline{9-11}\cline{13-13}
\hline
\label{tab:params}
\end{tabular}
\end{center}
Note: The above table gives the parameters for the linear fits shown in the figures indicated in the table. The fits were made using the outlier-resistant two-variable linear regression method implemented in the IDL routine `robust\_linefit.pro' of the IDL NASA Astronomy Library \citep{landsman93}. The $\sigma$\ column gives the standard deviation of the fit to the \atlas\/ data. The `Notes' column indicates the aperture size and relevant sub-samples used in the fit. If no aperture size is given, \Reff\/ is assumed.
\end{table*}

\begin{figure*}
 \begin{center}
  \epsfig{file=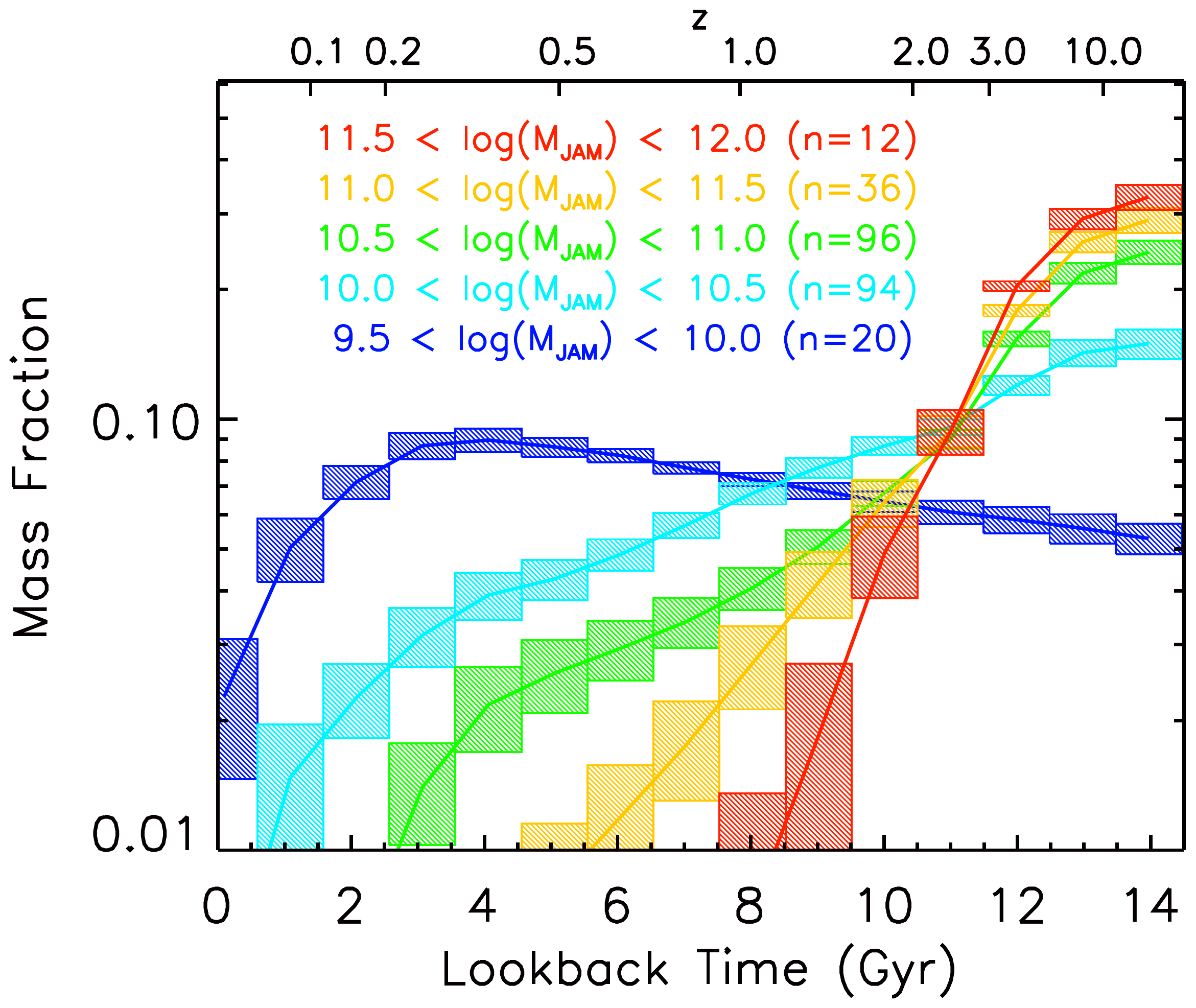, angle=0, width=13cm}
  \caption[]{The average star formation history of the corresponding bin in dynamical mass, \mjam. The star formation history is quantified here as the relative fraction of stellar mass formed at each epoch, derived from the regularised weighted combination of MIUSCAT SSP models fitted to the aperture spectrum. Hatched regions indicate the standard error on the mean at each lookback time, estimated from the standard deviation in each mass-bin of age divided by $\sqrt{n}$, where $n$ is the number of objects in the mass bin as indicated. The upper horizontal scale indicates corresponding redshift, assuming $\Omega_0=\Omega_m + \Omega_\Lambda = 1$, where $\Omega_m=0.27$ and $\Omega_\Lambda = 0.73$.}
  \label{fig:sfh_mass}
  \end{center}
\end{figure*}

It can be seen from Figure \ref{fig:ssp_mass_environment} that regions of constant stellar population parameters show varying dependency on both mass and environment. The population properties are essentially independent of environment over two orders of magnitude (within the range $-1 \leq \log(\Sigma_3) \leq 1$). The highest-density regions, dominated by Virgo cluster members (see vertical histograms for the distribution of members with $\log(\Sigma_3)$), shift markedly to older ages, higher abundance ratios, and earlier half-mass formation timescales for a given mass with respect to the lower density regions. Metallicity shows only a weak difference between high and low densities, in the sense that the highest densities show mildly higher metallicities. At the lowest densities ($\log(\Sigma_3)<-1$), there is a weak inflection towards even younger ages, lower abundances, and longer half-mass formation timescales.

Most of our sample (62\%) lie within the range $-1 \leq \log(\Sigma_3) \leq 1$, within which galaxy mass is the only significant driver of stellar population differences. As we have shown, mass-weighted properties are less sensitive to small amounts of recent star formation, and therefore trace the build up of stars over the entire lifetime of a galaxy. The consistent trends between the mass-weighted and SSP-equivalent indicators therefore imply that, on average, the influence of environment on the stellar populations of early-type galaxies has a long-term influence on the galaxies' evolution. The suppression of only {\it recent} star formation in cluster galaxies, for example, is not sufficient to explain the trends we see, which extend across all masses, and in galaxies where significant levels of star formation is long-since over. As we show in Section \ref{sec:sfh}, there must also have been a more rapid build of up massive galaxies in high density environments at early times, followed by subsequent preferential suppression of star formation in those regions. Outside of cluster-like over-densities, the mass of individual galaxies (or perhaps more accurately, the parent haloes of today's galaxies) seems to have been the dominant influence governing the star-formation history over the effects of the local galaxy density. This supports a picture of separable effects of mass and environment on the stellar populations of ETGs, consistent with \cite{peng10}. The weak influence of environment (as compared with mass) in low- and intermediate-density regimes is also similar to the findings of \citet{thomas10}.

%
%

\begin{figure*}
 \begin{center}
  \epsfig{file=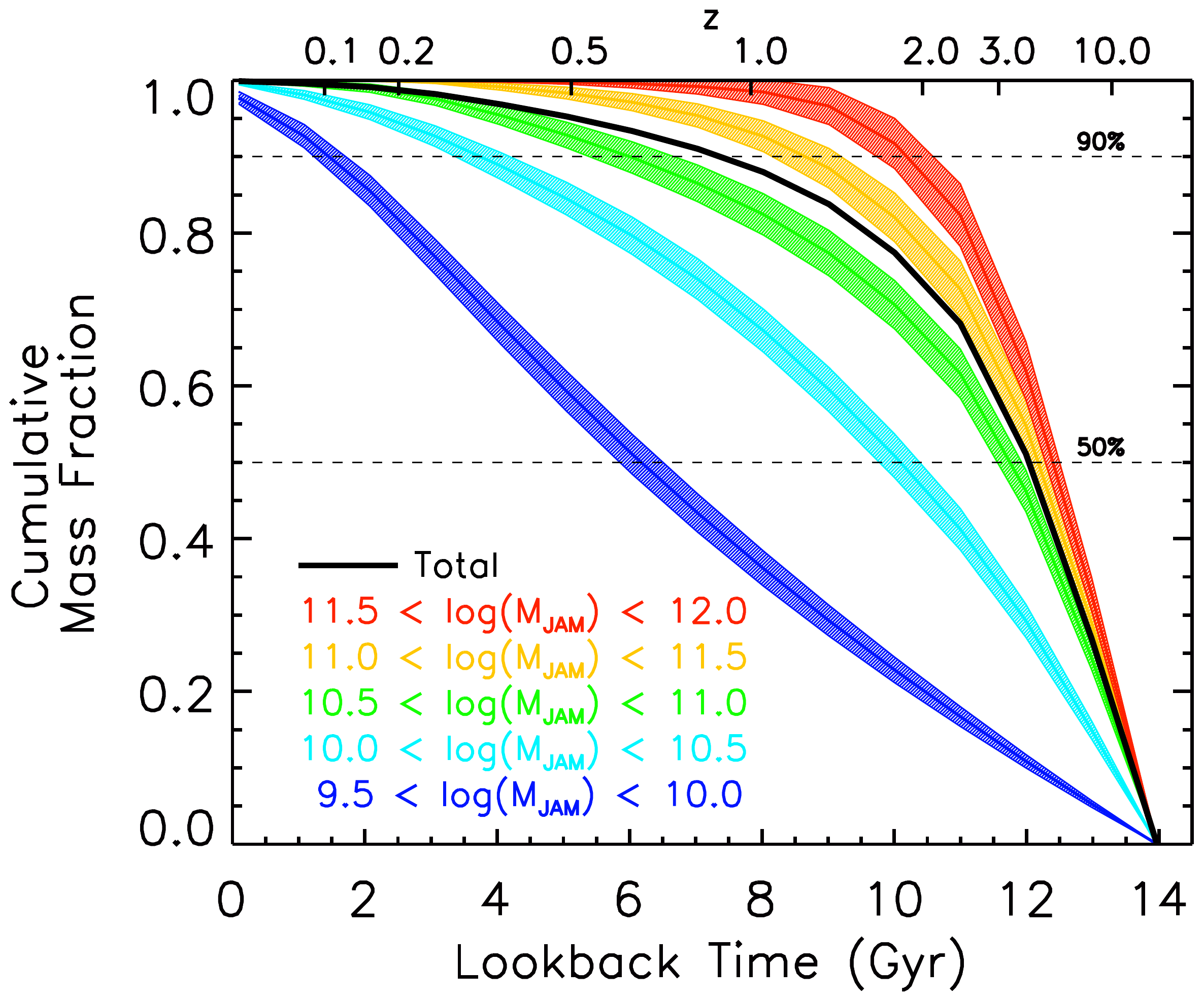, angle=0, width=13cm}
  \caption[]{Cumulative functions of the star formation histories shown in Figure \ref{fig:sfh_mass}, showing the integrated mass fraction existing at given time. Colours correspond to mass bins as in Figure \ref{fig:sfh_mass}. Hatched regions indicate the dispersion of values at each time step.}
    \label{fig:sfh_cum_mass}
  \end{center}
\end{figure*}

\section{Star formation histories}
\label{sec:sfh}

In this section, we use our spectral fits to the integrated \Reff\/ aperture spectra to investigate the history of star formation within our sample galaxies. The results which follow use the same spectral fits used to determine the mass-weighted population parameters presented above. As described in Section \ref{sec:specfit}, we make a regularized linear fit to our aperture spectra of SSP model spectra sampling a grid of age and metallicity. The resulting distribution of weights provides an estimate of the star formation history (SFH) of the galaxy in question, by describing the mass contributions of different populations with given formation epochs. The spectral fit is made using model spectra logarithmically-spaced in age. Here we rebin the resulting logarithmically-spaced weights onto a linear time axis, rigorously preserving the total mass fractions within each linear time step (set as 1\,Gyr).

\subsection{Star formation histories as a function of mass}
\label{sec:sfh}

Figure \ref{fig:sfh_mass} shows the measured SFHs of our sample averaged within five bins of dynamical mass (as measured at $z=0$), spanning the range of our sample. The SFH is presented as the mean fraction of stellar mass formed in each 1\,Gyr step, down to mass fractions of 1\%, for galaxies within the (present-day) mass bin. The variation of mass fraction formed within each 1\,Gyr bin is indicated by the hatched regions, quantified as the mean standard deviation within each 1\,Gyr bin.

On average, objects with \mjam$ > 10^{10}$\msun\/ ($>90$\% of our sample) had their peak in star formation at the earliest epochs, followed by a decline that is systematically more rapid at higher (present-day) masses. In our lowest mass bin, the behaviour is different: on average, these objects build a gradually increasing fraction of their stellar mass as the universe evolves, until a more rapid decline beginning around 4\,Gyr ago. This decline is most easily interpreted as a result of our sample selection, which targeted galaxies consistent with an early-type morphology, and thus preferentially selects objects with low present-day star formation rates (though this is not a selection requirement, and many objects show evidence for ongoing star formation). We note also that this lowest mass bin likely suffers from incompleteness, as the sample was selected based on K-band luminosity, which has some variation in mass-to-light ratio with stellar population properties. This may bias against old populations at this low mass extreme, as they will be slightly fainter than their younger counterparts. Nonetheless, the lowest mass bin is consistent with the overall trend of more extended star formation in lower mass early-type galaxies, with a truncation at more recent times than the high mass objects.

The star-formation histories shown in Fig. \ref{fig:sfh_mass} provide an alternative picture to that of the schematic view inferred from assuming linear trends of SSP-equivalent parameters, as in \cite{thomas05,thomas10}. In that case, galaxies of a given mass have star formation rates that rise and fall symmetrically in time, being described by a Gaussian centred on the mean SSP-equivalent age, with a width given by a conversion of abundance ratio to star-formation timescale. With that approach, the stellar population scaling relations arise from a systematic delay in the peak of star formation in lower mass galaxies towards later times, combined with a corresponding increase in the star-formation timescale. 

By comparison, our non-parametric approach shows that objects of all masses experience a significant epoch of star-formation at the earliest times, and that the subsequent variation in present-day stellar population properties is mainly driven by the differing rate at which star formation declines for galaxies of different masses. The overall picture is qualitatively similar between both approaches. However, by reproducing the observed spectra with a distribution of ages and metallicities directly, our non-parametric star-formation histories emphasise the significance of early star formation in all galaxies, and the subsequent decline in star formation activity that varies systematically with mass, independent of the timescale of the individual star formation events. This reflects the conceptual differences of fitting the full spectrum with a distribution of models to infer the star formation history directly, compared to using line-strength indices to estimate mean properties. Our findings are in good agreement, for example, with the previous studies of \cite{heavens04} and \cite{panter07}, who used comparable spectral-fitting methods on thousands of aperture spectra from SDSS (though without any morphological selection applied).

Figure \ref{fig:sfh_cum_mass} presents the associated mean cumulative mass fraction as a function of look-back time for the same mass bins as in Figure \ref{fig:sfh_mass}, as well as the total cumulative formed mass for the entire sample. Assuming that our sample represents most of the stellar content of today's local Universe, approximately 50\% of all stars formed within the first 2\,Gyr following the big bang. The vast majority of these stars reside today in the most massive galaxies, which themselves formed 90\% of their stars within the first 3\,Gyr (i.e. by $z \sim$2). The low mass objects, in contrast, have formed barely half their stars in this time interval, forming the remaining 50\% of their stellar mass in the last 6--10\,Gyr. We note that, from this analysis, we can only tell {\em when} the stars formed, and not whether they formed within the gravitational potential of the current-day galaxy, or instead were formed in separate, smaller systems that later combined to form the galaxy we see today.

\begin{figure*}
 \begin{center}
  \epsfig{file=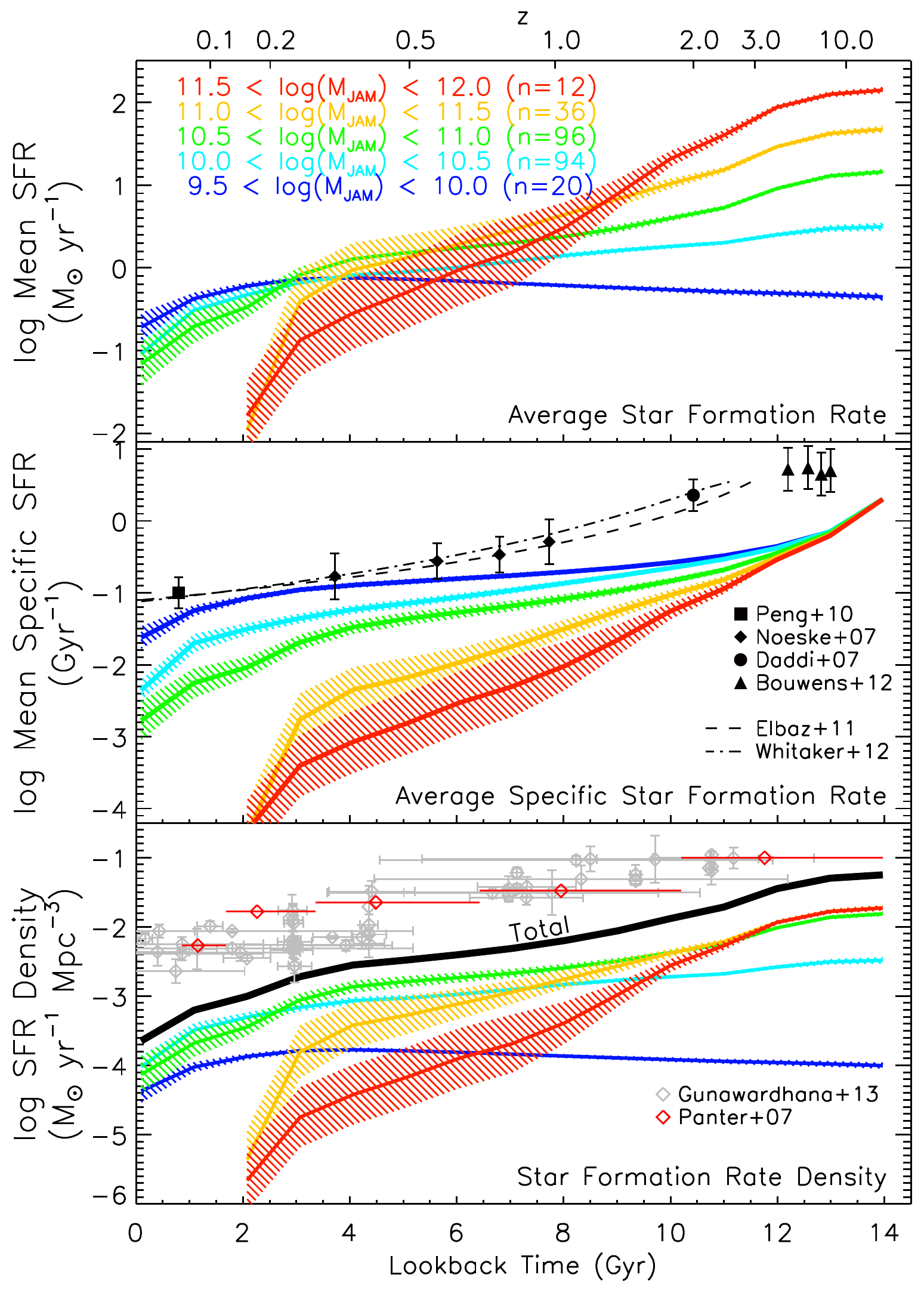, angle=0, width=14cm}
  \caption[]{{\em Top:} Star formation rates averaged within the indicated (present-day) mass bins. These are calculated by multiplying the mass fractions in Figure \ref{fig:sfh_mass} by the dynamical mass \mjam. {\em Middle:} Mean (time-averaged) specific star formation rate in each $\Delta{\rm t_{SFH}} = 1$\,Gyr time step (thick solid lines), obtained by normalising the star formation rate by the mass formed up to the epoch considered. Direct measurements of the instantaneous sSFR from the literature are also shown, either as discrete values (black data points) or empirically-constrained continuous relations (broken lines). {\em Bottom:} Integrated star formation rate density for our local volume, as a function of present day galaxy mass, and the integrated total for our sample (black line). Grey points show instantaneous star formation rate density measurements from the compilation of \cite{gunawardhana13}; red points show the measurements of \cite{panter07} using a similar approach to our analysis. In all panels, hatched regions show the standard deviation of properties within each age and mass bin.}
    \label{fig:sfr}
  \end{center}
\end{figure*}

Figure \ref{fig:sfr} shows the fractional formation masses in absolute terms (top panel), which allows comparison to other studies that measure star formation rates. We do this by assuming that our star formation histories apply to the galaxy's total mass as quantified by \mjam, scaling this present-day mass by the weights returned by pPXF. We also derive the history of {\em specific} star formation rates (middle panel) by normalising the star formation rate by the integrated mass formed by the epoch being considered (though the galaxy may not have assembled yet). In addition, since our sample is volume-limited, we also present the history of star formation rate density for our local volume implied by our analysis, both as a function of mass, and the combined total for our full sample (bottom panel).

The top panel of Figure \ref{fig:sfr} clearly shows that the most massive current-day galaxies had the highest absolute star formation rates of all such objects for the first 4\,Gyr following the big bang. The bottom panel of Figure \ref{fig:sfr}, however, shows that these objects do not dominate the history of integrated star formation rate density, due to their smaller number density. 

The middle panel of Figure \ref{fig:sfr} presents the time-averaged specific star formation rate (sSFR), defined as the mean star formation rate during the 1\,Gyr time bin divided by the mass formed up to that epoch (taken as the mass formed up to the centre of the time bin). For comparison, we show direct measurements of the instantaneous sSFR taken from the literature. Our derived sSFR values lie generally well below the direct measurements, which is most easily explained by the inconsistent timescales probed by the two measures of star formation rate. Our `archeological' values probe the mean time-averaged star formation during each $10^9$ year time step, whereas the direct measurements probe the star formation occurring on timescales of $10^6-10^8$ years. The super-solar abundance ratios of our most massive (and oldest) galaxies require star formation to have occurred on characteristic timescales of order $10^8$ year \citep{thomas05}, which is below the time resolution of our spectral fitting at these early epochs. Such shorter `bursts' of star formation would boost our effective star SFR and sSFR values by an order of magnitude, potentially making them more consistent with direct instantaneous measurements. However, we cannot directly constrain this timescale during the oldest age bins using our current approach.

\begin{figure}
  \epsfig{file=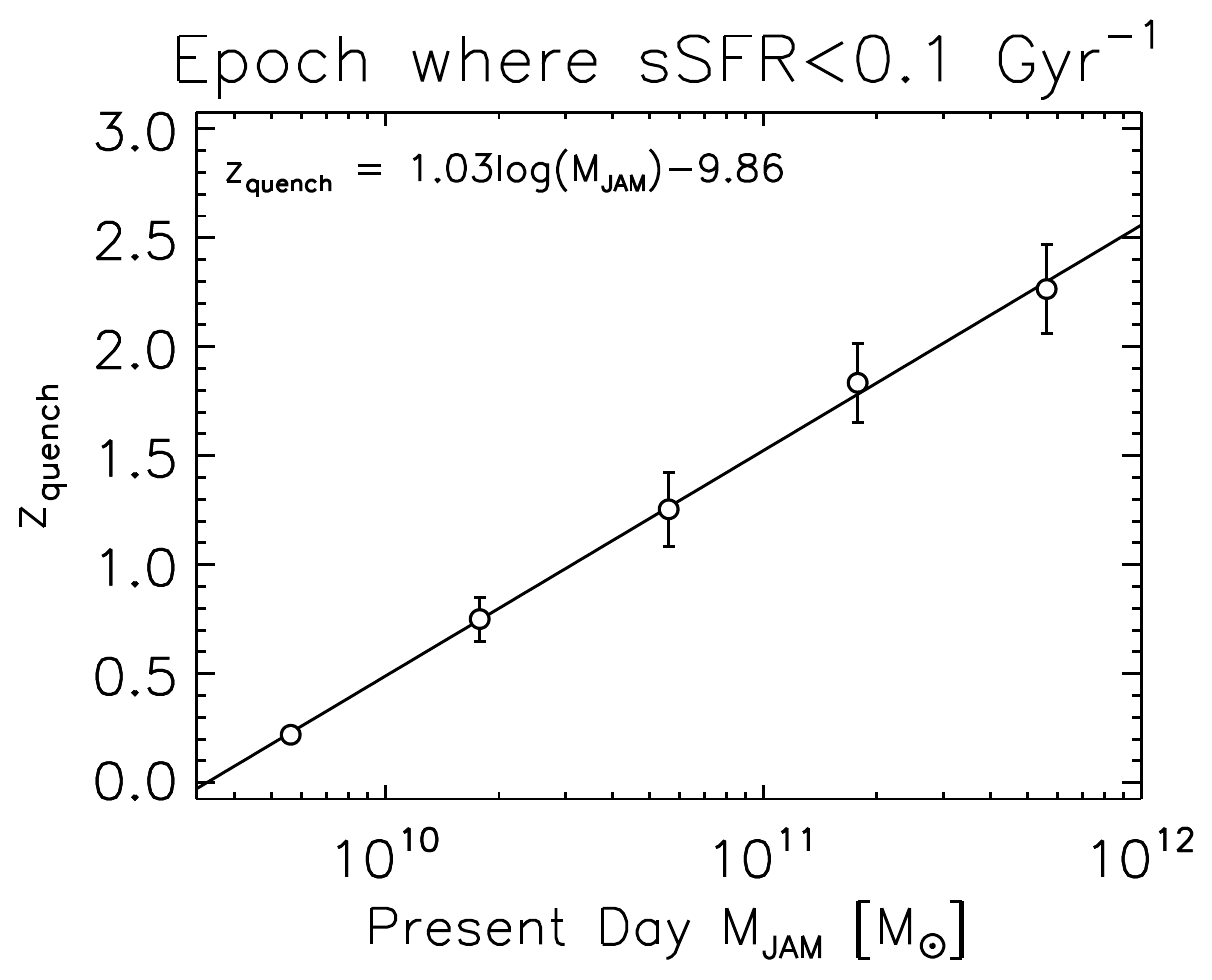, angle=0, width=8cm}
  \caption[]{Redshift at which the sSFR in different bins of galaxy mass, \mjam, crosses the value of 0.1\,Gyr$^{-1}$, corresponding to the local sSFR of main-sequence star-forming galaxies from \cite{peng10}. The quenching redshift follows an approximately linear relation with mass, as indicated by fit (the function is also given). Errors are computed using the spread within each mass bin, as indicated with hatched regions in Figure \ref{fig:sfr}.}
    \label{fig:quench}
\end{figure}

In addition to a redshift dependence, the sSFR histories show a strong mass-dependence of the sSFR at a given redshift, irrespective of the assumed star-formation timescale. This indicates the influence of a mass-dependent feedback mechanism that is relatively more efficient in higher-mass galaxies, reducing the sSFR more rapidly than at lower masses, and bringing higher-mass galaxies off of the star-formation main sequence at earlier times. Through our `archeological' approach, we can directly trace when galaxies turn off of the star-formation main sequence and become quiescent. Due to the problems of comparing instantaneous and archeological sSFR estimates, however, we cannot easily compare with the measured sSFR main sequence at high redshift. Instead, we adopt the sSFR of 0.1\,Gyr$^{-1}$ measured in local galaxies from \cite{peng10} as a robust threshold between `star forming' and `quiescent' states, and measure the epoch at which our galaxies cross this threshold. Figure \ref{fig:quench} presents this `quenching redshift' for the different mass bins. We find a near-linear dependence with $\log$(\mjam), with galaxies of mass above $10^{11}$\,\msun\/ quenching their star-formation at redshift $z=2-4$, whereas galaxies below $10^{10}$\,\msun\/ quench below redshift 1, with our lowest mass galaxies quenching below $z\sim0.2$ (in the last 1\,Gyr), on average.

The cosmic SFR density for \atlas\/ given in the bottom panel of Figure \ref{fig:sfr} also shows measurements from \cite{panter07} derived using a similar technique to ours using SDSS spectra of $\sim 3 \times 10^5$ galaxies, and from the compilation of \cite{gunawardhana13} using various measures of the (instantaneous) star formation rate. Our estimate of the total star formation rate density is significantly lower than these other studies, differing by an order of magnitude at most recent times, but with better agreement at the earliest epochs. We interpret this as a result of our morphological selection, which effectively excludes galaxies with strong total star formation rates at the current time, and which have likely contributed significantly to the integrated star formation budget of the universe over its lifetime. In contrast, the literature studies either impose no particular mass or morphological selection \citep[as in the case of][]{panter07}, or preferentially select galaxies with detectable star formation \citep[as in the data compiled by][]{gunawardhana13}, both of which will act to increase the inferred integrated star formation rate. Cosmic variance may also play a role given our modest survey volume, which for example lacks extremely massive clusters like Coma. In general, however, we see that early-type galaxies have experienced a similarly-paced decline in star formation rate as the total population, though with a more rapid drop off and reduction in contributions from galaxies of higher masses.

%
%

\subsection{Star formation histories as a function of environment}
\label{sec:sfh_env}

\begin{figure*}
  \epsfig{file=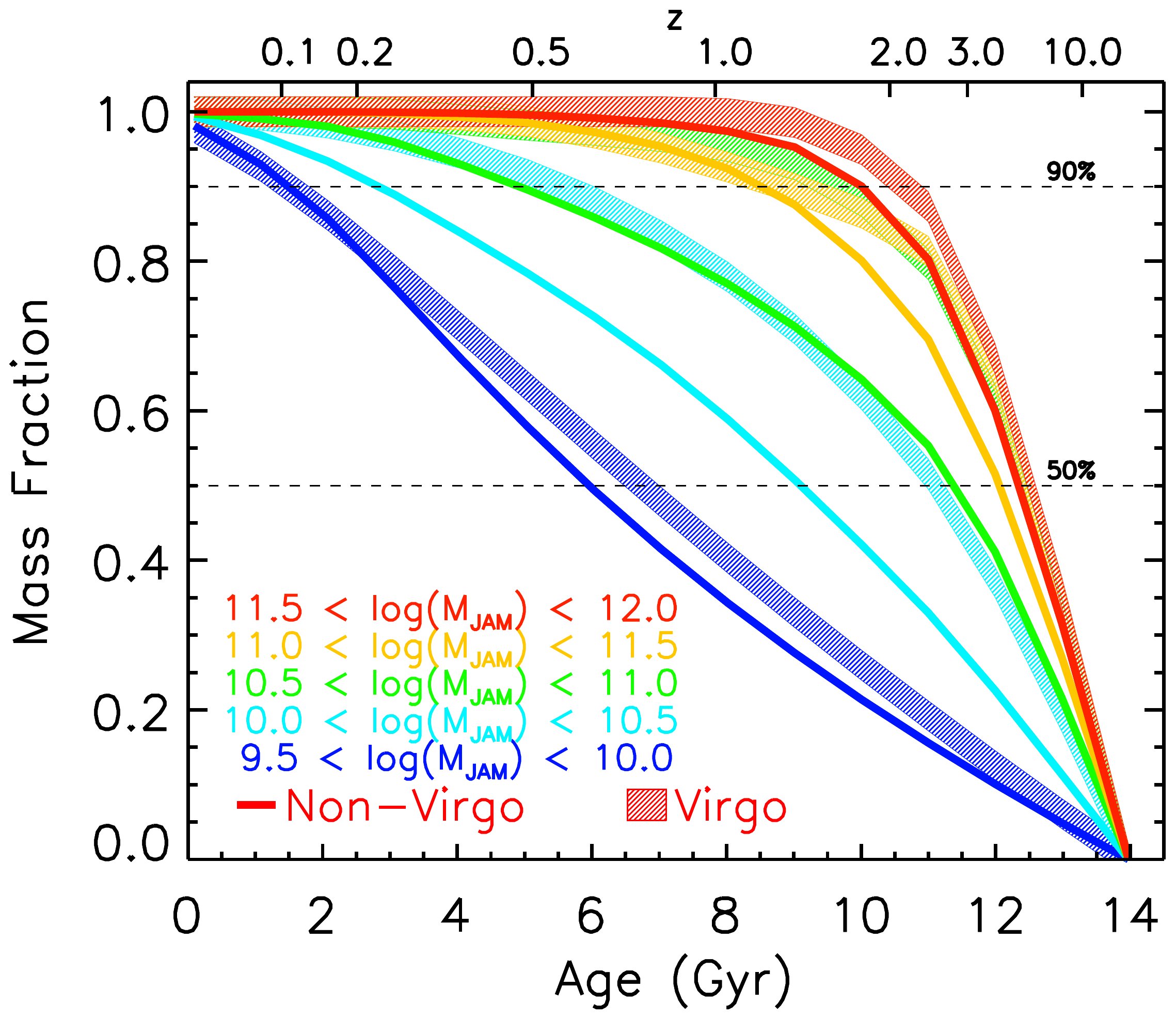, angle=0, width=13cm}
  \caption[]{Cumulative star formation history for Virgo members (thick hatched regions) compared to outside of Virgo (solid lines) for galaxies averaged in bins of dynamical mass. Line colours show the range of dynamical masses considered, as indicated.}
  \label{fig:sfh_virgo_mass}
\end{figure*}

In Section \ref{sec:environment} we presented evidence for differing stellar population properties in the different environments present in our sample. Here we investigate whether this is also recovered directly in the SFHs determined from spectral fitting. Figure \ref{fig:sfh_virgo_mass} shows the mean cumulative fraction of stellar mass formed as a function of time for members of the Virgo cluster and the rest of the sample, binned by present-day mass. The population outside of Virgo (solid lines) show consistently more extended star formation than the objects that are Virgo members (hatched regions), with the exception of the lowest mass bin, where the extended star formation is similar in both populations. This low-mass bin has just three Virgo objects, all of which are located in the outer regions of the cluster, and are presumably entering Virgo for the first time. Overall, these trends directly show the impact of environment on the mass growth and star-formation histories of early-type galaxies, in addition to the mass-dependent effects illustrated in the previous section.

The apparent suppression of extended star formation within Virgo compared to objects of the same mass outside of the cluster environment is qualitatively consistent with an extension of what is found in Virgo late-type galaxies, where the intra-cluster medium appears to truncate extended neutral gas disks with subsequent implications on star-formation activity \citep{koopmann04,crowl08,chung09}. Indeed, even though the detection rate of CO in our sample is not statistically different between Virgo members and non-members \citep{young11}, we have shown in \cite{davis13} that this present-day reservoir of molecular gas responsible for fuelling current star formation is spatially less extended in the Virgo galaxies, and often shows signs of recent interaction and/or truncation.

Further evidence comes from the kinematic alignment of gas in our sample. In \cite{davis11} we showed that gas in various phases (ionised, molecular or neutral) is almost always aligned with the stars in Virgo galaxies, in contrast to galaxies in low-density environments where more than half show misalignments. This strongly suggests that the gas supply in the cluster is limited to internal recycling. Without any additional external accretion of cold gas to maintain or periodically boost levels of star formation, galaxies in the cluster have less extended star formation histories compared to those of similar mass outside the cluster environment. In less dense environments, accretion of cold gas can still occur, as indicated by the high incidence of gas components misaligned to the stellar motions. Recall also that none of the extremely low metallicity, gas rich objects shown in Figure\,\ref{fig:co_pop} are found in Virgo.

Our star formation histories suggest that this is not restricted to a recent suppression of star formation. On average, cluster objects have {\em always} exhibited shorter formation timescales than field galaxies of comparable present-day mass. This finding is in good agreement with the predictions of \cite{khochfar08}, where a combination of ram pressure stripping and gravitational heating act to reduce the duration of star formation in dense environments. The corollary to this is that, at early times when cluster galaxies were rapidly forming the bulk of their stars, their effective star formation rate would have been higher than that of galaxies in the field that would eventually match their mass in the present day. Such a picture is consistent with direct observations at high redshift, where an apparent `reversal' of the low redshift behaviour (where star formation is lower in high density regions) occurs, and dense environments exhibit enhanced star formation compared to the field \citep{elbaz07,tran10}, potentially driven by cold flows \citep{khochfar09}. This phase would be largely completed by redshift $z \sim 2$, by which time the most massive galaxies have formed their stars, and galaxies of lower mass and in less dense environments become the primary sources of star formation for the remaining 10\,Gyr.

%
%
\section{Discussion and Conclusions}
\label{sec:conc}

Our integral-field spectroscopy provides a unique opportunity to study the mass-scaling properties of stellar populations in early-type galaxies with high accuracy, by virtue of minimising the effects of aperture size and slit losses. For a large range of galaxy masses and apparent sizes, we implicitly include the effects of line-strength gradients in our measurements of population properties by matching scaled apertures with effectively no slit loss effects. Our study also uses \sigmae: the velocity dispersion integrated within \Reff, which is empirically close to the true second velocity moment $\langle v^2_{\rm los}\rangle_\infty$ that appears in the scalar virial theorem \citep{binney08}, and is directly proportional to mass. Moreover, the IFU data allows the dynamical mass to be tightly constrained by fitting general dynamical models that account for non-homology and dark matter. Our sample selection is also based on a classical prescription of determining galaxy type, based on the presence of spiral arms and/or a significant dust disk. In this way, we do not exclude galaxies that have ongoing star-formation on anything other than morphological grounds. Finally, we have a large collection of high quality measurements on individual galaxies, allowing the intrinsic scatter of our relations to be accurately assessed.

From consideration of the classic Fundamental Plane, in \cite{cappellari13a} we showed that the intrinsic scatter and tilt of this relation is driven by stellar population variations, including the stellar initial mass function, and that accounting for these effects in the stellar mass-to-light ratio brings galaxy masses into very close agreement with the scalar virial relation \mjam$\propto \sigma_e^2 R_e^{\rm maj}$. Here we explored the distribution of stellar population parameters on the thin plane of ($M_{\rm JAM}, \sigma_e, R_e^{\rm maj}$) - or the `Mass Plane' - and show that there are indeed smooth variations of age, metallicity and abundance ratio within this plane. By consideration of empirical star formation histories derived from spectral fitting, we confirm that this reflects trends in the mass-weighted properties and star formation timescales across the Mass Plane. At fixed mass, galaxies with smaller sizes are on average older, more metal rich, and more alpha-enhanced than their larger counterparts, showing that they formed earlier and in a shorter time. These trends tie directly to a higher mass normalisation of the IMF in these more compact objects \citep{cappellari12,mcdermid14}, as well as a larger bulge fraction and lower molecular gas fractions \citep{cappellari13b}.

Why are today's massive galaxies older, more metal rich, and more enhanced in alpha elements compared to their lower-mass counterparts? There seem to be two processes at work. All of today's early-type galaxies share an early period of intense star formation, making the systems compact, metal rich and alpha enhanced. This is consistent with the picture of \cite{khochfar06a}, who argue that high levels of dissipation at high redshift leads to more compact galaxies, with consequently older ages and enhanced abundance ratios. Lower-mass systems can continue to accrete gas from the inter-galactic medium, thereby increasing their apparent size and forming new stars, resulting in younger average ages when viewed today, with lower average metallicities. Dissipation of accreted gas results in a extended and regular stellar rotation field in these objects, which remains in tact in the absence of any major merger events, producing a `regular' or `fast' rotator \citep{khochfar11,naab14}. In the case where a major merger does occur, depending on the orbital parameters of the interaction, the resulting system can remain as a `fast' rotator, or in the case of misaligned spin vectors of the progenitors, give rise to a `two-sigma' galaxy with a significant fraction of stars on counter-rotating orbits \citep{bois11}.

Massive systems, in contrast, quench early, and thereafter predominantly accrete stars \citep{khochfar06b,oser10}. Most of these stars have also formed early, but in smaller systems with lower metallicity. Via this stellar accretion, these galaxies will also increase in size, but remain old on average. The accreted low mass systems are dispersed at a radius proportional to where the local potential approaches the binding energy of the accreted system \citep{villumsen83}, resulting in a gradient of metallicity, decreasing to large radii \citep{hirschmann13,navarro13}. Moreover, the collisionless accretion also reduces the average angular momentum of the system, resulting in massive `non-rotating' galaxies \citep{naab14}. Where neither accretion of gas nor stars is efficient, galaxies remain compact, with characteristic stellar populations that are relatively old, metal-rich and alpha-enhanced for their mass. We note that in both cases (gas accretion at lower mass, and stellar accretion at higher mass), the stellar metallicity must remain closely related to the local potential, as traced by the spatially-resolved relationship between stellar population parameters and the local escape velocity \citep{scott13a}.

\citet{cappellari13b} discuss evolution in the Mass Plane at length, describing the processes by which objects may evolve within that plane. In summary, the evolutionary picture described there is one of increasing bulge fraction (or decreasing disk fraction) as mass increases, bringing galaxies to higher \sigmae\/ and more compact sizes. This process may be driven by dissipative mergers, which encourage central star formation and increase the effective concentration of stellar mass. Above a characteristic mass of $\log$(\mjam)=11.2, the range of early-type galaxy sizes is dramatically reduced, together with the scatter in stellar population properties. The process driving mass growth in this regime is effective at increasing galaxy size without significantly changing the stellar population properties or velocity dispersion. Simulations of collisionless (or `dry') mergers show behaviour that is largely consistent with this picture \citep{khochfar03,naab09,hilz13}, and given the low gas fractions associated with the oldest and/or most massive early-type galaxies \citep{young11}, are a suitable mechanism for driving the mass growth along the upper mass envelope of the mass-size plane. Moreover, we present evidence that the mean scaling relations for the metallicity in non-regularly rotating galaxies is offset to higher mass and velocity dispersion for a given metallicity, consistent with a major merger origin for these objects.

We also derived empirical, non-parameteric star formation histories (SFHs) for our sample, in order to derive mass-weighted mean ages and metallicities. As expected, the mass-weighted ages are significantly older than the SSP-equivalent ages, which are strongly biased by the presence of bright young stars in galaxies with recent star formation. 

In addition to providing mass-weighted (instead of SSP-equivalent) stellar population properties, our derived star formation histories themselves provide unique insights into the formation of our sample galaxies. We find a very regular dependence of star formation history on present-day mass, consistent with similar results using the SDSS \citep{heavens04,panter07} and at higher redshift \citep{juneau05}. Within bins of galaxy mass, the peak of star formation is, on average, within the first billion years following the big bang for all galaxies with \mjam$>10^{10}$\msun. This provides an alternative view to the picture of staggered Gaussian star formation that is implied assuming linear relations of SSP-equivalent parameters and abundance-ratios to derive formation timescales, as in \cite{thomas05,thomas10}, though the qualitative picture is similar.

Through consideration of the mean absolute star formation rate, we demonstrate that the classic `downsizing' scenario of \cite{cowie96} applies to early-type galaxies, showing that less massive galaxies have more extended star formation histories, and as a result, the characteristic mass of the objects dominating the star formation rate decreases as the Universe evolves. This trend is driven by the increasingly rapid fall off in star formation for galaxies with higher masses.

Our most massive galaxies ($11.5<\log$(\mjam)$<12$) have formed 50\% of their stars no more than 2\,Gyr after the big bang ($z\sim3$). By comparison, the lowest-mass galaxies ($9.5<\log$(\mjam)$<10$) reach the same fraction after around 8\,Gyr ($z\sim 0.6$). This is in reasonable agreement with recent simulation results \citep[e.g.][]{oser10}, where models are indeed characterized by early intense star formation fed by cold flows and subsequent assembly of the formed stars, with ongoing `in situ' star formation more prevalent in lower-mass galaxies. Although we cannot constrain {\it  where} the stars have formed from our integrated measurements, it is certainly the case that the recent \citep[and even ongoing:][]{shapiro2010,davis14} star formation is more present in the central regions of our lower-mass galaxies, supporting the view that this is `in situ'.

Finally, we explore the environmental dependence of our star formation histories, and find that, on average, galaxies in Virgo have an additional suppression of extended-duration star formation with respect to galaxies outside the cluster environment. As a result, comparing galaxies of similar mass, Virgo members build their stellar mass on a shorter timescale than non-members - an effect that is most pronounced at intermediate masses ($10^{10} < $\mjam$ < 10^{11}$\msun). The highest mass (\mjam $> 10^{11.5}$\msun) galaxies are very alike in their star formation histories, forming stars on comparably short timescales (90\% of stellar mass formed within 3\,Gyr). This suggests that the star formation properties of these galaxies are relatively unaffected by their current-day environment. Interestingly, galaxies with the lowest masses (\mjam $<10^{10}$\msun) also share very similar star formation histories, with protracted formation of stars until recent times. These lowest-mass Virgo objects are at large cluster-centric distances, suggesting they have only recently entered the dense cluster environment for the first time, and the effective suppression of star formation by the environment has not yet made an impact.

%
%

\section*{Acknowledgements}

We thank the referee, Daniel Thomas, for a constructive report, which helped improve this paper. RMcD would like to thank Inger J{\o}rgensen for a careful reading of early drafts of this paper, and Ricardo Schiavon for numerous instructive discussions. RMcD and RLD gratefully acknowledge the hospitality of the ESO visitor programme during the preparation of this work. MC acknowledges support from a Royal Society University Research Fellowship. This work was supported by the rolling grants `Astrophysics at Oxford' PP/E001114/1 and ST/H002456/1 and visitor grants PPA/V/S/2002/00553, PP/E001564/1 and ST/H504862/1 from the UK Research Councils. RLD acknowledges travel and computer grants from Christ Church, Oxford and support from the Royal Society in the form of a Wolfson Merit Award 502011.K502/jd. TN acknowledges support from the DFG Cluster of Excellence Origin and Structure of the Universe. MS acknowledges support from a STFC Advanced Fellowship ST/F009186/1. TAD acknowledges the support provided by an ESO fellowship. The research leading to these results has received funding from the European Community's Seventh Framework Programme (/FP7/2007-2013/) under grant agreement No 229517. The authors acknowledge financial support from ESO. SK acknowledges support from the Royal Society Joint Projects Grant JP0869822. NS acknowledges support of Australian Research Council grant DP110103509.

Funding for SDSS-III has been provided by the Alfred P. Sloan Foundation, the Participating Institutions, the National Science Foundation, and the U.S. Department of Energy Office of Science. The SDSS-III web site is http://www.sdss3.org/.

SDSS-III is managed by the Astrophysical Research Consortium for the Participating Institutions of the SDSS-III Collaboration including the University of Arizona, the Brazilian Participation Group, Brookhaven National Laboratory, Carnegie Mellon University, University of Florida, the French Participation Group, the German Participation Group, Harvard University, the Instituto de Astrofisica de Canarias, the Michigan State/Notre Dame/JINA Participation Group, Johns Hopkins University, Lawrence Berkeley National Laboratory, Max Planck Institute for Astrophysics, Max Planck Institute for Extraterrestrial Physics, New Mexico State University, New York University, Ohio State University, Pennsylvania State University, University of Portsmouth, Princeton University, the Spanish Participation Group, University of Tokyo, University of Utah, Vanderbilt University, University of Virginia, University of Washington, and Yale University.
%
%
%

\bibliographystyle{mn2e}

\begin{thebibliography}{}
\bibitem[Planck Collaboration, Ade et al.(2013)]{planckXVI} Planck Collaboration, Ade, P.~A.~R., Aghanim, N., et al.\ 2013, arXiv:1303.5062 
\bibitem[Auger et al.(2010)]{auger10} Auger, M.~W., Treu, T., Bolton, A.~S., et al.\ 2010, ApJ, 724, 511 
\bibitem[\protect\citeauthoryear{Bacon et al.}{2001}]{bacon01} Bacon R., et al., 2001, MNRAS, 326, 23
\bibitem[Balzano(1983)]{balzano83} Balzano, V.~A.\ 1983, ApJ, 268, 602 
\bibitem[Bernardi et al.(2006)]{bernardi06} Bernardi, M., Nichol, R.~C., Sheth, R.~K., Miller, C.~J., \& Brinkmann, J.\ 2006, AJ, 131, 1288
\bibitem[Binney \& Tremaine(2008)]{binney08} Binney, J., \& Tremaine, S.\ 2008, Galactic Dynamics: Second Edition, Princeton University Press
\bibitem[Bois et al.(2011)]{bois11} Bois, M., Emsellem, E., Bournaud, F., et al.\ 2011, MNRAS, 416, 1654 
\bibitem[Bouwens et al.(2012)]{bouwens12} Bouwens, R.~J., Illingworth, G.~D., Oesch, P.~A., et al.\ 2012, ApJ, 754, 83 
\bibitem[Cappellari \& Copin(2003)]{cappellari03} Cappellari, M., \& Copin, Y.\ 2003, MNRAS, 342, 345 
\bibitem[Cappellari \& Emsellem(2004)]{cappellari04} Cappellari, M., \& Emsellem, E.\ 2004, PASP, 116, 138 
\bibitem[Cappellari et al.(2006)]{cappellari06} Cappellari, M., et al.\ 2006, MNRAS, 366, 1126 
\bibitem[Cappellari et al.(2011a)]{cappellari11a} Cappellari, M., Emsellem, E., Krajnovi{\'c}, D., et al.\ 2011, MNRAS, 413, 813 
\bibitem[Cappellari et al.(2011b)]{cappellari11b} Cappellari, M., Emsellem, E., Krajnovi{\'c}, D., et al.\ 2011, MNRAS, 416, 1680 
\bibitem[Cappellari et al.(2012)]{cappellari12} Cappellari, M., McDermid, R.~M., Alatalo, K., et al.\ 2012, Nature, 484, 485 
\bibitem[Cappellari et al.(2013a)]{cappellari13a} Cappellari, M., Scott, N., Alatalo, K., et al.\ 2013, MNRAS, 432, 1709 
\bibitem[Cappellari et al.(2013b)]{cappellari13b} Cappellari, M., McDermid, R.~M., Alatalo, K., et al.\ 2013, MNRAS, 432, 1862 
\bibitem[Cassisi et al.(1997)]{cassisi97} Cassisi, S., Castellani, M., \& Castellani, V.\ 1997, A\&A, 317, 108 
\bibitem[Cattaneo et al.(2008)]{cattaneo08} Cattaneo, A., Dekel, A., Faber, S.~M., \& Guiderdoni, B.\ 2008, MNRAS, 389, 567 
\bibitem[Chung et al.(2009)]{chung09} Chung, A., van Gorkom, J.~H., Kenney, J.~D.~P., Crowl, H., \& Vollmer, B.\ 2009, AJ, 138, 1741
\bibitem[Cid Fernandes et al.(2005)]{cidfernandes05} Cid Fernandes, R., Mateus, A., Sodr{\'e}, L., Stasi{\'n}ska, G., \& Gomes, J.~M.\ 2005, MNRAS, 358, 363 
\bibitem[Cleveland \& Devlin(1988)]{cleveland88} Cleveland W., \& Devlin S., 1988, Journal of the American Statistical
Association, 596
\bibitem[Cowie et al.(1996)]{cowie96} Cowie, L.~L., Songaila, A., Hu, E.~M., \& Cohen, J.~G.\ 1996, AJ, 112, 839 
\bibitem[Crowl \& Kenney(2008)]{crowl08} Crowl, H.~H., \& Kenney, J.~D.~P.\ 2008, AJ, 136, 1623 
\bibitem[Daddi et al.(2007)]{daddi07} Daddi, E., Dickinson, M., Morrison, G., et al.\ 2007, ApJ, 670, 156 
\bibitem[Davis et al.(2011)]{davis11} Davis, T.~A., Alatalo, K., Sarzi, M., et al.\ 2011, MNRAS, 417, 882 
\bibitem[Davis et al.(2013)]{davis13} Davis, T.~A., Alatalo, K., Bureau, M., et al.\ 2013, MNRAS, 429, 534 
\bibitem[Davis et al.(2014)]{davis14} Davis, T.~A., Young, L.~M., Crocker, A.~F., et al.\ 2014, MNRAS, 444, 3427
\bibitem[Dekel \& Birnboim(2006)]{dekel06} Dekel, A., \& Birnboim, Y.\ 2006, MNRAS, 368, 2 
\bibitem[de La Rosa et al.(2011)]{delarosa11} de La Rosa, I.~G., La Barbera, F., Ferreras, I., \& de Carvalho, R.~R.\ 2011, MNRAS, 418, L74 
\bibitem[de Zeeuw et al.(2002)]{dezeeuw02} de Zeeuw, P.~T., Bureau, M., Emsellem, E., et al.\ 2002, MNRAS, 329, 513 
\bibitem[Elbaz et al.(2007)]{elbaz07} Elbaz, D., Daddi, E., Le Borgne, D., et al.\ 2007, A\&A, 468, 33 
\bibitem[Emsellem et al.(2007)]{emsellem07} Emsellem, E., Cappellari, M., Krajnovi{\'c}, D., et al.\ 2007, MNRAS, 379, 401 
\bibitem[Emsellem et al.(2011)]{emsellem11} Emsellem, E., et al.\ 2011, MNRAS, 414, 888
\bibitem[Franx et al.(2008)]{franx08} Franx, M., van Dokkum, P.~G., Schreiber, N.~M.~F., et al.\ 2008, ApJ, 688, 770 
\bibitem[Gallazzi et al.(2006)]{gallazzi06} Gallazzi, A., Charlot, S., Brinchmann, J., \& White, S.~D.~M.\ 2006, MNRAS, 370, 1106 
\bibitem[Girardi et al.(2000)]{girardi00} Girardi, L., Bressan, A., Bertelli, G., \& Chiosi, C.\ 2000, A\&AS, 141, 371 
\bibitem[Gonz{\'a}lez(1993)]{gonzalez93} Gonz{\'a}lez, J.~J.\ Ph.D.~Thesis, University of California, Santa Cruz, 1993 (Source: Dissertation Abstracts International, Volume: 54-05, Section: B, page: 2551)
\bibitem[Graves \& Schiavon(2008)]{graves08} Graves, G.~J., \& Schiavon, R.~P.\ 2008, ApJS, 177, 446 
\bibitem[Graves et al.(2009a)]{graves09a} Graves, G.~J., Faber, S.~M., \& Schiavon, R.~P.\ 2009a, ApJ, 693, 486 
\bibitem[Graves et al.(2009b)]{graves09b} Graves, G.~J., Faber, S.~M., \& Schiavon, R.~P.\ 2009b, ApJ, 698, 1590
\bibitem[Graves \& Faber(2010)]{graves10} Graves, G.~J., \& Faber, S.~M.\ 2010, ApJ, 717, 803 
\bibitem[Gunawardhana et al.(2013)]{gunawardhana13} Gunawardhana, M.~L.~P., Hopkins, A.~M., Bland-Hawthorn, J., et al.\ 2013, MNRAS, 433, 2764 
\bibitem[Gunn \& Gott(1972)]{gunn72} Gunn, J.~E., \& Gott, J.~R., III 1972, ApJ, 176, 1 
\bibitem[Heavens et al.(2004)]{heavens04} Heavens, A., Panter, B., Jimenez, R., \& Dunlop, J.\ 2004, Nature, 428, 625 
\bibitem[Hilz et al.(2012)]{hilz12} Hilz, M., Naab, T., Ostriker, J.~P., et al.\ 2012, MNRAS, 425, 3119 
\bibitem[Hilz et al.(2013)]{hilz13} Hilz, M., Naab, T., \& Ostriker, J.~P.\ 2013, MNRAS, 429, 2924 
\bibitem[Hirschmann et al.(2013)]{hirschmann13} Hirschmann, M., Naab, T., Dav{\'e}, R., et al.\ 2013, MNRAS, 436, 2929 
\bibitem[Hopkins et al.(2010)]{hopkins10} Hopkins, P.~F., Bundy, K., Hernquist, L., Wuyts, S., \& Cox, T.~J.\ 2010, MNRAS, 401, 1099 
\bibitem[Johansson et al.(2012)]{johansson12} Johansson, J., Thomas, D., \& Maraston, C.\ 2012, MNRAS, 421, 1908
\bibitem[Jones \& Worthey(1995)]{jones95} Jones, L.~A., \& Worthey, G., 1995, ApJL, 446, L31
\bibitem[J{\o}rgensen(1999)]{jorgensen99} J{\o}rgensen, I.\ 1999, MNRAS, 306, 607 
\bibitem[Juneau et al.(2005)]{juneau05} Juneau, S., Glazebrook, K., Crampton, D., et al.\ 2005, ApJL, 619, L135 
\bibitem[Kauffmann et al.(2003)]{kauffmann03} Kauffmann, G., Heckman, T.~M., White, S.~D.~M., et al.\ 2003, MNRAS, 341, 54 
\bibitem[Kaviraj et al.(2007)]{kaviraj07} Kaviraj, S., Schawinski, K., Devriendt, J.~E.~G., et al.\ 2007, ApJS, 173, 619 
\bibitem[Khochfar \& Burkert(2003)]{khochfar03} Khochfar, S., \& Burkert, A.\ 2003, ApJL, 597, L117 
\bibitem[Khochfar \& Silk(2006a)]{khochfar06a} Khochfar, S., \& Silk, J.\ 2006a, ApJL, 648, L21 
\bibitem[Khochfar \& Silk(2006)b]{khochfar06b} Khochfar, S., \& Silk, J.\ 2006b, MNRAS, 370, 902 
\bibitem[Khochfar \& Ostriker(2008)]{khochfar08} Khochfar, S., \& Ostriker, J.~P.\ 2008, ApJ, 680, 54 
\bibitem[Khochfar \& Silk(2009)]{khochfar09} Khochfar, S., \& Silk, J.\ 2009, ApJL, 700, L21 
\bibitem[Khochfar et al.(2011)]{khochfar11} Khochfar, S., Emsellem, E., Serra, P., et al.\ 2011, MNRAS, 417, 845 
\bibitem[King (1985)]{king85} King D. L., 1985, Technical Note 31, Atmospheric Extinction at the Roque de los Muchachos Observatory, La Palma. The Royal Greenwich Observatory (RGO)
\bibitem[Koopmann \& Kenney(2004)]{koopmann04} Koopmann, R.~A., \& Kenney, J.~D.~P.\ 2004, ApJ, 613, 866  
\bibitem[Krajnovi{\'c} et al.(2006)]{krajnovic06} Krajnovi{\'c}, D., Cappellari, M., de Zeeuw, P.~T., \& Copin, Y.\ 2006, MNRAS, 366, 787 
\bibitem[Krajnovi{\'c} et al.(2011)]{krajnovic11} Krajnovi{\'c}, D., Emsellem, E., Cappellari, M., et al.\ 2011, MNRAS, 414, 2923 
\bibitem[Kuntschner et al.(2006)]{kuntschner06} Kuntschner, H., et al. 2006, MNRAS, 369, 497
\bibitem[Kuntschner et al.(2010)]{kuntschner10} Kuntschner, H., Emsellem, E., Bacon, R., et al.\ 2010, MNRAS, 408, 97 
\bibitem[Landsman(1993)]{landsman93} Landsman, W.~B.\ 1993, Astronomical Data Analysis Software and Systems II, 52, 246 
\bibitem[Lauer et al.(1995)]{lauer95} Lauer, T.~R., Ajhar, E.~A., Byun, Y.-I., et al.\ 1995, AJ, 110, 2622
\bibitem[McDermid et al.(2006)]{mcdermid06} McDermid, R.~M., et al.\ 2006, MNRAS, 373, 906 
\bibitem[McDermid et al.(2014)]{mcdermid14} McDermid, R.~M., Cappellari, M., Alatalo, K., et al.\ 2014, ApJL, 792, L37
\bibitem[Naab et al.(2009)]{naab09} Naab, T., Johansson, P.~H., \& Ostriker, J.~P.\ 2009, ApJL, 699, L178 
\bibitem[Naab et al.(2014)]{naab14} Naab, T., Oser, L., Emsellem, E., et al.\ 2014, MNRAS, 444, 3357 
\bibitem[Navarro-Gonz{\'a}lez et al.(2013)]{navarro13} Navarro-Gonz{\'a}lez, J., Ricciardelli, E., Quilis, V., \& Vazdekis, A.\ 2013, MNRAS, 436, 3507 
\bibitem[Noeske et al.(2007)]{noeske07} Noeske, K.~G., Weiner, B.~J., Faber, S.~M., et al.\ 2007, ApJL, 660, L43 
\bibitem[Oser et al.(2010)]{oser10} Oser, L., Ostriker, J.~P., Naab, T., Johansson, P.~H., \& Burkert, A.\ 2010, ApJ, 725, 2312 
\bibitem[Panter et al.(2007)]{panter07} Panter, B., Jimenez, R., Heavens, A.~F., \& Charlot, S.\ 2007, MNRAS, 378, 1550 
\bibitem[Peng et al.(2010)]{peng10} Peng, Y.-j., Lilly, S.~J., Kova{\v c}, K., et al.\ 2010, ApJ, 721, 193 
\bibitem[Press et al.(1992)]{press92} Press, W.~H., Teukolsky, S.~A., Vetterling, W.~T., \& Flannery, B.~P.\ 1992, Cambridge: University Press
\bibitem[Proctor et al.(2004)]{proctor04} Proctor, R.~N., Forbes, D.~A., \& Beasley, M.~A.\ 2004, MNRAS, 355, 1327 
\bibitem[S{\'a}nchez-Bl{\'a}zquez et al.(2006)]{blazquez06} S{\'a}nchez-Bl{\'a}zquez, P., et al., 2006, MNRAS, 371, 703 
\bibitem[\protect\citeauthoryear{Sarzi et al.}{2006}]{sarzi06} Sarzi M., et al., 2006, MNRAS, 366, 1151
\bibitem[Scott et al.(2013)]{scott13a} Scott, N., Cappellari, M., Davies, R.~L., et al.\ 2013, MNRAS, 432, 1894 
\bibitem[Scott et al.(2013)]{scott13b} Scott, N., Graham, A.~W., \& Schombert, J.\ 2013, ApJ, 768, 76 
\bibitem[Schawinski et al.(2007)]{schawinski07} Schawinski, K., Kaviraj, S., Khochfar, S., et al.\ 2007, ApJS, 173, 512 
\bibitem[Schiavon(2007)]{schiavon07} Schiavon, R.~P.\ 2007, ApJS, 171, 146 
\bibitem[Serra \& Trager(2007)]{serra07} Serra, P., \& Trager, S.~C.\ 2007, MNRAS, 374, 769 
\bibitem[Serra et al.(2012)]{serra12} Serra, P., Oosterloo, T., Morganti, R., et al.\ 2012, MNRAS, 422, 1835 
\bibitem[Shankar et al.(2010)]{shankar10} Shankar, F., Marulli, F., Bernardi, M., Dai, X., Hyde, J.~B., \& Sheth, R.~K.\ 2010, MNRAS, 403, 117 
\bibitem[Shapiro et al.(2010)]{shapiro2010} Shapiro, K.~L., et al.\ 2010, MNRAS, 402, 2140 
\bibitem[Smith et al.(2008)]{smith2008} Smith, R.~J., et al.\  2008, MNRAS, 386, L96 
\bibitem[Thomas et al.(2003)]{thomas03} Thomas, D., Maraston, C., \& Bender, R.\ 2003, MNRAS, 339, 897 
\bibitem[Thomas et al.(2004)]{thomas04} Thomas, D., Maraston, C., \& Korn, A.\ 2004, MNRAS, 351, L19 
\bibitem[Thomas et al.(2005)]{thomas05} Thomas, D., Maraston, C., Bender, R., \& Mendes de Oliveira, C.\ 2005, ApJ, 621, 673 
\bibitem[Thomas et al.(2010)]{thomas10} Thomas, D., Maraston, C., Schawinski, K., Sarzi, M., \& Silk, J.\ 2010, MNRAS, 404, 1775 
\bibitem[Thomas et al.(2011)]{thomas11} Thomas, D., Maraston, C., \& Johansson, J.\ 2011, MNRAS, 412, 2183 
\bibitem[Trager et al.(1998)]{trager98} Trager, S.~C., Worthey, G., Faber, S.~M., Burstein, D., \& Gonzalez, J.~J.\ 1998, ApJS, 116, 1 
\bibitem[Trager et al.(2000)]{trager2000} Trager, S.~C., Faber, S.~M., Worthey, G., \& Gonz{\'a}lez, J.~J.\ 2000, AJ, 119, 1645 
\bibitem[Tran et al.(2010)]{tran10} Tran, K.-V.~H., Papovich, C., Saintonge, A., et al.\ 2010, ApJL, 719, L126 
\bibitem[van der Wel et al.(2009)]{vdwel09} van der Wel, A., Bell, E.~F., van den Bosch, F.~C., Gallazzi, A., \& Rix, H.-W.\ 2009, ApJ, 698, 1232 
\bibitem[van Dokkum \& Conroy(2010)]{vandokkum10} van Dokkum, P.~G., \& Conroy, C.\ 2010, Nature, 468, 940 
\bibitem[Vazdekis et al.(2012)]{vazdekis12} Vazdekis, A., Ricciardelli, E., Cenarro, A.~J., et al.\ 2012, MNRAS, 424, 157 
\bibitem[V{\'e}ron-Cetty \& V{\'e}ron(2010)]{veroncetty10} V{\'e}ron-Cetty, M.-P., \& V{\'e}ron, P.\ 2010, A\&A, 518, A10 
\bibitem[Villumsen(1983)]{villumsen83} Villumsen, J.~V.\ 1983, MNRAS, 204, 219 
\bibitem[Wake et al.(2012)]{wake12} Wake, D.~A., van Dokkum, P.~G., \& Franx, M.\ 2012, ApJL, 751, L44 
\bibitem[Whitaker et al.(2012)]{whitaker12} Whitaker, K.~E., van Dokkum, P.~G., Brammer, G., \& Franx, M.\ 2012, ApJL, 754, L29 
\bibitem[\protect\citeauthoryear{Worthey et al.}{1994}]{worthey94} Worthey G., Faber S.~M., Gonzalez J.~J., Burstein D., 1994, ApJS, 94, 687
\bibitem[York et al.(2000)]{york2000} York, D.~G., Adelman, J., Anderson, J.~E., Jr., et al.\ 2000, AJ, 120, 1579 
\bibitem[Young et al.(2008)]{young2008} Young, L.~M., Bureau, M., \& Cappellari, M.\ 2008, ApJ, 676, 317 
\bibitem[Young et al.(2011)]{young11} Young, L.~M., Bureau, M., Davis, T.~A., et al.\ 2011, MNRAS, 414, 940 
\bibitem[Young et al.(2014)]{young14} Young, L.~M., Scott, N., Serra, P., et al.\ 2014, MNRAS, 444, 3408 

\end{thebibliography}
{}

\appendix

%
%

\section[]{Line-strength calibration and errors from standard stars}
\label{app:stars}

For observations made before the installation of the VPH grating in \sauron\/ in 2004, we refer the reader to \cite{kuntschner06} for details. An extensive set of line-strength standard star observations were also obtained post-VPH installation to provide a calibration to the Lick system of index measurements, and to monitor run-to-run and inter-run variations. The stars were chosen to overlap with several well-known stellar libraries, including \cite{worthey94}, \cite{jones95} and MILES \citep{blazquez06}. In total, 67 stars were observed, with several repeat observations, giving a total number of 160 stellar spectra. Wherever possible, velocity measurements for the stars were derived from the matching MILES library spectrum, otherwise the closest spectral type from the MILES library was used. This minimized template mismatch effects from introducing large velocity errors, and subsequent errors in the index measurements. The stars do not suffer from the field coverage issue mentioned previously, so we include it here for reference.

\begin{figure}
 \begin{center}
  \epsfig{file=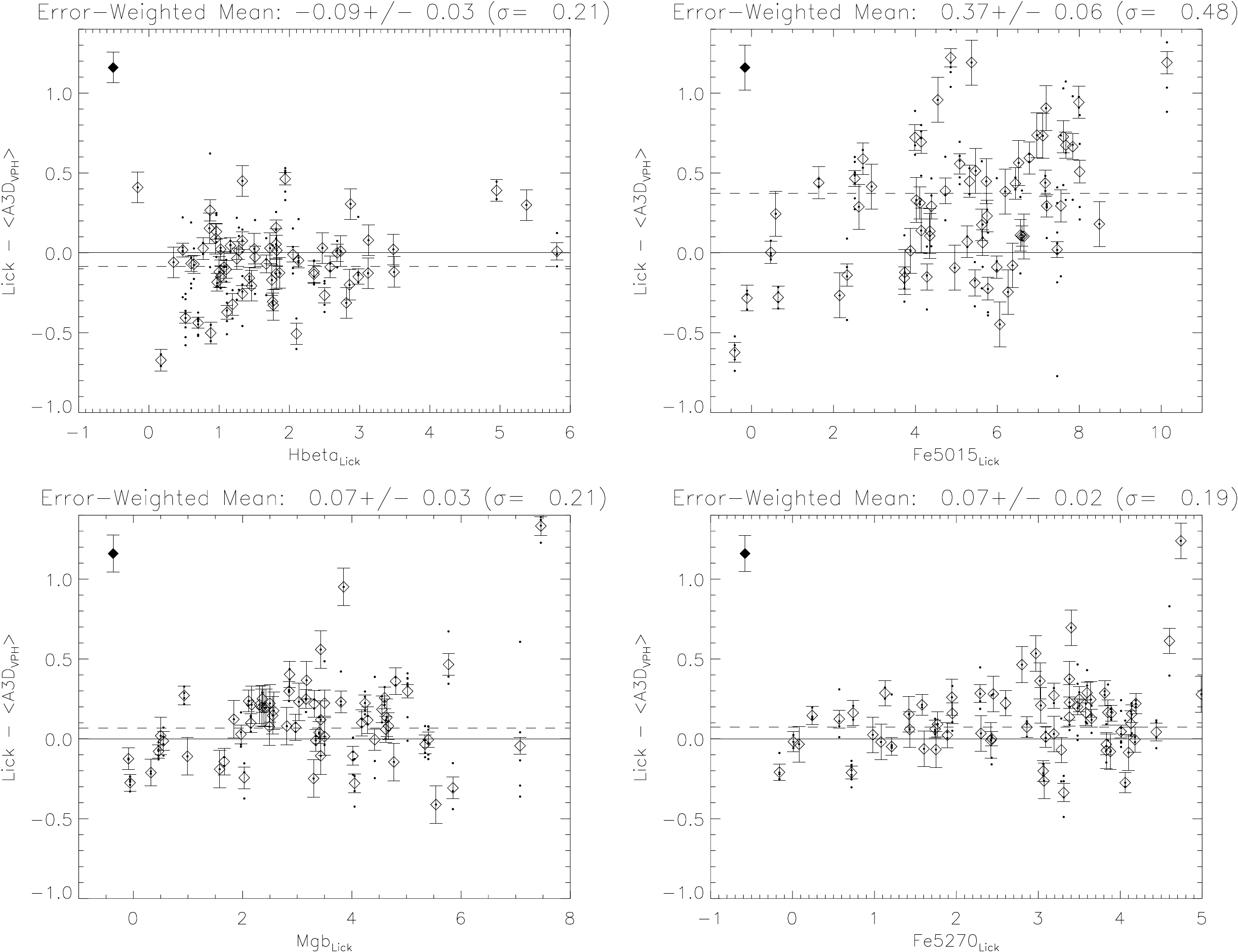, angle=0, width=9cm}
  \caption[]{Comparison of Lick indices measured from the 67 standard stars overlapping between the \atlas\/ survey and those from the Lick/IDS system \citep{worthey94}. The derived biweight-mean offsets (shown as a dashed line) and errors on the mean are given in Table \ref{tab:lick_offsets}. Dots show individual measurements; diamonds show individual and (where available) combined measurements. A representative error bar is plotted in the upper left of each panel showing the empirical scatter in repeat measurements from six stars having four or more observations over the four observing runs. This value is also tabulated in Table \ref{tab:lick_offsets}, and shows that the scatter in this comparison is largely driven by scatter in the original Lick measurements, which have an average uncertainty around twice as large.}
  \label{fig:lick_offsets}
  \end{center}
\end{figure}

Figure \ref{fig:lick_offsets} presents the difference between our measurements and the standard reference values of the Lick/IDS system \citep{worthey94}. Since some stars have several repeat measurements, and others do not, we use an error-weighted mean to derive the average offsets. Six stars were observed four times or more, upon which we base estimates of our intrinsic measurement errors in Figure \ref{fig:lick_offsets} and given in Table \ref{tab:errors}. This characteristic error was adopted for each individual measurement, and repeat measurements are averaged, with the characteristic error scaled by $(n-1)^{-1/2}$, where $n$ is the number of repeat measurements. We effectively ignore the intrinsic errors of the reference values themselves, which in fact are the dominant source of scatter (compare the errors in Table \ref{tab:errors} with the scatter in Figure \ref{fig:lick_offsets}, quantified by $\sigma$ in the panel titles). As with all figures in this section, we employ outlier-resistant `robust' estimators of the standard deviation.

\begin{figure}
 \begin{center}
  \epsfig{file=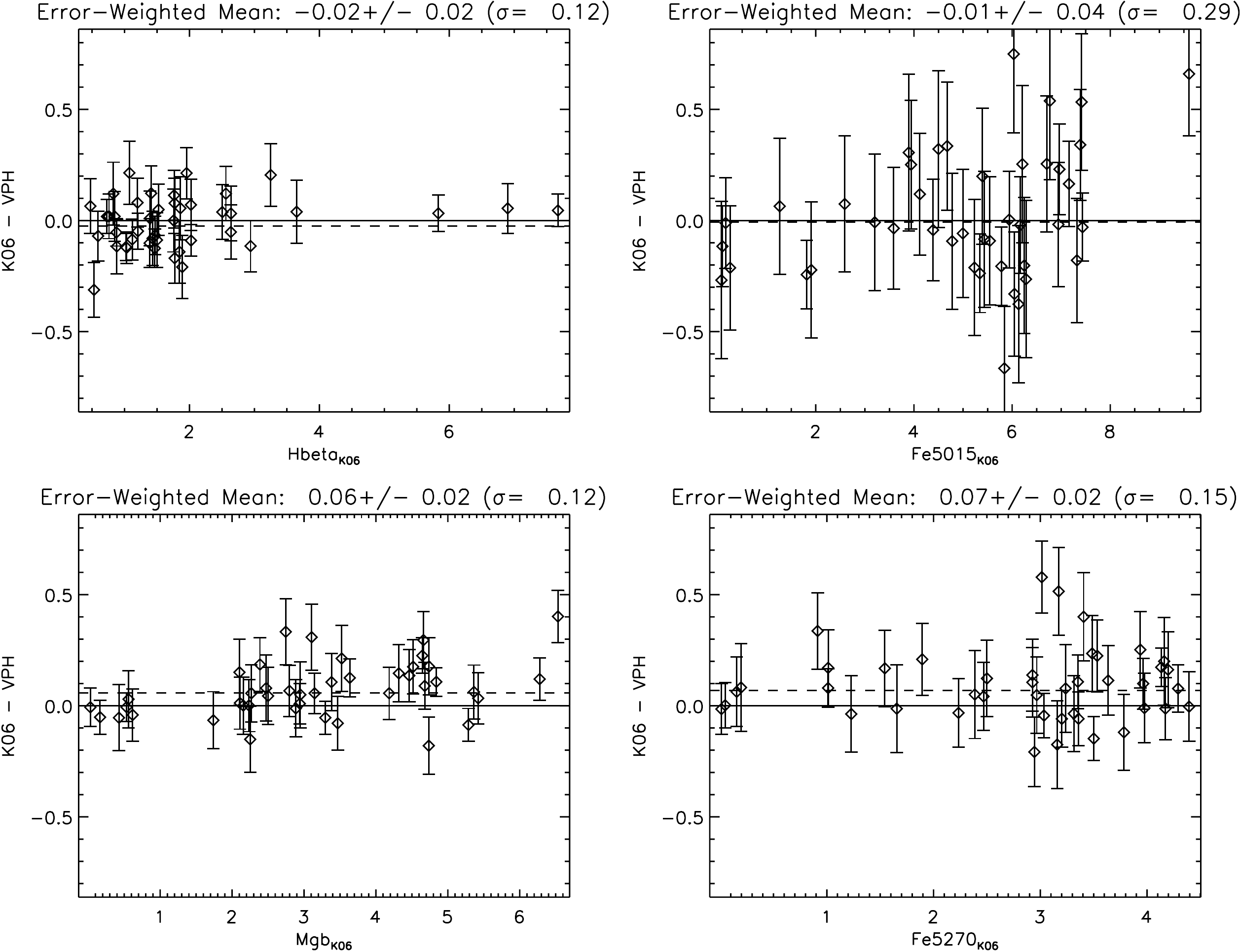, angle=0, width=9cm}
  \caption[]{Comparison of Lick indices measured from the 43 stars observed with \sauron\/ before \cite[from][]{kuntschner06} and after the installation of the VPH grating. Repeat measurements were averaged in both data sets, and corresponding errors combined. The plotted errors assume the errors come equally from both data sets, giving $\sigma_{\rm Overlap}$ in Table \ref{tab:errors}. Dashed line indicates the biweight mean.}
  \label{fig:compare_harald}
  \end{center}
\end{figure}

The average offsets from our post-VPH standard stars are consistent within the $1\sigma$ uncertainties of the pre-VPH `stars only' values of \citet[column 2 of their table 3]{kuntschner06}, with the exception of \mgb, for which we derive	a larger offset, indicating slightly weaker index values with the VPH grating. To further compare the pre- and post-VPH consistency of our \sauron\/ data, we compare measurements of the {\it same} stars measured with the two different gratings. Figure \ref{fig:compare_harald} shows this comparison, where the values have been averaged wherever possible. The mean offsets in this comparison are fully consistent with the differences found in the comparison with the Lick/IDS values, indicating that the small differences between the two gratings are real. We therefore apply {\it different} offsets for the pre- and post-VPH data. The final applied offsets are given in Table \ref{tab:lick_offsets}. We also derive an estimate of our uncertainties from this differential comparison, assuming that the scatter comes from equally-sized errors from both observations. The inferred values are also given in Table \ref{tab:errors}, and are in good agreement with the estimates from repeated stars.

\begin{figure}
 \begin{center}
  \epsfig{file=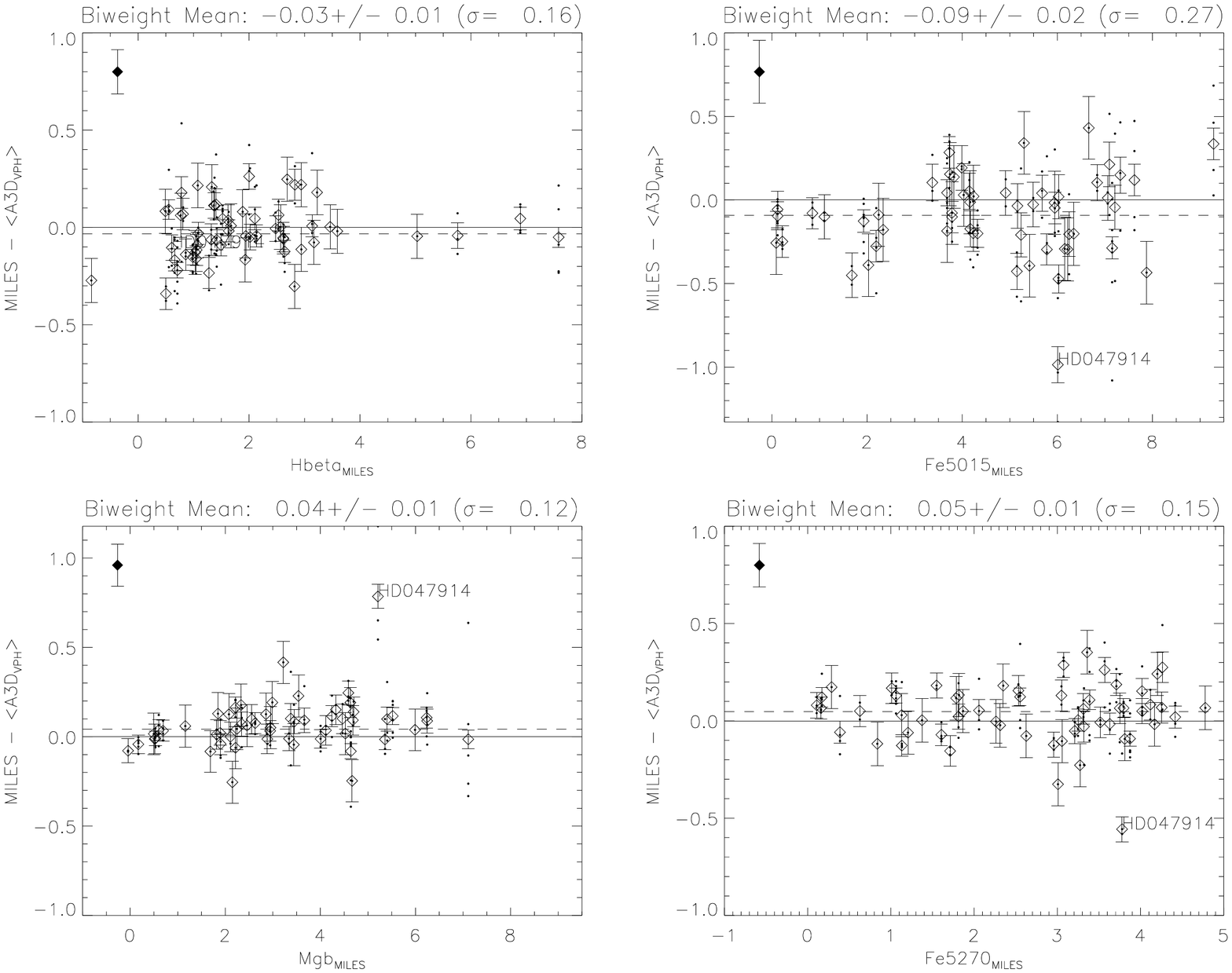, angle=0, width=9cm}
  \caption[]{Comparison of Lick indices measured from the 57 standard stars in common with the MILES library of \citet{blazquez06}. Small dots show all 147 individual measurements, and diamonds show individual and averaged values where repeat measurements are available. The mean, error on the mean and dispersion are given in the plot titles, which are computed from the {\it individual} values to provide a measure of our uncertainties. The representative error bar (filled symbol) assumes the scatter comes equally from both data sets.}
  \label{fig:compare_miles}
  \end{center}
\end{figure}

Whilst these error estimates are inferred from observations spanning several years, they are still 'internal' error estimates, coming from the same instrument and similar data reduction etc. As a way to check our 'external' errors, i.e. how we may compare with others, we present in Figure \ref{fig:compare_miles} a comparison of the Lick indices derived from our measurements (post-VPH) and from the MILES spectra for the same stars. The scatter is consistent with our internal comparison, suggesting that our error estimates are reasonable. Moreover, the mean offset and scatter are comparable to those found between targeted stellar libraries \cite[e.g. see][]{blazquez06}.

\begin{table}
 \begin{center}
  \caption{Estimate of Lick offsets derived from all stars.}
  \begin{tabular}{lcc}
  \hline
   Index  & Pre-VPH Offset  & Post-VPH Offset     \\
   \hline
\hb    & -0.13 $\pm$  0.05~\AA & -0.09 $\pm$  0.03~\AA \\
Fe5015 & +0.28 $\pm$  0.05~\AA & +0.37 $\pm$  0.06~\AA \\
\mgb   & +0.00 $\pm$  0.05~\AA & +0.07 $\pm$  0.03~\AA \\
Fe5270 & +0.00 $\pm$  0.05~\AA & +0.07 $\pm$  0.02~\AA \\
   \hline
  \label{tab:lick_offsets}
  \end{tabular}
 \end{center}
 Note: For each index, the biweight mean and dispersion estimate was used to reduce the influence of outlying values. The error on the mean offset derived from the $N$ observations is given as the dispersion scaled by $1/\sqrt{N}$. Pre-VPH offsets are taken from \citep{kuntschner06}.
\end{table}

\begin{table}
 \begin{center}
  \caption{Error estimates derived from the standard star measurements. $\sigma_{\rm Repeat}$ is derived from the mean standard deviation of six stars with four or more measurements. $\sigma_{\rm Overlap}$ is derived by assuming constant and equal errors between pre- and post-VPH observations of stars in common to both set-ups. $\sigma_{\rm MILES}$ is derived by comparing our Lick indices with those from MILES spectra for the same stars, and assuming the scatter comes from both data sets equally. The last column gives the average deviation, which we take as our minimum uncertainty in each index.}
  \begin{tabular}{lcccc}
  \hline
   Index  & $\sigma_{\rm Repeat}$ & $\sigma_{\rm Overlap}$  & $\sigma_{\rm MILES}$  & ${\bar \sigma}$ \\
   \hline
\hb         &  0.10~\AA  & 0.08~\AA  & 0.11~\AA & 0.10~\AA \\
Fe5015 &  0.14~\AA  & 0.21~\AA  & 0.19~\AA & 0.18~\AA \\
\mgb      &  0.12~\AA  & 0.08~\AA  & 0.08~\AA & 0.09~\AA \\
Fe5270 &  0.11~\AA  & 0.11~\AA  & 0.11~\AA & 0.11~\AA \\
   \hline
  \label{tab:errors}
  \end{tabular}
 \end{center}
\end{table}

%
%
\section{Comparison with other authors}
\label{app:literature}

\subsection{Comparison of line-strengths}

\begin{figure}
 \begin{center}
  \epsfig{file=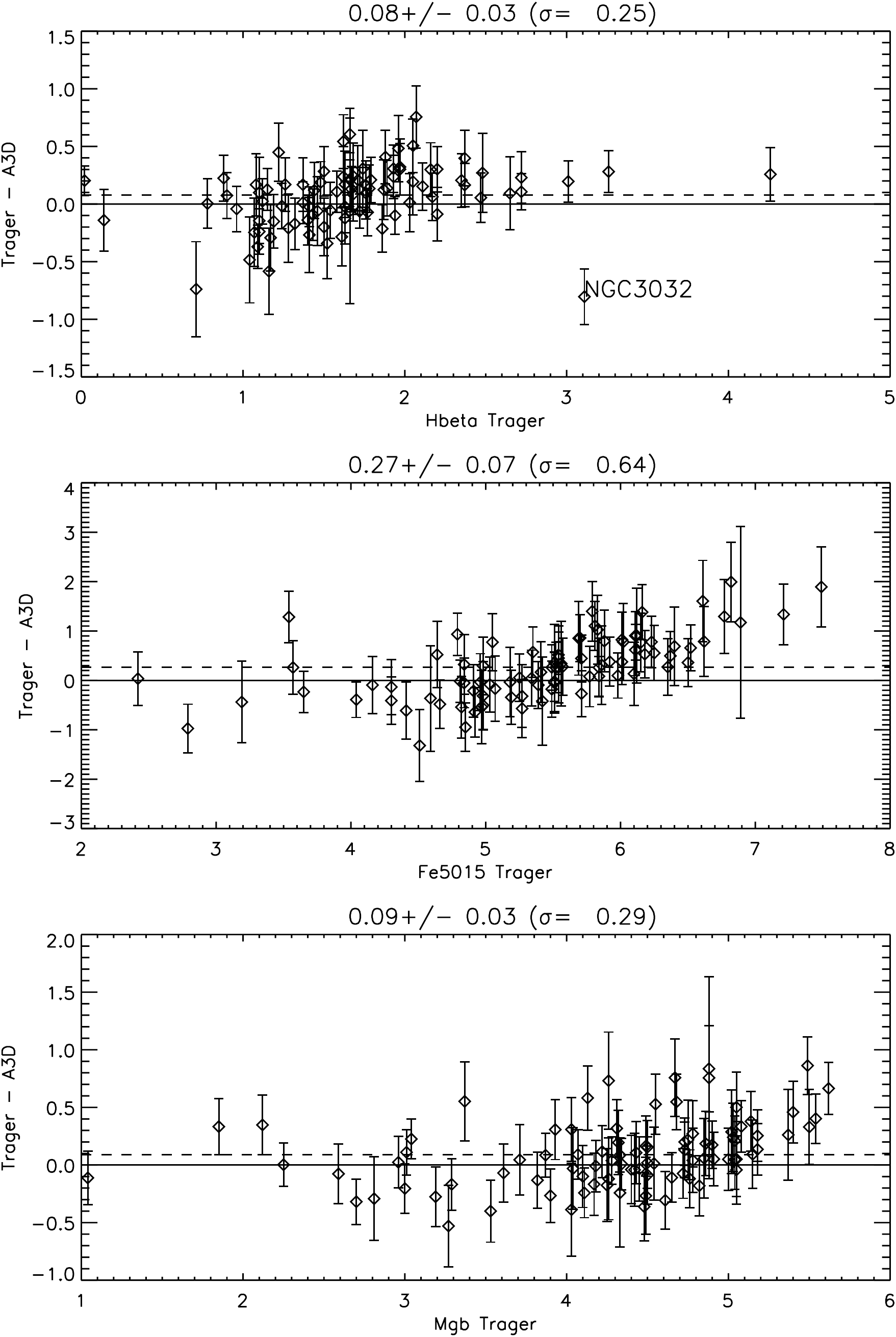, angle=0, width=8cm}
  \caption[]{Comparison with the Lick indices of \citet[T98]{trager98}. No correction for emission is made in either data set. The T98 data were measured in 1.4\arcsec $\times$ 4\arcsec\/ apertures of unknown orientation. The \atlas\/ data (A3D) are measured in circles of equivalent area (i.e. of radius 1.335\arcsec). Errors show the quadrature sum of both data set errors. Plot titles show the robust mean (also shown as a dashed line), error on the mean, and robust dispersion (in parentheses).}
  \label{fig:compare_trager98}
  \end{center}
\end{figure}

\begin{figure}
 \begin{center}
  \epsfig{file=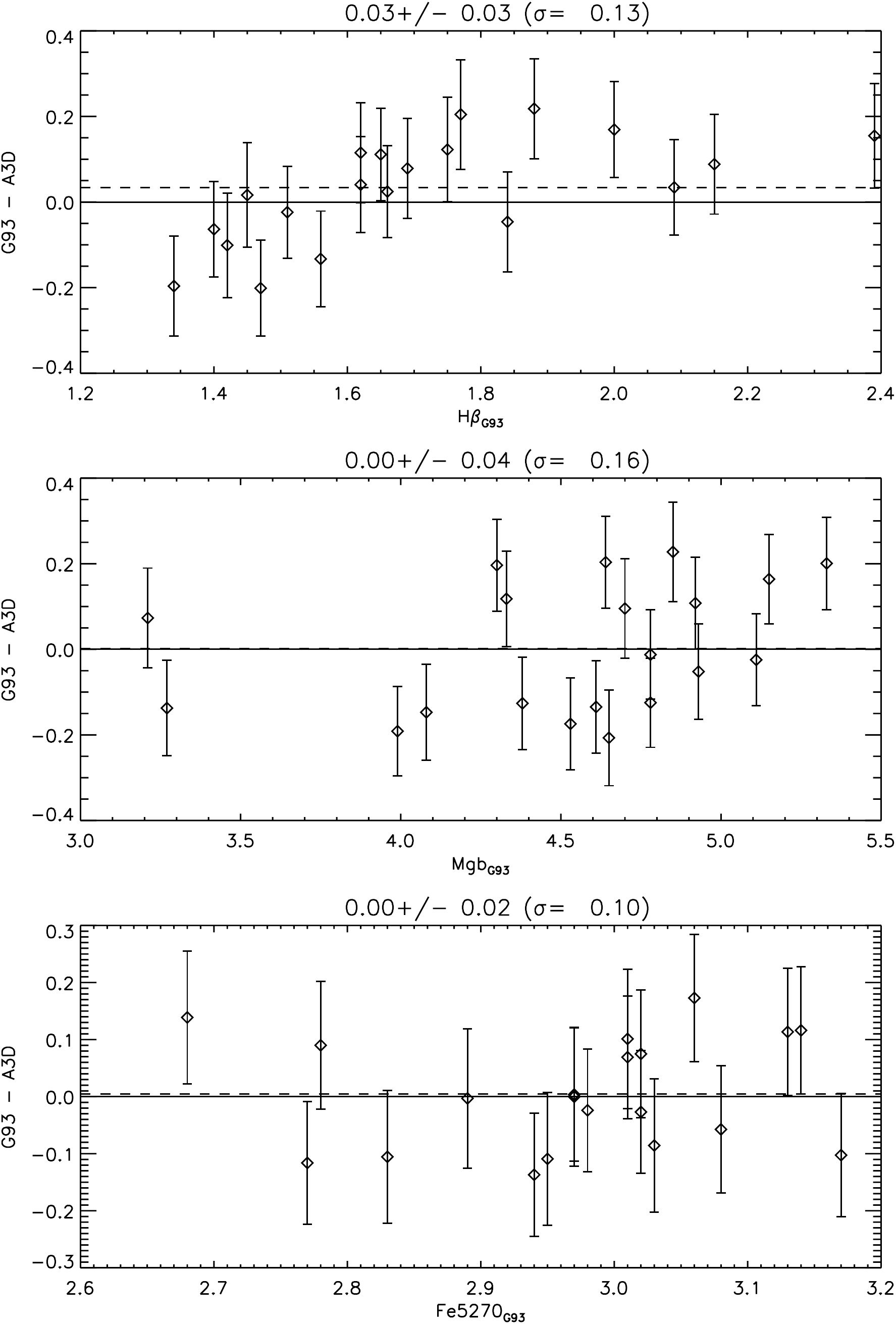, angle=0, width=8cm}
  \caption[]{Comparison with the indices of \citet[G93]{gonzalez93}. In both cases, \hb\/ is corrected for emission by adding 0.7 times the \oiii\/ emission. The G93 values were measured from one or two long-slit position angles. The A3D values were measured on circular apertures using the radii given in table 4.7 of G93, which give the luminosity-weighted circular aperture radius equivalent to the long-slit apertures. Errors, dashed lines and plot titles are as per Figure \ref{fig:compare_trager98}.}
  \label{fig:compare_gonzalez}
  \end{center}
\end{figure}

The dominant uncertainty in line strength measurements from high signal-to-noise ratio data such as is available for nearby galaxies are the {\it systematic} errors from data calibration, rather than statistical errors from noise. We therefore choose from the literature two studies that have reasonable overlap in sample galaxies, namely the extensive Lick/IDS sample of \citet{trager98} (90 galaxies in common), and the higher-quality data from \citet{gonzalez93} (20 galaxies in common).

For the \citet{trager98} data, as per \cite{kuntschner06} we extract a circular aperture with the same area as their $1.4$\arcsec$ \times 4$\arcsec (i.e. with a radius of 1.335\arcsec), since we do not know the position angle of the slits used. The \citet{trager98} data is also not corrected for emission, which is a major difference to our data. We therefore measure the comparison values with no correction for emission. Figure \ref{fig:compare_trager98} presents the comparison, with robust mean, error on the mean, and dispersion (in parenthesis) shown in the title. Observational errors for both data sets have been added in quadrature. The comparison is reasonable, though systematic differences can be seen, in particular an apparent trend with increasing Fe5015. Corrections for velocity dispersion are comparable between the two studies to $\sim 1$\%. The outlier in the \hb\/ comparison, NGC3032, demonstrates the sensitivity to such comparisons to PSF and aperture differences. This particular galaxy has an unresolved blue nucleus where \hb\/ strengthens considerably \citep{mcdermid06}, making the PSF a dominant uncertainty.

Figure \ref{fig:compare_gonzalez} presents the comparison with the \Reffe\/ aperture values from \cite{gonzalez93}, with titles and errors as per Figure \ref{fig:compare_trager98}. In this work, careful consideration of the aperture definitions was made, and the translation between the slit radial length and the corresponding circular aperture size are conveniently tabulated there (his table 4.7). Using this radius, we extract spectra within circles, and compare the measured line-strengths. In this case, a correction for \hb\/ emission was determined from the measured \oiii\/ emission strength, as described earlier. We therefore apply the same correction procedure, adding $0.7 \times$ \oiii\/ as measured using the GANDALF code to the measured \hb\/ index, with no additional emission correction.

Instead of comparing Fe5015 (which is not in the Gonzalez table), we compare Fe5270, which has the advantage of being free from emission corrections. To avoid the filter truncation effects described earlier, we convert our measured Fe5270$_S$ values to the Lick Fe5270 index following \cite{kuntschner06}: Fe5270 $= 1.28 \times $Fe5270$_S + 0.03$.

The comparison of all three comparable indices is rather good, with the mean difference in all three indices consistent with zero within the scatter. The effective reduced $\chi^2$ values are also close to unity, implying that the assumed errors are representative. The higher degree of consistency with \citet{gonzalez93} over that with \citet{trager98} likely reflects more the more accurate aperture matching with the latter study, rather than some intrinsic difference in the measurement quality (with respect to the errors).

\subsection{Comparison with MOSES sample}

\begin{figure}
 \begin{center}
  \epsfig{file=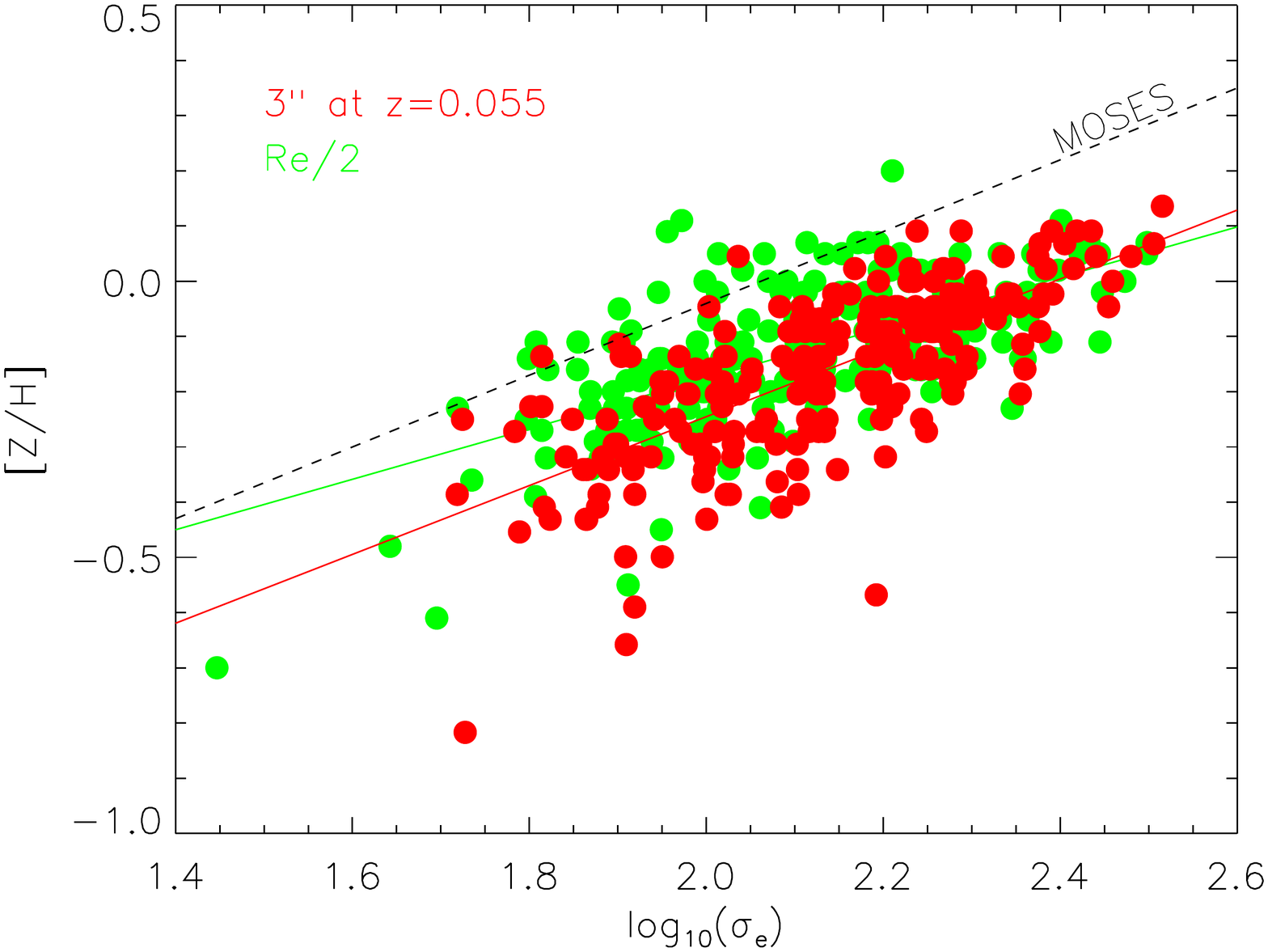, angle=0, width=9cm}
  \caption[]{Simulation of the effect of a fixed 3\arcsec\/ aperture on the slope of the relation between metallicity and velocity dispersion. Green points show the \Reff/2 values presented in Figure \ref{fig:ssp_sigma}. The red points were measured assuming a fixed aperture, scaled to mimic the redshift of the MOSES sample \cite{thomas10}. The effect of the fixed aperture is to include more of the lower metallicity regions in the outskirts of smaller, lower dispersion galaxies, thus steepening the slope compared to our values. Red and green lines show robust fits to the points of corresponding colour. The black dashed line is the relation from \citet{thomas10}. The slopes match within the uncertainties, but the offsets differ by 0.2\,dex.}
  \label{fig:moses}
  \end{center}
\end{figure}

In Section \ref{sec:ssp_sigma} we compared our relations of SSP parameters and velocity dispersion to those of \cite{thomas10}, and found that their metallicity relation was significantly steeper than ours. We can understand the difference in the metallicity trend by considering that no corrections were made by \citet{thomas10} for the fixed 3\arcsec\/ diameter fiber used. In Figure \ref{fig:moses} we simulate the effect of measuring metallicity for our sample with a 3\arcsec\/ fiber at a redshift matching the mean redshift of the MOSES study (0.055). We overplot robust linear fits, excluding objects younger than 3.2\,Gyr, which accounts for the redshift difference of our sample. The slopes agree within the errors, but a 0.2 dex offset in the relations remains. As in \citet{kuntschner06}, we find that applying the models of either \citet{schiavon07} or \citet{thomas05} yield very similar results, thus cannot explain the large metallicity offset (indeed, the offset becomes even slightly larger). Most likely it arises from the different combination of indices used by the two studies, but it is beyond the scope of this paper to explore this further.

%
%
\section{Generic aperture corrections from the \atlas\/ sample}
\label{app:appcor}

\begin{figure}
 \begin{center}
  \epsfig{file=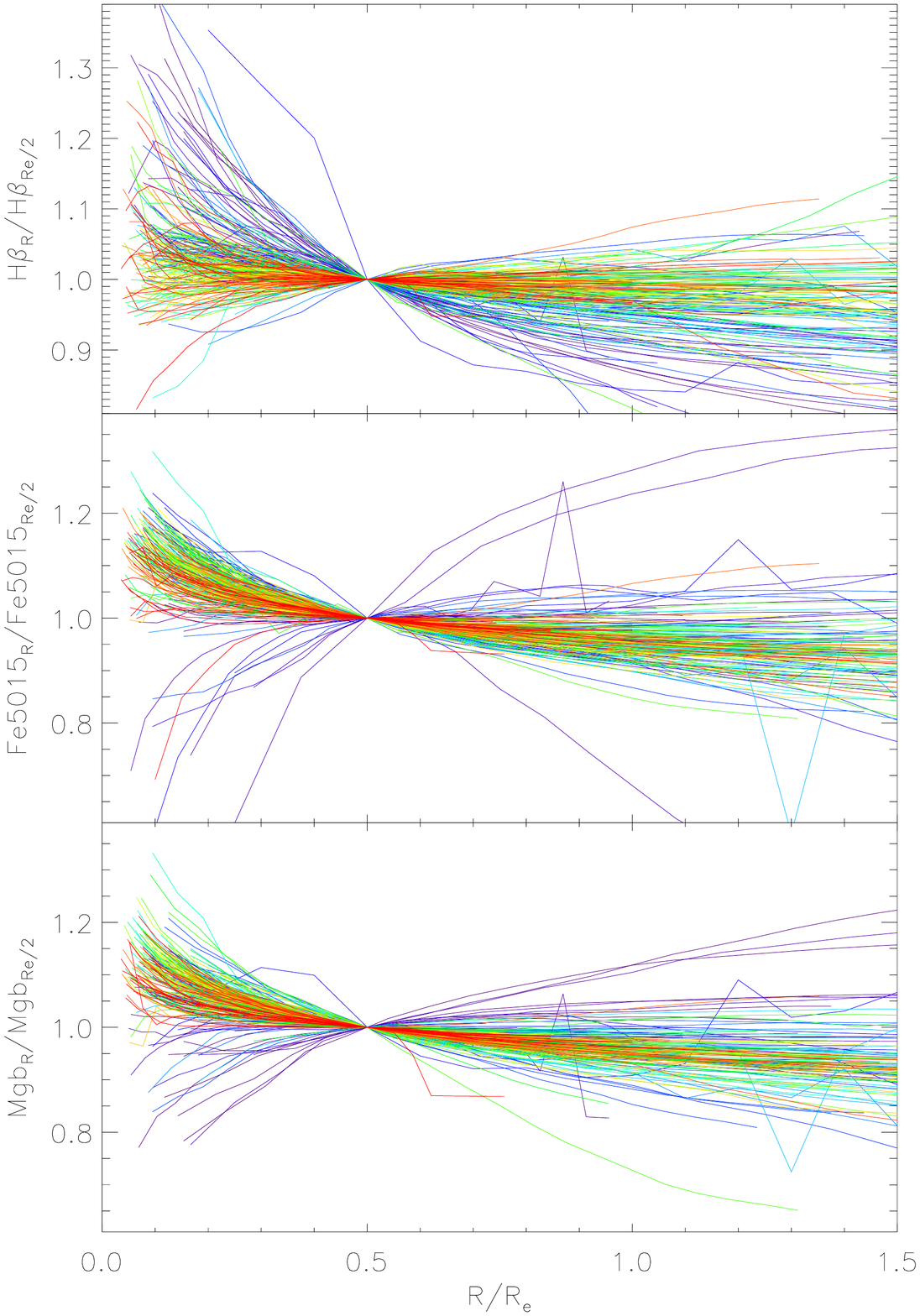, angle=0, width=9cm}
  \caption[]{Curves of growth for each of our line indices, measured within circular apertures of increasing radius, plotted as a function of the effective radius. The curves are normalized by the aperture value at \Reff/2. Colours indicate the velocity dispersion within the normalization radius, ordered low (blue) to high (red). The central 1.2\arcsec\/ are excluded, as are aperture sizes that are less than 85\% filled with spectra.}
  \label{fig:ap_corr}
  \end{center}
\end{figure}

Since we have the privilege of extensive two-dimensional information for a complete sample of galaxies, it is useful to investigate aperture corrections more generally, for use by studies without access to this spatial information. Figure \ref{fig:ap_corr} presents index growth curves as a function of \Reff\/ for all objects with maximal radial coverage of \Reff/2 or greater. The curves are normalized by the value at \Reff/2. The colour of the curves are ordered in velocity dispersion from low (blue) to high (red). We do not include Fe5270 in this analysis due to the limited field coverage issue mentioned above.

From this one can see differences between the red and blue curves, especially in the case of \hb\/ and \mgb. Comparing this figure to the almost equivalent Figure 16 of \citet{kuntschner06}, we see that there are many more galaxies with rising central \hb\/ strengths in our larger sample, and these galaxies tend to have lower velocity dispersions, although for a given velocity dispersion, there is a wide variety of slopes.

We attribute this behaviour to the presence of a centrally-concentrated population of young stars that give rise to a steep gradient in \hb. If this is not taken into account in the aperture correction, the strength of \hb\/ may be over-estimated. Figure \ref{fig:ap_corr_hb} plots the curve-of-growth slopes (measured from robust line fits to the coloured curves in Figure \ref{fig:ap_corr}) for all galaxies as a function of the \hb\/ line-strength measured within \Reff/8, for each of our three indices. There is clearly a relationship between the slopes of \hb\/ and \mgb\/ with the \Reff/8 \hb\/ line-strength, presumably reflecting the known age-sensitivity of these two lines. Thus, instead of providing a single aperture correction value for our three indices, we construct robust linear fits to the data in Figure \ref{fig:ap_corr_hb}, giving the corrections as a function of the \hb\/ lines-strength within an \Reff/8 aperture. Table \ref{tab:ap_corr} gives the coefficients of these linear fits, which can be used to derive the aperture correction for each index such that:

\begin{figure}
 \begin{center}
  \epsfig{file=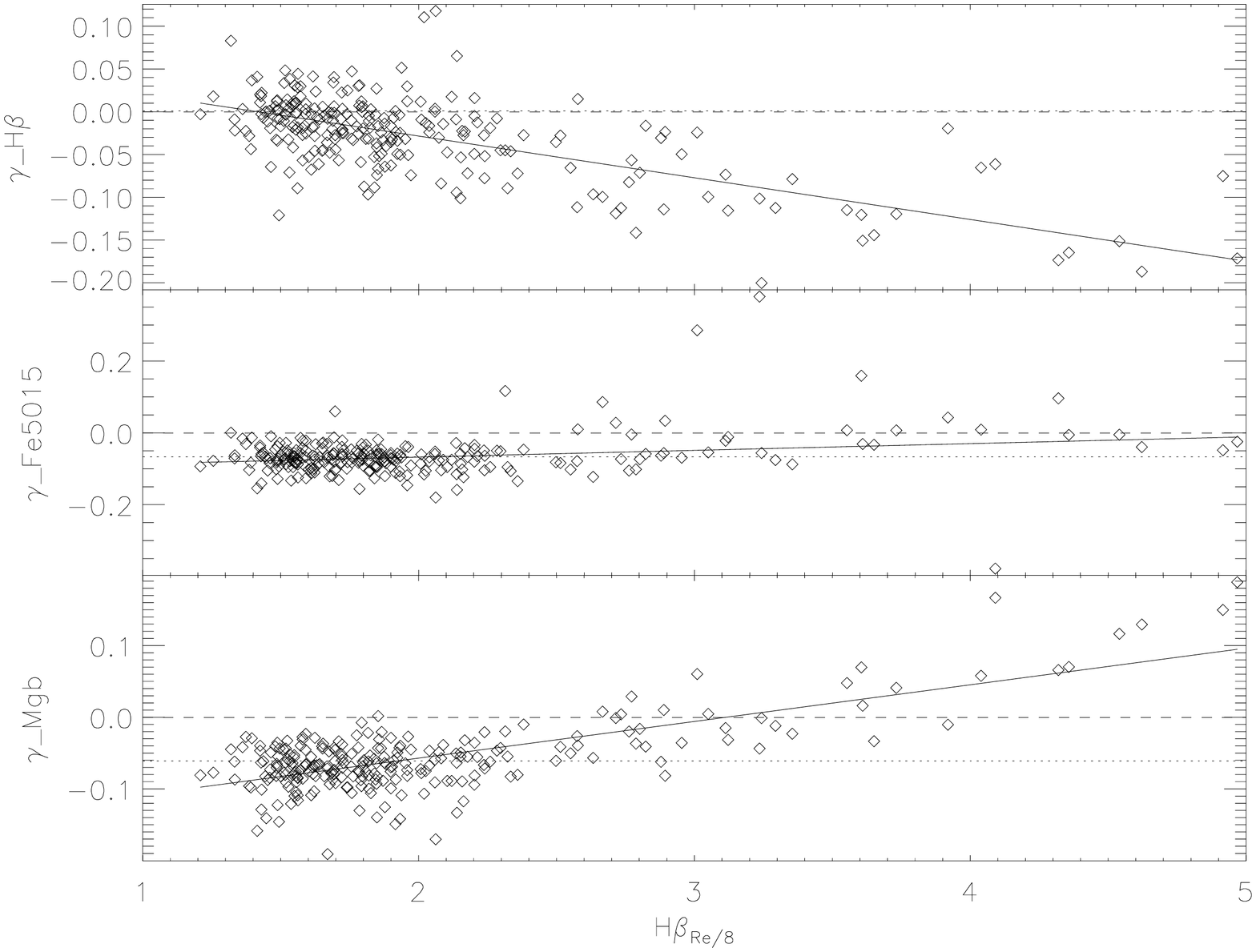, angle=90, width=9cm}
  \caption[]{The power-law slope values for fits to all the galaxies in Figure \ref{fig:ap_corr}, plotted as a function of \hb\/ line-strength within an \Reff/8 aperture. This illustrates how the size of aperture correction changes as a function of the central \hb\/ line-strength, due to the typically centrally-concentrated nature of the young stars. Dotted lines show the fixed corrections from \cite{kuntschner06}.}
  \label{fig:ap_corr_hb}
  \end{center}
\end{figure}

\begin{equation}
\mathrm{Index}_{R} / \mathrm{Index}_{norm} = (R/R_{norm})^{\gamma}
\end{equation}

\noindent where:

\begin{equation}
\gamma = a H\beta_{R_{e}/8} + b
\label{equ:ap_corr_hb}
\end{equation}

\noindent and $a$ and $b$ are given in Table \ref{tab:ap_corr}. The \hb\/ dependence is important for central \hb\/ values larger than around 2~\AA, and neglecting the effect in this regime will lead to over-estimated Balmer line strengths, and under-estimated \mgb. Below this threshold, however, the dependence is marginal, and adopting a constant correction (such as that of \cite{kuntschner06}, shown as dotted lines) gives similar results.

\begin{table}
 \begin{center}
  \caption{Coefficients of linear fits to aperture correction factors as a function of central \hb\/ strength, for use with equation \ref{equ:ap_corr_hb}.}
  \begin{tabular}{lcc}
  \hline
   Index & $a$  & $b$    \\
   \hline
   \hb         &   -0.052 &  0.069 \\
Fe5015 &   0.021 & -0.105 \\
\mgb      &    0.047 & -0.151 \\
   \hline
  \label{tab:ap_corr}
  \end{tabular}
 \end{center}
\end{table}

%
%
\section{Galaxies older than the Universe?}
\label{app:old_ages}

A number of galaxies show \hb\/ absorption values that imply ages older than the currently accepted age of the Universe \cite[$13.798\pm0.037$\,Gyr;][]{planckXVI}. Here we ask the question: Are these objects consistent with this upper limit given our observational uncertainties? Even though the SSP models make predictions up to 17\,Gyr, some data points still lie just outside the model grid. For this reason, we determine instead the differences of the measured \hb\/ to the \hb\/ predicted by the stellar population models for the age of the universe.

\begin{figure}
 \begin{center}
  \epsfig{file=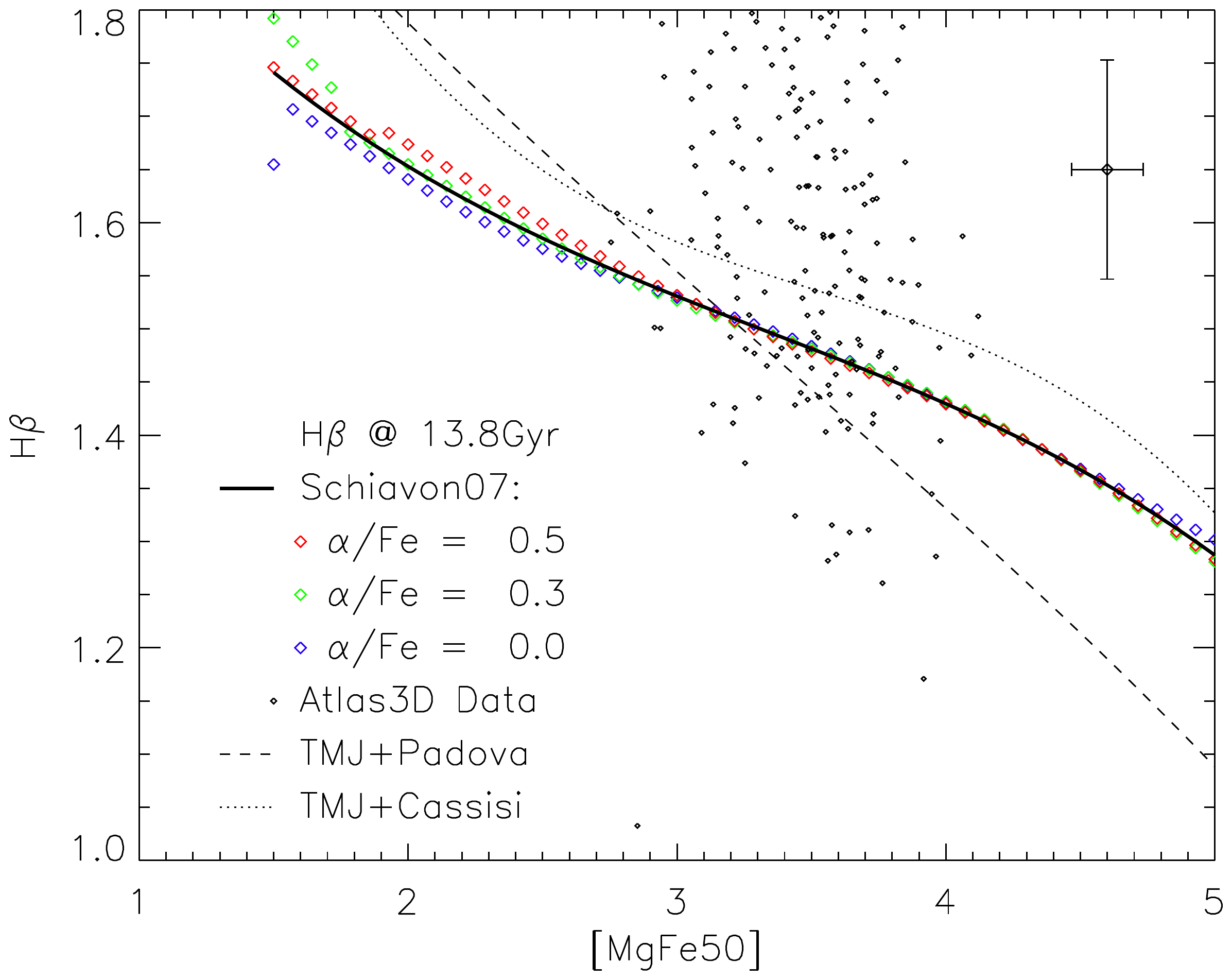, angle=0, width=8cm}
  \caption[]{Large open diamonds show the \citet{schiavon07} SSP model index predictions for a fixed age of 13.8\,Gyr, as a function of metallicity (increasing left to right) and abundance ratio (indicated by colour). The solid black line is a cubic fit to these predictions which describes the \hb\/ at this age as a function of \mgfe. This line is used to find the offset in \hb\/ between the measured points (small diamonds) and the canonical ages of the Universe, as shown in Figure \ref{fig:delta_hb}. For comparison, we also show the equivalent predictions from the models of \citet{thomas11}, using the stellar evolutionary tracks of \citet[dotted line]{cassisi97} and the Padova group \cite[dashed line]{girardi00}. The mean measurement error is also indicated in the upper-right corner.}
  \label{fig:wmap_grid}
  \end{center}
\end{figure}

We note that \hb\/ and age are not completely orthogonal, and there is a small but significant metallicity dependence. Abundance ratio dependences are small for the \citet{schiavon07} models used here. Figure \ref{fig:wmap_grid} shows the line-strength predictions at the WMAP age for all metallicities and abundance ratios in the model grid (coloured diamonds). The solid line shows a cubic fit to the model points, giving the mean predicted \hb\/ value as a function of \mgfe. For comparison, we also show equivalent cubic fits derived from the ``TMJ" models of \citet{thomas11} using the stellar evolutionary tracks of \citet[][``TMJ+Cassisi'']{cassisi97} and the Padova group \cite[][``TMJ+Padova'']{girardi00}. We shift all model predictions to the Lick system using the offsets from \citet{schiavon07}. This comparison shows that the data extend below predictions for the fiducial age for all models, though the issue is somewhat less for the TMJ+Padova models, and worse for the TMJ+Cassisi models.

\begin{figure}
 \begin{center}
  \epsfig{file=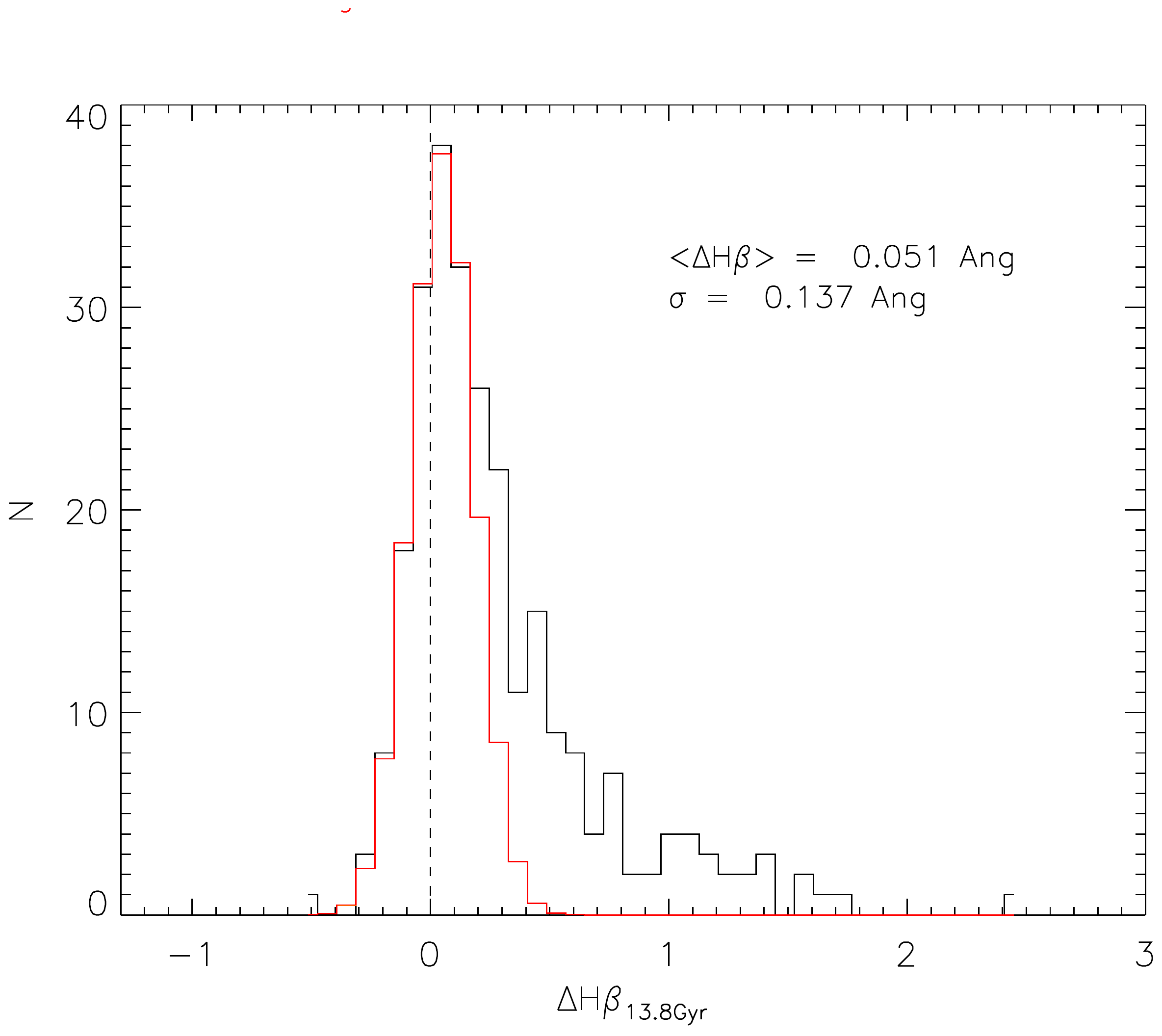, angle=0, width=8cm}
  \caption[]{Histogram showing the distribution of differences between the measured \hb\/ values and the model predictions for the canonical age of the Universe (13.8\,Gyr). The red line shows a Gaussian fitted to the lower portion of the distribution (fitted region is indicated), with a mean close to zero, and a dispersion similar to the observational error in \hb. From this we conclude that the galaxies appearing apparently older than the Universe are consistent with our measurement uncertainties.}
  \label{fig:delta_hb}
  \end{center}
\end{figure}

Figure \ref{fig:delta_hb} shows the distribution of differences between the observed \hb\/ values and the prediction for 13.8\,Gyr from the \citet{schiavon07} models used in this paper. Galaxies with negative differences have implied ages older than the Universe. We then fit a Gaussian to the lower portion of this distribution, and find that the tail of negative points can be described by objects with a slightly positive mean offset (i.e. younger than 13.8\,Gyr), and a dispersion of 0.137\AA. This is quite consistent with our estimate of the \hb\/ measurement error, which is assumed to have a minimum of 0.1\AA. Part of the additional observed dispersion may arise from uncertainties in the emission-line correction, in the form of emission lines below our detection limit, which is typically a few hundredths of an Angstrom at the signal-to-noise ratios considered here \citep{sarzi06}. We therefore conclude that the distribution of ages we measure are consistent with the fiducial age of the Universe after taking into account our measurement uncertainties, as well as systematic uncertainties in the models.

\label{lastpage}
\end{document}